\documentclass[sn-mathphys,Numbered]{sn-jnl}

\usepackage[latin1]{inputenc}
\usepackage{graphicx}%
\usepackage{multirow}%
\usepackage{amsmath,amssymb,amsfonts}%
\usepackage{amsthm}%
\usepackage{mathrsfs}%
\usepackage[title]{appendix}%
\usepackage[svgnames]{xcolor}
\usepackage{textcomp}%
\usepackage{manyfoot}%
\usepackage{booktabs}%
\usepackage{algorithm}%
\usepackage{algorithmicx}%
\usepackage{algpseudocode}%
\usepackage{listings}%
\usepackage{ulem}

\theoremstyle{thmstyleone}%
\theoremstyle{thmstyletwo}%
\newtheorem{remark}{Remark}%

\theoremstyle{thmstylethree}%
\newtheorem{definition}{Definition}%

\begin{document}

\title[The Resident Space Objects Network]{The Resident Space Objects Network: a complex system approach for shaping space sustainability}


\author*[2,3]{\fnm{Matteo} \sur{Romano}}\email{matteo.romano@unamur.be, matteo.romano@ext.esa.int}

\author[1,2]{\fnm{Timoteo} \sur{Carletti}}\email{timoteo.carletti@unamur.be}

\author[1,2]{\fnm{Jérôme} \sur{Daquin}}\email{jerome.daquin@unamur.be}

\affil[1]{\orgdiv{Department of Mathematics}, \orgname{Université de Namur}, \orgaddress{\street{Rue Grafé 2}, \city{Namur}, \postcode{5000}, \country{Belgium}}}

\affil*[2]{\orgdiv{Namur Institute of Complex Systems (naXys)}, \orgname{Université de Namur}, \orgaddress{\street{Rue Grafé 2}, \city{Namur}, \postcode{5000}, \country{Belgium}}}

\affil*[3]{\orgdiv{ESA Academy - Training and Learning Program}, \orgname{Space Applications Services NV/SA}, \orgaddress{\street{Leuvensesteenweg 325}, \city{Bruxelles, Zaventem}, \postcode{1932}, \country{Belgium}}}


\abstract{Near-Earth space continues to be the focus of critical services and capabilities provided to the society. With the steady increase of space traffic, the number of Resident Space Objects (RSOs) has recently boomed in the context of growing concern due to space debris. The need of a holistic and unified approach for addressing orbital collisions, assess the global in-orbit risk, and define sustainable practices for space traffic management has emerged as a major societal challenge. Here, we introduce and discuss a versatile framework  {rooted on the use of} the complex network paradigm to introduce  {a novel risk index} for space sustainability criteria.  {With an entirely data-driven, but flexible, formulation,} we introduce the Resident Space Object Network (RSONet) by connecting RSOs that experience near-collisions events over a finite-time window. The structural collisional properties of RSOs are thus encoded into the RSONet and analysed with the tools of network science. We formulate a geometrical index highlighting the key role of specific RSOs in building up the risk of collisions with respect to the rest of the population. Practical applications based on Two-Line Elements and Conjunction Data Message databases are presented.}




\maketitle

\section{Introduction}

\subsection{The problem of space traffic}

At the current date, the estimated number of objects orbiting Earth is of the order of millions, most of which smaller than 10 cm in size. Only about 35,000 of them are large enough to be regularly tracked by the US Space Surveillance Network and maintained in their catalogue, which includes objects larger than about 5-10 cm in low-Earth orbit (LEO) and 30 cm to 1 m at geostationary (GEO) altitudes\footnote{\url{https://www.esa.int/Space_Safety/Space_Debris/Space_debris_by_the_numbers} (Online; accessed 10-Dec-2023).}. About 9,000 objects represent intact and operational satellites\footnote{\url{https://www.esa.int/Space_Safety/Space_Debris/About_space_debris} (Online; accessed 10-Dec-2023).}. 

\begin{figure}[!tp]
\centering
\includegraphics[width=\textwidth,trim={0.75cm 0cm 1.75cm 0cm},clip]{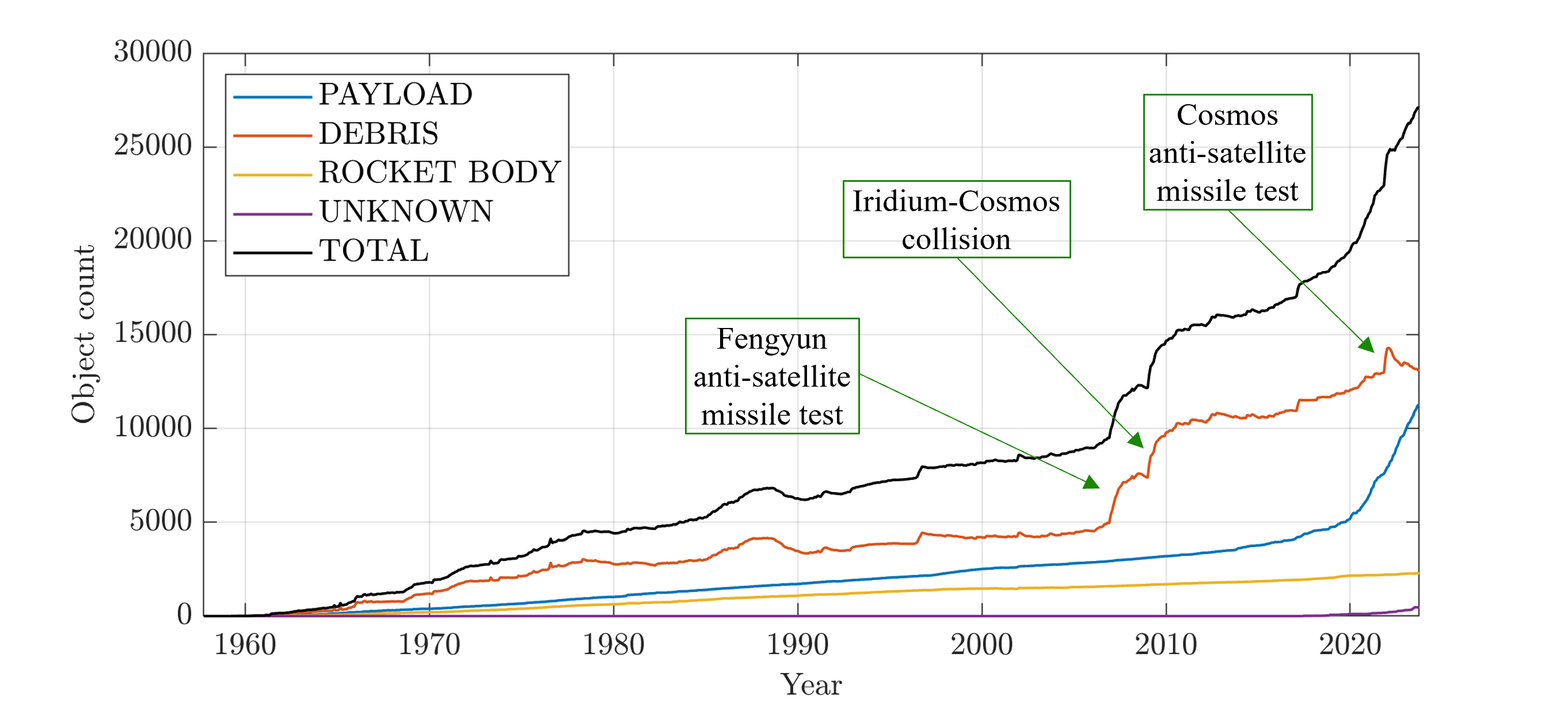}
\caption{Evolution in time of the amount and type of artificial objects orbiting Earth that can be detected from ground: around $30,000$ objects larger than $5-10$ cm are catalogued and tracked every day, while smaller objects are estimated to be orders of magnitude larger (data extracted from the Space-Track catalogue).}
\label{fig:debris}
\end{figure}

In the past decades, there has been a steady increase of the amount of objects in space, {as reported in \cite{ESA2023a}}.  The early 2000s saw a boom of fragmentation debris due to two main events: the Chinese anti-satellite test which lead to the voluntary destruction of the Fengyun-1C weather satellite in January 2007 \citep{kelso2007,pardini2007,lambert2018}, and the accidental collision of the U.S. Iridium 33 {and} Russian Cosmos 2251 satellites in February 2009 \citep{kelso2009,anselmo2009,pardini2017}. {While the launch rate of new satellites had maintained a slow, steady growth until a few years ago, it has recently accelerated to achieve an exponential growth, as visible in Fig.\,\ref{fig:debris}. Thanks to the ever more affordable access to space and miniaturisation of spacecraft components, commercial exploitation of space has become more affordable. At the same time, many of the satellites that were launched in the past and those that are operative now are left in space around the Earth, slowly breaking apart, exploding, or colliding with other objects.}

 {While} the amount of fragmentation debris has remained the major contributor to the population {until recently, currently active satellites represent the largest share of space traffic, so much that high-risk conjunctions with active and inactive payloads in lower altitudes exceed conjunctions with fragments \cite{ESA2023a,weber2023}.} In the latest years, space became more and more affordable to exploit as a resource, both scientifically and commercially as a platform to offer communication services. The amount of active satellites is quickly growing, with the resulting increase of complexity of the space environment around Earth. The most obvious example of this fact is the rise of megaconstellations, with planned projects such as OneWeb and Starlink where hundreds up to thousands of light satellites are being launched in LEO to provide high speed internet coverage to the whole globe \citep{mcdowell2020}.

 {In LEO, multiple} close approaches between catalogued space resident objects (RSOs) are notified to satellite operators per week\footnote{\url{https://leolabs-space.medium.com/quantifying-conjunction-risk-in-leo-e6eee8134211}}, in possible collision scenarios, while several fragmentation events occur spontaneously every year\footnote{\url{https://www.esa.int/Space_Safety/Space_Debris/Space_debris_by_the_numbers} (Online; accessed 10-Oct-2023).}, {further contributing to the debris population and thus to the risk of collisions with active satellites}. Following the current trends, the projected increase in conjunction rate and collision risk is expected to reach levels orders of magnitude higher in the next decade \citep{massey2020,acciarini2023}.

\subsection{State of the art}

{Space debris and space traffic management have become a concern from multiple viewpoints. On the one hand, it is a matter of safety, as the growth in space debris and space traffic makes orbit operations more hazardous and costly if frequent maneuvers are required to avoid other objects. On the other hand, it is a matter of sustainability, that is ensuring that space as a resource remains usable for future generations as it is for us now \cite{martinez2021copuos}.} The various approaches to {tackle} the problem of space debris and space traffic management can be split into two main groups: prevention, mitigation, and remediation procedures on the one hand \citep{wormnes2013,cattani2021,letizia2023a,letizia2023b}, and the use of the long-term dynamics of space objects on the other hand \citep{ely1996,breiter1999,gkolias2018,skoulidou2018,rossi2018,rossi2019,gondelach2019}.

As for the first approach, it includes all solutions aimed at reducing the number of space debris, either indirectly by preventing the formation of new ones, or directly by employing active debris removal (ADR) strategies. {Similarly, we can cast into this category studies aimed at predicting and preventing catastrophic collisions between existing objects, grounded on the development of methods to quantify collision risk \cite{maclay2021}.}

As for the second approach, understanding the secular dynamics of objects subject to Earth's gravitational influence and other natural orbital perturbations is the focus of numerous studies aiming at exploiting the natural evolution of orbits under perturbations {and resonances \cite{liou2020,esa2023b}} to either achieve an eventual re-entry \citep{valk2009,mLa11,mLa14,aCe18,emAl18,jDa22} or to permanently contain objects in long-term stable graveyard orbits \citep{skinner2022,IADC2023,ESA2023a} {and reduce the creation of and impact associated with space debris}.\\

This manuscript exploits network theory to represent the ``interactions" among RSOs.
Networks are used as descriptive and representative tools for several applications involving complex systems. Social and animal interactions \citep{landi2018}, contagion patterns \citep{dekemmeter2023}, reactions among chemical species, pattern emergence  \citep{muolo2019}, are only a few examples of systems falling in this framework and whose dynamics depends on the interactions between the members (i.e., the nodes) of these networks \citep{newman2003}.\\

The study of the interactions between objects in space becomes relevant for not only visualising the intricate patterns arising due to the possible collisions between them, but also to analyse how the collision risk evolves in time and propagates across the population when fragmentations occur. 
Thus, shifting the perspective from the individual events to the whole population, can provide new insight in the identification of the main actors when studying collision risk and of the possible targets to act upon for reducing this risk.

A first attempt to represent the space population as a network was done by Lewis et al. \cite{Lewis2010}. In their work, they establish a link between two nodes, i.e., two RSOs, if the collision probability during a conjunction between the corresponding objects exceeds a given threshold. They analyse the network statistics to identify objects which affect the most the rest of the network in terms of conjunction and collision probability, representing a threat for the population. Moreover, they establish a multi-relational network by defining different types of links and nodes (representing the relationships ``conjunction'', ``is a fragment of'', or ``is a member of'') to extract more information from those statistics and identify candidate objects whose removal from the network would benefit the overall in-orbit situation \citep{newland2012}.

An evolution of this model was carried out in the works by Acciarini et al. \cite{Acciarini2020,Acciarini2021} and Wang et al. \cite{wang2023}, who establish the network of RSOs using a two-layer temporal model, where a first layer captures the physical effects of collisions propagating across the network, while a secondary layer models the exchange of information among satellites via telecommunication. Both layers model dynamically the disruption of the network as nodes become inactive due to collisions, malfunctions, or deliberate attacks, with the goal of identifying liabilities and improve resilience of the space environment. Stevenson et al. \cite{stevenson2022} continue along the network path, by constructing a network using a three-filter approach similarly to Casanova et al. \cite{Casanova2014} and applying machine learning techniques to obtain efficient conjunction assessment and predict the existence of upcoming conjunction links over a given screening period, to eventually reduce the number of collision avoidance manoeuvres.

\subsection{A novel approach}

While the prediction of in-orbit conjunctions is already efficiently carried out by several organisms dedicated to the surveillance of Earth's orbital environment and collision avoidance practices, existing approaches  {generally} focus on the conjunction events individually, in an effort to prevent collisions with active satellites{, or on studying the space debris environment via analytical models to evaluate risk of collisions on a statistical level}. 

This manuscript introduces a novel measure of the risk of collisions within the population {of RSOs}. The close encounters between space residents are embedded into a network, whose topology and properties are analysed with the tools from graph theory and combined to define a metric to quantify how each object contributes to the overall risk of collisions with respect to the others. This framework is used {to study the patterns of interactions between the objects, which drive the risk of generating new debris, and} to eventually evaluate the sustainability of the current state of the population.

\begin{figure}[!ht]
\centering
\includegraphics[width=\textwidth,trim={2.5cm 1.75cm 1.75cm 1cm},clip]{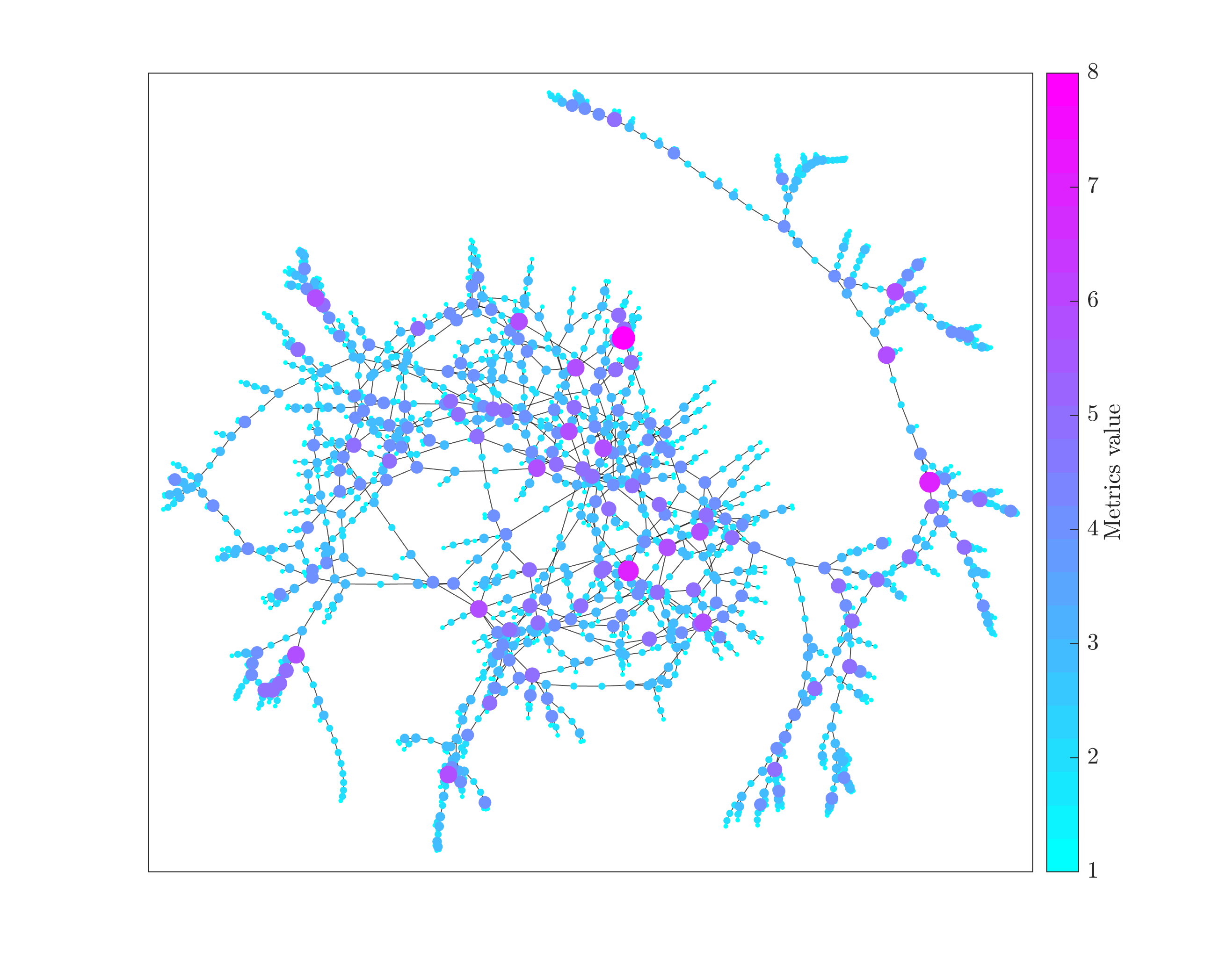}
\caption{Largest connected component of the network of RSOs. The size and the colour of the nodes  {correspond to} the values of the relevance metric introduced in Sect.\,\ref{sec:rating}.}
\label{fig:network}
\end{figure}

{Table\,\ref{table:existing_metrics} presents a selection of metrics that have been proposed to evaluate in-orbit collision risk, taken among those presented by McKnight et al. \cite{mcknight2021}. Notice that many existing models rely on the definition of risk as the combination of probability and consequences of the collision, e.g., McKnight, Anselmo \& Pardini, Letizia \& Lemmens, Dolado Perez \& Ruch, Colombo et al., where the probability of a collision is estimated statistically by considering the average density of object in a selected orbital region and the average lifetime for the specific object, while consequences of the collision are expressed by including the debris-generating mass.}

\begin{table*}[!h]
\begin{center}
\caption{{An overview of some of the existing metrics to evaluate the risk of collisions in orbit.}}
\label{table:existing_metrics} 
\begin{tabular}{ll} 
\hline
Authors & Description \\
\hline
McKnight et al. \cite{mcknight2017,mcknight2018} & \begin{minipage}[l]{0.65\textwidth} SMC rank, combines probability factors (collision rate, area, etc.) with consequence factors (mass, lifetime, satellite density) \end{minipage} \\
\hline
Anselmo \& Pardini \cite{anselmo2017} & \begin{minipage}[l]{0.65\textwidth} Normalised ranking combining probability factors (orbital debris flux, lifetime, mass) with consequence factors (fragment mass, decay time) \end{minipage} \\
\hline
Letizia et al. \cite{letizia2017} & \begin{minipage}[l]{0.65\textwidth} ECOB, combines the probability of a catastrophic collision involving an object across its lifetime due to simulated debris flux with a severity term \end{minipage} \\
\hline
Rossi et al. \cite{rossi2009} & \begin{minipage}[l]{0.65\textwidth} CSI, normalised index combining properties of the object (mass and lifetime) with spatial statistics of a selected orbital shell (density and inclination)  \end{minipage} \\
\hline
Lewis \cite{lewis2020} & \begin{minipage}[l]{0.65\textwidth} Ranking based on multiple metrics defined according to the average simulated collision probability and various properties of the object \end{minipage} \\
\hline
Dolado Perez \& Ruch \cite{ruch2020} & \begin{minipage}[l]{0.65\textwidth} Index combining ECOB and CSI with importance weights to rank objects  \end{minipage} \\
\hline
Jing, Dan \& Wang & \begin{minipage}[l]{0.65\textwidth} ADR selection based on mass, collision probability, and objects spatial density \end{minipage} \\
\hline
Colombo et al. \cite{muciaccia2023}  & \begin{minipage}[l]{0.65\textwidth} Index following the ECOB formulation to assess the effect of collisions and fragmentations along an object lifetime \end{minipage} \\
\hline
\end{tabular}
\end{center}
\end{table*}

The  approach {proposed here}  {differs from the ones cited above in multiple aspects. Firstly, it is data-driven, namely it does not rely on statistical models to estimate the rate of collisions, rather it uses already available information (CDMs) or computes the conjunctions between RSOs directly from orbital data (propagation of TLEs or otherwise) of individual objects. Secondly it does not consider the properties of the RSOs such as mass or lifetime, rather it relies only on the relationships existing between the RSOs to draw qualitative and quantitative conclusions about the state of risk of the population. Let us observe that this assumption can be easily relaxed to allow for a risk index taking into account also some RSO features. Lastly, while networks have been employed in the past, the proposed approach is novel} as it produces a network based on conjunctions before including the probability of collision, which is then used to weight the links. Moreover, instead of analysing the statistics of the network separately to judge which objects are the most dangerous or vulnerable, and thus possible candidates for removal, here the statistics are combined into a unique metric which summarises the relevance of each object with respect to the whole population.

An example of the resulting RSO network is represented in Fig.\,\ref{fig:network}, which illustrates the largest connected component of the RSO network; {let us notice that nodes size and shades of colors encode the proposed relevance score, allowing thus a straightforward identification of the most relevant objects}.

{The purpose of the new framework is twofold: on the one hand, it shifts the perspective for analysing in-orbit collision risk from individual events to a more global viewpoint related to the interactions between RSOs within the population; on the other hand, it provides a guide to define new metrics, such as the one proposed here, to quantify the risk of collisions via an entirely data-driven approach adaptable to various data formats and propagation tools, unlike other methods based on statistical models.} \\

The manuscript is structured as follows. Section\,\ref{sec:theory} presents the tools used for data collection, orbital propagation, and the network embedding process. Sect.\,\ref{sec:rating} introduces the definition of the new ranking score in an effort to assess space sustainability. Sect.\,\ref{sec:results} presents test cases to apply the ranking score to the network and comments upon the results. Finally, the conclusive Sect.\,\ref{sec:conclusions} sums up the main results and discusses our future work directions.

\section{The Resident Space Objects Network}
\label{sec:theory}
This section introduces the definition of the network of RSOs and the main steps and assumptions in building it, from the network embedding procedure up to the description of the initial data and the numerical propagator.\\

Let us consider two RSOs $x_{i}$ and $x_{j}$ of the chosen database of $N$ objects. We define the binary collision coefficient
\begin{align}
\label{eq:CollCoeff}
 c_{ij}(\epsilon,T) = \Theta\big(
\epsilon-\min_{0 \le t \le T}
\Vert
\mathbf{r}_{i}(t)-\mathbf{r}_{j}(t) \Vert_{2}
\big),
\end{align}
where $\Theta$ is the Heaviside step function, $\epsilon$ a positive real parameter, $T$ a time horizon, and $\Vert \bullet \Vert_{2}$ the Euclidean norm. This coefficient captures $\epsilon$-close encounter between two Cartesian geocentric RSOs' time-dependent position vectors $\mathbf{r}_{i}(t)$ and $\mathbf{r}_{j}(t)$ when moving on their respective orbits during the finite time window $[0,T]$.
The key-point consists in introducing a network from the near-collision matrix $\mathbf{C}(\epsilon,T)=(c_{ij})_{i,j}$. For this, we define the (binary) matrix 
\begin{align}
\label{eq:AdjMatrix}
\mathbf{A}(\epsilon,T) = \mathbf{C}(\epsilon,T) - \mathbf{I},
\end{align}
where $\mathbf{I}$ is the $N \times N$ identity matrix. We interpret the matrix $\mathbf{A}$ as the adjacency matrix of an undirected, free of self-loops, and static network $G=(V,E)$, where $V=\{x_{1},\cdots,x_{N}\}$ denotes the set of vertices corresponding to $N$ distinct RSOs, and $E \subseteq V \times V$ defines the set of edges between vertices. Distinct RSOs are connected if they {are} $\epsilon$-close {over $(0,T]$}, {this implies that} edges are defined by solely identifying close conjunctions between the objects in the data set. We call this network the \textit{resident space object network}, hereafter denoted RSONet.\\

\begin{figure}[!htp]
\centering
\begin{tabular}{cc}
\includegraphics[width=0.50\textwidth]{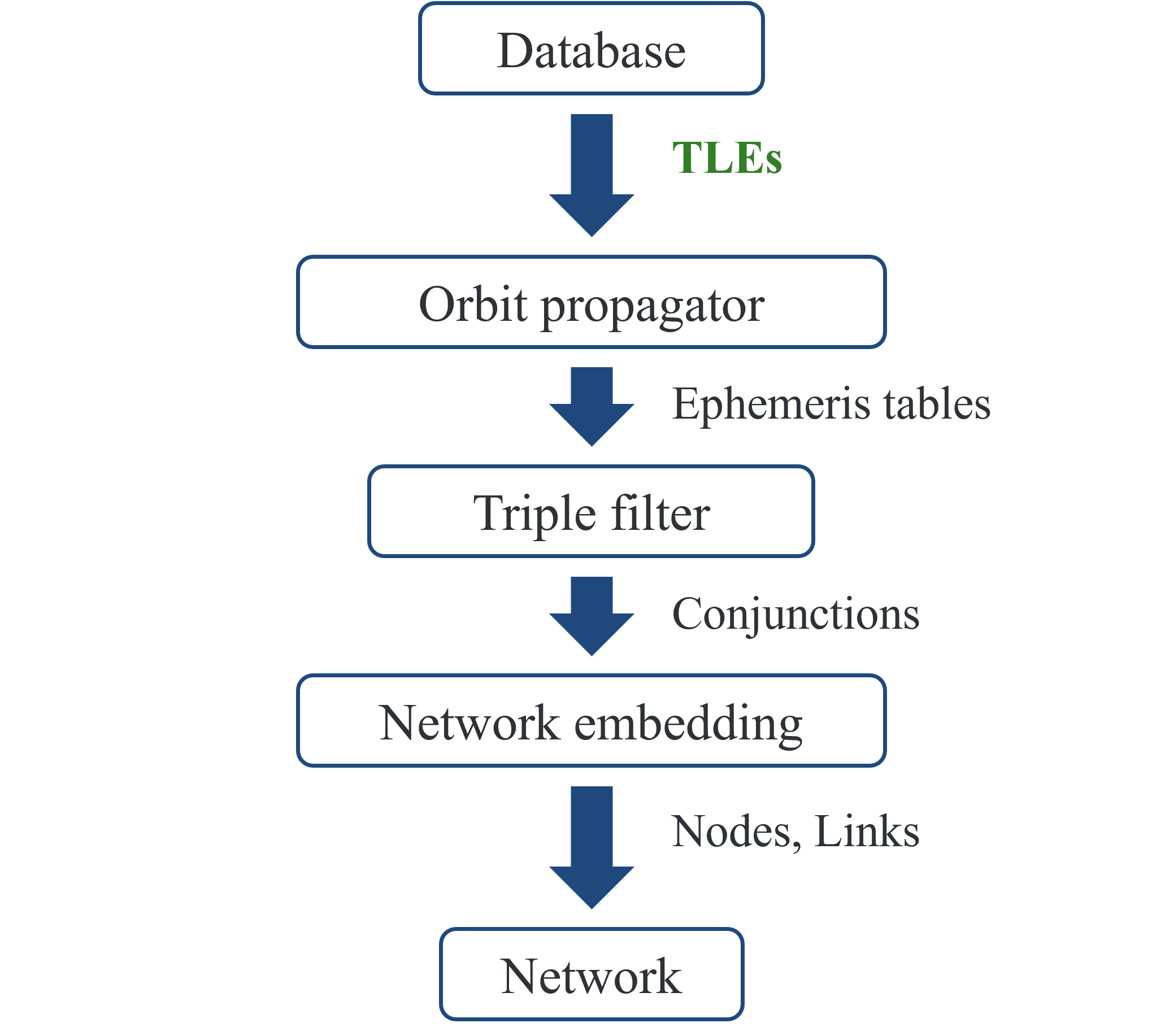} & \includegraphics[width=0.37\textwidth]{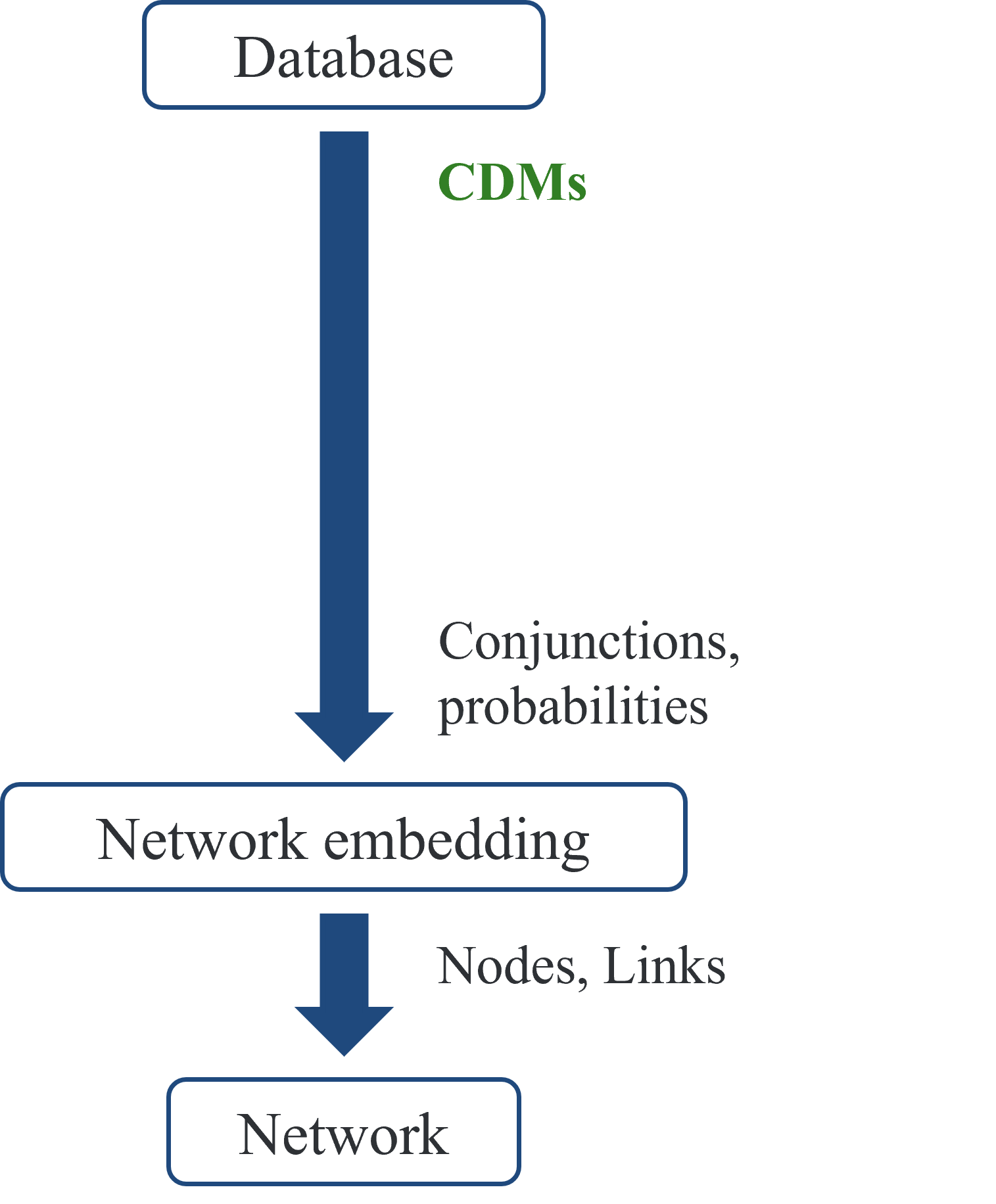}\\
\end{tabular}
\caption{Flowchart of the network embedding process: on the left, the embedding starting from TLE data, which includes the propagation and filtering of the trajectories to detect conjunctions; on the right, the embedding starting from CDM data, which contain already a list of conjunctions.}
\label{fig:embedding}
\end{figure}

We build the RSONet by starting either from Two-Line Element (TLE) data \citep{Vallado2012,kelso1988} or from Conjunction Data Messages (CDMs) \citep{berry2014,berry2018} {(a breakdown of the CDM format is provided in Appendix \ref{app:D})}. The workflow of the two approaches is represented in Fig.\,\ref{fig:embedding}: the key difference between the two approaches lays in the computation of the minimum distance between each couple of RSOs, which is expressed as $\min_{0 \le t \le T} \Vert \mathbf{r}_{i}(t)-\mathbf{r}_{j}(t) \Vert_{2}$ in Eq.\,\eqref{eq:CollCoeff} and through which conjunctions are identified. {It has to be noted that Space-Track pre-screens the detected events to remove debris-on-debris conjunctions, among others. Nonetheless, the strategy to build the RSONet is independent from the input data, TLEs or CDMs; observe however that in the former case we have to consider the temporal propagation of RSOs. To conclude we would like to emphasise that the framework is solely data-driven and thus flexible enough to be adapted to different kinds of formats and propagation methods.}

\subsection{RSONet based on TLEs}
\label{sec:theory_RSOnetwork}

The minimum distance between RSOs is computed via the TLE-based propagation of the objects' orbits and a process to detect conjunctions.

At the beginning of the analysis, TLEs {referring to a specific date} are read from an input file. {Afterwards, a first pre-screening checks for the presence of multiple TLEs for the same object,} retaining only the most recent, thus more precise, entry. All elements, then, are propagated forward within a given period for $m$ time steps to build an ephemeris table, which will be used later to compute the states of the objects at any time $t_1 \leq t \leq t_m$ via interpolation, where $t_k=k \Delta$ and $\Delta$ is the fixed time step duration.

To retain maximum prediction accuracy, we use the analytical Standard General Perturbations 4 (SGP4) model to propagate the set of TLEs \citep{Hoots1984,kelso1988,Vallado2006b,vallado2008}. The SGP4 propagator generates ephemeris in the True Equator Mean Equinox (TEME) coordinate system based on the epoch of the specified TLE \citep{seago2000}. The implementation used for this work is the one provided by Vallado on the Celestrak portal\footnote{\url{https://celestrak.org/publications/AIAA/2006-6753/} (Online; accessed 10-Oct-2023).}. The analytical modelling of SGP4 allows the rapid propagation of space objects, making it a desirable option at this stage of the work, although the simplifications of the perturbation model of SGP4 limit the accuracy of the propagation to intervals of a few days \citep{Vallado2012}. {Let us however stress that} the methodology introduced here {is general enough and it} applies {beyond} the choice of the propagator {made here}. \\

At the current stage, the links between the nodes of the network are built by solely identifying close conjunctions between the various objects in the data set. Close approaches are identified by following the three-filter approach developed by Casanova et al. \cite{Casanova2014}, that improves the original triple-loop filter proposed by Hoots et al. \cite{Hoots1984}.
As per its name, the approach consists of three filters which are applied in series to compare the relative geometry of each {pair} of orbits contained in the TLE set. At each step, the data pool is pruned to remove  RSOs which do not satisfy any of the filters and are, thus, unable to experience close encounters with each other. This pruning process allows to avoid comparing {pairwise elements}, reducing thus the computational load of the problem (initially of the order of $N^2/2$); because the number of unique objects appearing in the daily TLE sets is of the order of $\sim 20,000$, the rough pairwise check would result in $\sim 200,000,000$ comparisons. The breakdown of the three filters is shown in App.\,\ref{app:A}.
{In case multiple conjunctions between the same two objects are found in the given time window, only the one with the lowest distance is considered for the network embedding, in order to avoid creating a non-simple graph (i.e., a graph with multiple links between two nodes) and keep the model as simple as possible.} \\

\subsection{RSONet based on CDMs}

The network embedding can also be generated directly from CDMs, summarised reports containing the main characteristics of a predicted conjunction between two catalogued objects, provided by surveillance agents to satellite operators. The message contains data such as the IDs of the two involved objects, the epoch, the minimum predicted distance, and the estimated probability of collision between the bodies.

The minimum distance between RSOs is obtained directly from the CDM data, thus skipping the propagation process as well as the application of filters, ascribing it to the class of data-driven methods. While faster, this method is less flexible regarding the choice of the duration window, $T$, and collision threshold, $\epsilon$, because CDMs are generated by using fixed values, which change depending on the provider and upon which we do not have any control.

{Similarly to the TLE-based case, in case of multiple conjunctions between the same two RSOs, only the one with the closest distance is kept to build the RSONet.}

\section{Definition of the relevance score}
\label{sec:rating}

The network embedding is exploited to introduce a new ranking score to estimate the importance of each object in relation to the rest of the population in the possibility to take part of a {possible collision}. {This way, it is possible to both quantify the likelihood of a collision during a close approach and to measure how each RSO contributes to this likelihood relative to one another}.
This score is based on statistics of network such as the degree, clustering coefficient, the closeness and betweenness centralities (see App.\,\ref{app:B} for the definition of those metrics). The following assumptions are made:

\begin{enumerate}
\item No distinction based on the type of objects (debris, payload, etc.) involved in the conjunctions is considered;
\item The conjunctions are all assumed to be independent from time, meaning that they are considered as if they occurred at the same time;
\item The conjunctions are all assumed to be mutually exclusive, meaning that the combined probability of either of two events $X$ and $Y$ occurring is $p(X \cup Y) = p(X) + p(Y)$, where $p(X)$ and $p(Y)$ are the probabilities of the two events;
\item No importance is given to the actual collision probability in each conjunction, thus a fixed value $p$ is assigned to all links as a weighting factor useful for the definition of the score;
\item The fixed probability value $p$ used to weight the links is assigned a value small enough (e.g., 10$^{-4}$) to be able to combine them in an intuitive way without the risk of obtaining a total value larger than 1;
\item The probability of collision between object $i$ and any debris produced by object $j$ is assumed to be the same probability of a collision between $i$ and $j$, the rationale being that during the short enough observation period the debris cloud originating from $j$ will remain very close to the latter object {(thus well within the accuracy limits of SGP4)} .
\end{enumerate}

\begin{remark}
\label{rem:assumptions}
The first assumption is necessary due to the nature of the problem itself. Since all conjunction events are pairwise interactions, as it will be explained in Sect.\,\ref{sec:results}, the network obtained by considering encounters over short periods would result in a small number of connected components {encompassing two nodes} each. Combining the conjunctions detected over longer periods of time produces more structured networks with larger components, not only allowing for better visualisation of the complexity of the interactions within the space resident population, but also to gain insight on the way risk changes according to various parameters. Thus, the network so obtained is a \text{cumulative} representation of the conjunction events, a superposition of the networks that would be obtained by searching over shorter periods of time. {Moreover, by using the same value $p$ to equally weight all links allows to factor it out to allow for the network topological properties to clearly emerge.}
\end{remark}

The combined probability of an object to collide with any other can be divided into different contributions, which account for the probability of direct collisions (local interactions) and of indirect collisions with debris produced by other objects (non-local interactions). This brings us to introduce three scalar quantities $\mathcal{C}_1$, $\mathcal{C}_2$, and $\mathcal{C}_3$, that are used to define the final relevance score $\mathcal{S}$, and that we are going to discuss. \\

\begin{figure}[!htp]
\centering
\begin{tabular}{ccl}
a) & \parbox[c]{0.5\textwidth}{\includegraphics[width=0.5\textwidth]{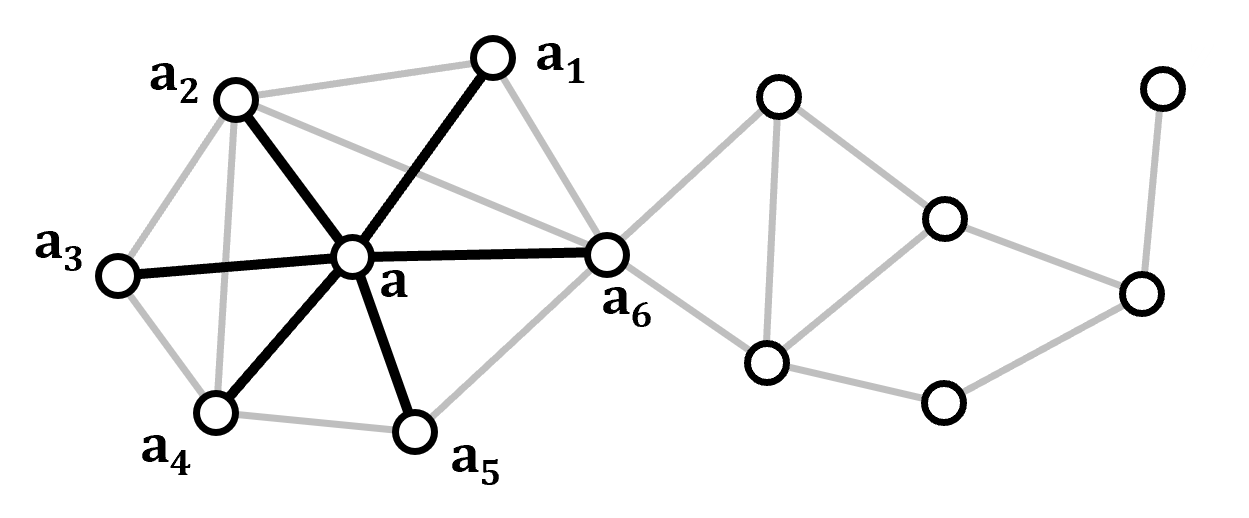}} & \begin{minipage}[c]{0.4\textwidth} $\hat{p}_a = p_{a a_1} + p_{a a_2} + p_{a a_3} + p_{a a_4} + p_{a a_5} + p_{a a_6} = \sum_{i=1}^{6}{p_{a a_{i}}}$ \end{minipage} \\
b) & \parbox[c]{0.5\textwidth}{\includegraphics[width=0.5\textwidth]{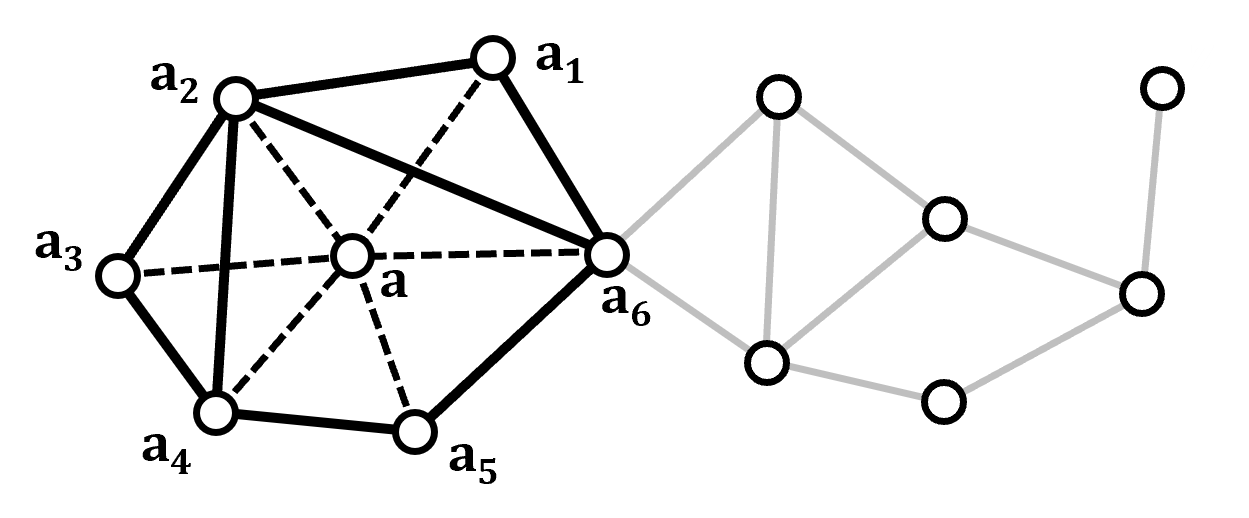}} & \begin{minipage}[c]{0.4\textwidth} $\tilde{p}_a = p_{a a_1}(p_{a_1 a_2}+p_{a_1 a_6}) + p_{a a_2}(p_{a_2 a_1}+p_{a_2 a_3}+p_{a_2 a_4}+p_{a_2 a_6}) + p_{a a_3}(p_{a_3 a_2}+p_{a_3 a_4}) + p_{a a_4}(p_{a_4 a_2}+p_{a_4 a_3}+p_{a_4 a_5}) + p_{a a_5}(p_{a_5 a_4}+p_{a_5 a_6}) + p_{a a_6}(p_{a_6 a_1}+p_{a_6 a_2}+p_{a_6 a_5}) = \sum_{i=1}^{6}{ p_{a a_{i}} \left( \sum_{\substack{j \neq i \\ j=1}}^{D_{a_i}-1} {p_{a_{i}a_{j}} } \right) }$ \end{minipage} \\
c) & \parbox[c]{0.5\textwidth}{\includegraphics[width=0.5\textwidth]{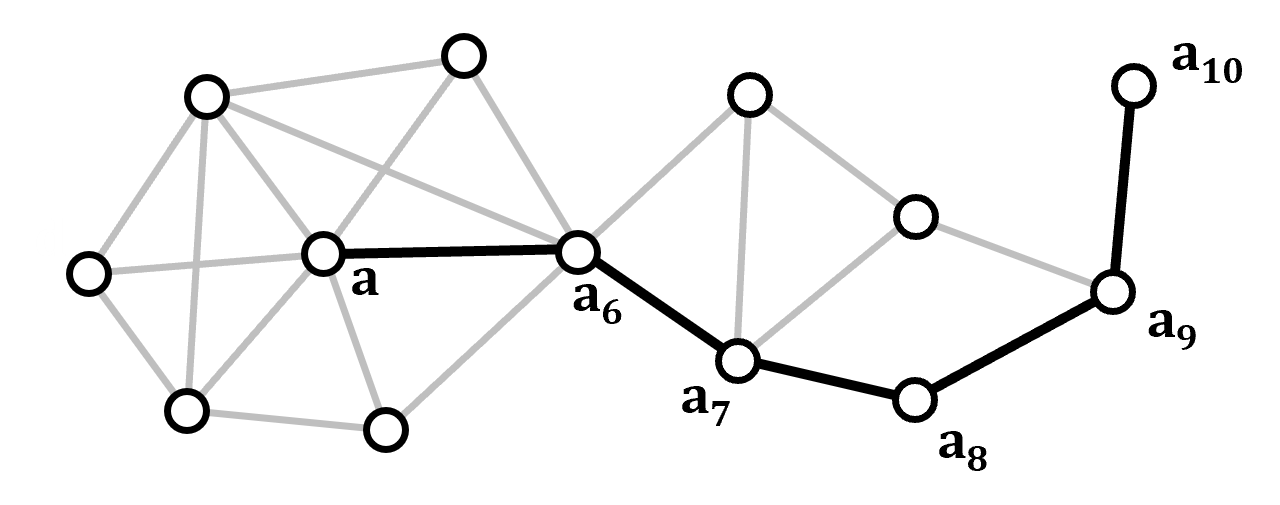}} & \begin{minipage}[c]{0.4\textwidth} $\stackrel{\frown}{p}_a = p_{a a_6} p_{a_6 a_7} p_{a_7 a_8} p_{a_8 a_9} p_{a_9 a_{10}} = p_{a a_6} \prod_{j=7}^{10}{p_{a_{j-1} a_j}}$ \end{minipage}
\end{tabular}
\caption{Illustration of the concept behind the definition of the three contributions to the ranking score.}
\label{fig:modelscore}
\end{figure}

With reference to the {diagrams} in Fig.\,\ref{fig:modelscore} that represent a network with $n=13$ nodes, let us consider node $a$ having $D_a=6$ neighbours denoted by $\{a_{1},a_{2},\dots,a_{6}\}$, namely equal to the the number of conjunctions it experiences. We hence denote by $p_{a a_{i}}$ the probability of a direct collision between any two distinct objects $a$ and, say, $a_{i}$, $1 \le i \le D_{a}$. We derive the definition of each contribution specifically for node $a$ first, to show how they are constructed, and then we provide a generalised definition valid for any node.

\paragraph*{Contribution 1}{The first contribution $\mathcal{C}_{1}$ arises by considering the conjunctions of an object with its direct neighbours (Fig.\,\ref{fig:modelscore}a), reflecting the overall probability of a direct pairwise collision with any of them. Considering node $a$, which has degree $D_a$, the combined probability $\hat{p}_a$ of a collision between $a$ and any of those neighbouring objects can be written as the sum of the individual probabilities. By using the assumption that all probabilities are identically given by $p$, the sum is written as:
\begin{equation*}
\hat{p}_a = \sum_{i=1}^{D_{a}}{p_{a a_{i}}} = \sum_{i=1}^{D_{a}}{p} = p D_{a} .
\end{equation*}

\noindent Generalising the previous expression for a generic node $i$ having degree $D_i$ and recalling the assumption of equal probability for any close approach, $\mathcal{C}_{1}^{\,(i)}$ can be expressed as
\begin{equation}
\mathcal{C}_{1}^{\,(i)} = p D_i .
\end{equation}
}

\paragraph*{Contribution 2}{The second contribution $\mathcal{C}_2$ considers the indirect effects on $a$ of the conjunctions between its neighbours (see Fig.\,\ref{fig:modelscore}b), aiming to quantifying the probability $\tilde{p}_a$ of $a$ encountering the debris produced by one of its neighbours; let us remember that, because of the previous assumption, this probability is the same as a collision of $a$ with the latter object {because we assume the debris cloud to remain close to the originating RSO}. Such probability can thus be written as:
\begin{align}
\begin{split}
\tilde{p}_a &= \sum_{i=1}^{D_{a}}{ p_{a a_{i}} \left( 
\sum_{\substack{j \neq i \\ j=1}}^{D_{a_j}} {
p_{a_{i}a_{j}} } 
\right) } = \sum_{i=1}^{D_{a}}{ p \left( 
\sum_{\substack{j \neq i \\ j=1}}^{D_{a_j}-1} {
p } 
\right) } \\
&= \sum_{i=1}^{D_{a}}{ p \left[ p (D_{a_i}-1) \right] } = p^2 \sum_{i=1}^{D_{a}}{ (D_{a_i}-1) } \\
&= p^2 \cdot 2 N_a ,
\end{split}
\end{align}
where $N_a$ {is the} number of pairs connected to $a$, which depends on the number of connections between the neighbours of $a$. This number can be derived from the numerator of the expression of the clustering coefficient given in App.\,\ref{app:B}:
\begin{equation}
N_a = \frac{1}{2} C_a D_a (D_a-1) .
\end{equation}
The generalised expression for {the second contribution for} a node $i$ can be thus obtained as 
\begin{equation}
\mathcal{C}_{2}^{\,(i)} = p^2 C_i D_i (D_i-1) .
\end{equation}
}

\paragraph*{Contribution 3}{The third contribution $\mathcal{C}_3$ considers the non-local interactions with distant, in the network topology, nodes. In this case, the intent is to estimate the probability $\stackrel{\frown}{p}_a$ of $a$ encountering debris produced in a chain reaction starting from a collision between distant nodes: the first collision produces debris, which strikes another object producing new fragments, propagating the chain, but still remaining close enough to the initial object from which they originate. Considering the chain from $a$ to $a_{j}$ as pictured in Fig.\,\ref{fig:modelscore}c, this probability can be defined as the product of the individual probabilities:
\begin{equation}
\stackrel{\frown}{p}_a = p^{d_{a a_{j}}} ,
\end{equation}
where $d_{a a_{j}}$ is the length of the shortest-path between nodes $a$ and $a_{j}$. The extension of the latter expression to all chains from $a$ to the other $n-1$ nodes in the network is straightforward and requires to consider the summation over the different chains:
\begin{equation}
\stackrel{\frown}{p}_a = \sum_{j=1}^{n-1}{p^{d_{a a_j}}} \, .
\end{equation}

\noindent Let us assume the shortest-paths {do} not vary too much across the network and thus to be well approximated by the average shortest-path length, $\ell_a$, between node $a$ and the other nodes in the network, as defined in Eq.\,\eqref{eq:meandistance} in App.\,\ref{app:B}. Thus, the previous expression can be approximated and simplified as follows:
\begin{equation}
\stackrel{\frown}{p}_a = \sum_{j=1}^{n-1}{p^{d_{a a_j}}} \approx (n-1) p^{\ell_a} = (n-1) p^{1/K_a} ,
\end{equation}
where $K_a$ is the closeness centrality as defined in Eq.\,\eqref{eq:closeness} in App.\,\ref{app:B}.

\noindent However, the expression above does not capture the whole picture, since it takes into account solely the probability of chain reactions directed to a given node. The importance of a node in this regard does not come only from the number of chains it can connect to, but also from the number of chains it can propagate, representing an object which can continue multiple fragmentation cascades. In terms of network properties, this translates to the concept of betweenness.

\noindent For this reason, the expression above has to be modified to accommodate for the possibility of node $a$ to propagate cascades. With reference to Fig.\,\ref{fig:modelscore_b}, let us modify the model network of Fig.\,\ref{fig:modelscore} by removing the edges between the neighbours of $a$, $a_{i}$, $1 \le i \le 6$, such that these nodes are only connected to $a$. If we perform the same construction as above, by computing the probability that each of these nodes is reached by a cascade from node $a_{10}$, we obtain
\begin{equation*}
\stackrel{\frown}{p}_{a_i} = p_{a a_i} \prod_{j=6}^{10}{p_{a a_j}} \, , \forall \, 1 \le i \le 6 .
\end{equation*}

\begin{figure}
\centering
{\includegraphics[width=0.5\textwidth]{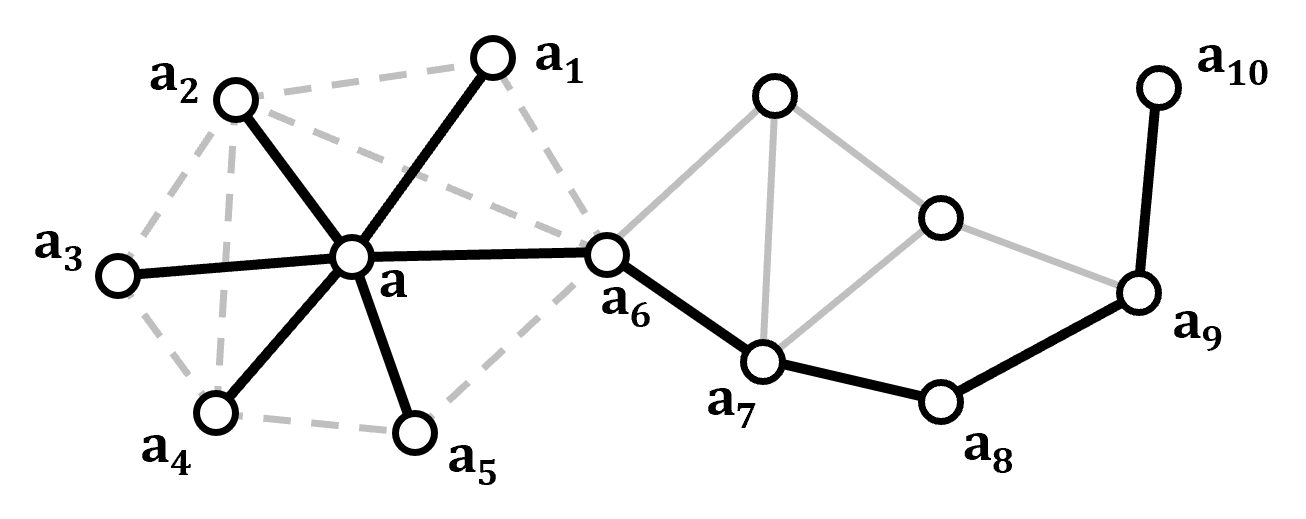}}
\caption{Modified model network used to illustrate the definition of contribution $\mathcal{C}_3$.}
\label{fig:modelscore_b}
\end{figure}

\noindent Since all the shortest paths between the neighbours of $a$ and $a_{10}$ must go through $a$, this nodes contributes further to the risk of collisions by propagating cascades; we can define another measure $\stackrel{\frown}{p}_{a}^{\,\prime}$ by summing all shortest paths passing through $a$:
\begin{equation*}
\stackrel{\frown}{p}_{a}^{\,\prime} = \sum_{i=1}^{n-1}{\sum_{j=1}^{n-1}{p^{d_{a a_j}}}} \approx B_a p^{\ell_a} = B_a p^{1/K_a} ,
\end{equation*}
which recalls the definition of betweenness given in App.\,\ref{app:B}.

\noindent The generalised expression for $\mathcal{C}_{3}^{\,(i)}$ can be finally obtained as
\begin{equation}
\mathcal{C}_{3}^{\,(i)} = B_i p^{1/K_i} .
\end{equation}
}

Finally, by summing the three contributions, one can obtain the formulation of a ranking score based exclusively on the topology and statistics of the network.

\begin{definition}[Relevance score]
Based on the previous discussion, the relevance score for each node $i$ is introduced as the scalar
\begin{equation*}
\label{eq:score}
\mathcal{S}^{\,(i)} = \mathcal{C}_{1}^{\,(i)} + \mathcal{C}_{2}^{\,(i)} + \mathcal{C}_{3}^{\,(i)} =  p D_i + p^2 C_i D_i (D_i-1) + B_i p^{1/K_i} .
\end{equation*}
\end{definition}

\begin{remark}
The definition introduced above combines two main aspects: the first one takes into account local properties, i.e., by weighting nodes highly interconnected with each other, thus experiencing a high risk of direct collisions, while the second one considers nodes connecting to a large number of chains on a less local scale, highlights which nodes can drive fragmentation cascades. The latter contributions, containing powers of $p$ with higher exponents, was observed to be orders of magnitude lower than the first contribution. On the other hand, nodes with high degree or betweennes may skew the relevance score towards the indirect contributions.
\end{remark} 

\begin{remark}
Let us observe that, by its very first definition, $\mathcal{S}^{\,(i)}>0$, and the lower bound can never be achieved, indeed in the smallest possible network made of two nodes and one link, the nodes will have degree equal to $1$ and clustering equal to $0$, resulting in a strictly positive score. The same holds true for any chain-like network.
\end{remark}

\section{Application and analysis}
\label{sec:results}

In this section, the use of networks to represent the RSOs environment and to evaluate the risk of collisions is showcased. In the first part, examples are made to highlight the characteristics of the networks and their dependency on the source of the data. In the second part, the application of the relevance score defined in Sect.\,\ref{sec:rating} is presented, to show the use of the properties of the network to provide some measure of the risk of collision. In the third part, a sensitivity analysis is made to show the dependency of the main network statistics and of the score values on the simulation parameters, that is the propagation time and the conjunction distance threshold.\\

{The proposed test cases use either TLEs or CDMs as the source of initial data. While the strategy to construct the RSONet is independent from the choice of the input data, as explained in Sect.\,\ref{sec:theory}, we want here to show that the latter choice can however affects the structure and composition of the RSONet, given that different objects appear in the two datasets as published by Space-Track and also that the propagation of TLEs and the three-filters process could return different encounter probability.}

All TLEs and CDMs used here are obtained from Space-Track\footnote{\url{https://www.space-track.org} (Online; accessed 10-Oct-2023).}, a public catalogue made available by NORAD and updated daily. Despite the ease of access, covariance matrices tied to the orbital state estimations are not published, so no uncertainties over the initial state of the objects are considered in the current model. CDMs contain conjunction predictions up to three days after their creation date, but no information about the uncertainty over the geometry of the conjunction. Due to the assumptions made in Sect.\,\ref{sec:rating}, in this work conjunctions are studied only in terms of distance between the objects rather than collision probability. TLEs were propagated using a time step $\Delta$ of $1$ minute. The data sets were gathered by using python interfaces to connect to Space-Track\footnote{\url{https://pypi.org/project/spacetrack/} (Online; accessed 10-Dec-2023).}, while post-processing and plotting of the results was performed using MATLAB native libraries.

\subsection{Results of the network embedding}

Both TLE and CDM data sets refer to the month of May 2023 and both analyses define conjunctions using a threshold distance $\epsilon= 1\, \mathrm{km}$ within a $T=3$ days long propagation interval, observing that those values are the standard accepted parameters used to generate the CDMs on Space-Track for conjunction monitoring\footnote{\url{https://www.space-track.org/\#conjunctions} (Online; accessed 10-Oct-2023).}. The TLE sets used in these simulations contained the whole catalogue, with no focus on a particular orbital region. \\

\begin{figure}[!htp]
\centering
{\includegraphics[width=\linewidth,trim={10cm 7cm 13cm 5cm},clip]{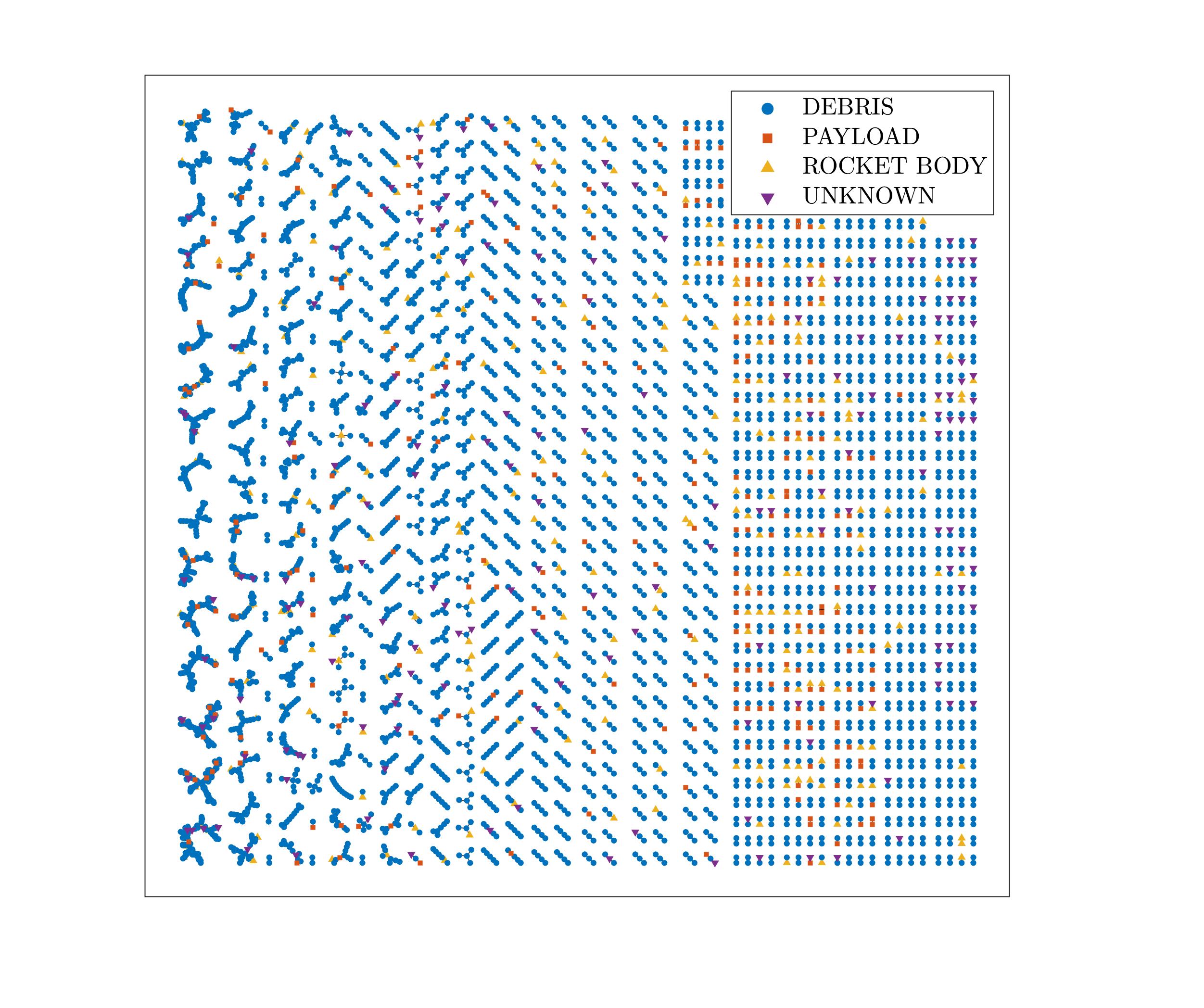}}
\caption{Complete CDM-based RSONet (nodes are color-coded by object classification).}
\label{fig:CMD_whole}
\end{figure}

\begin{figure}[!htp]
\centering
{\includegraphics[width=\linewidth,trim={10cm 7cm 7.5cm 5cm},clip]{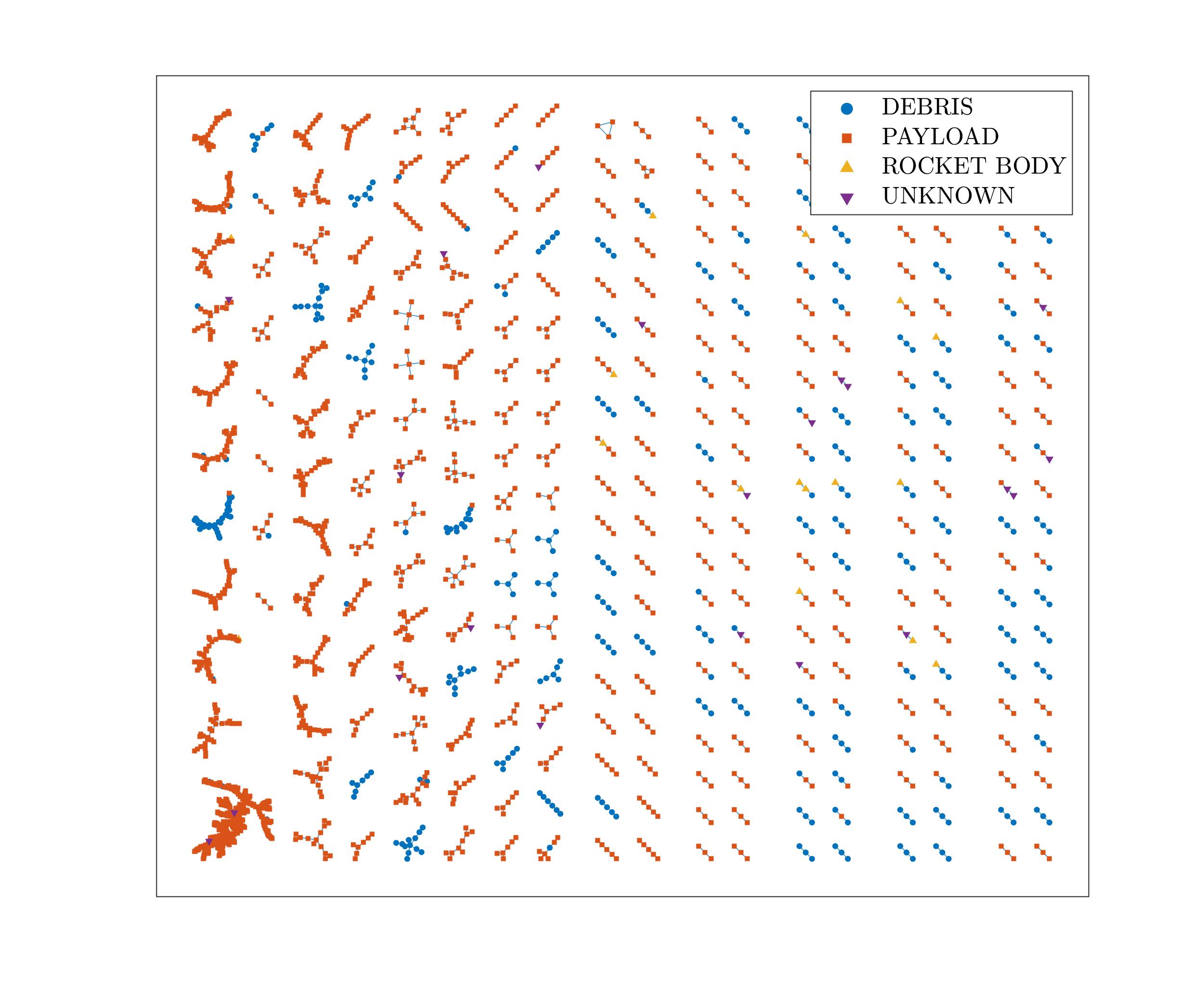}}
\caption{Complete TLE-based RSONet (nodes are color-coded by object classification).}
\label{fig:TLE_whole}
\end{figure}

\begin{figure}[!htp]
\centering
\begin{tabular}{ll}
a) & {\includegraphics[width=0.7\linewidth,trim={9cm 7cm 11cm 5cm},clip]{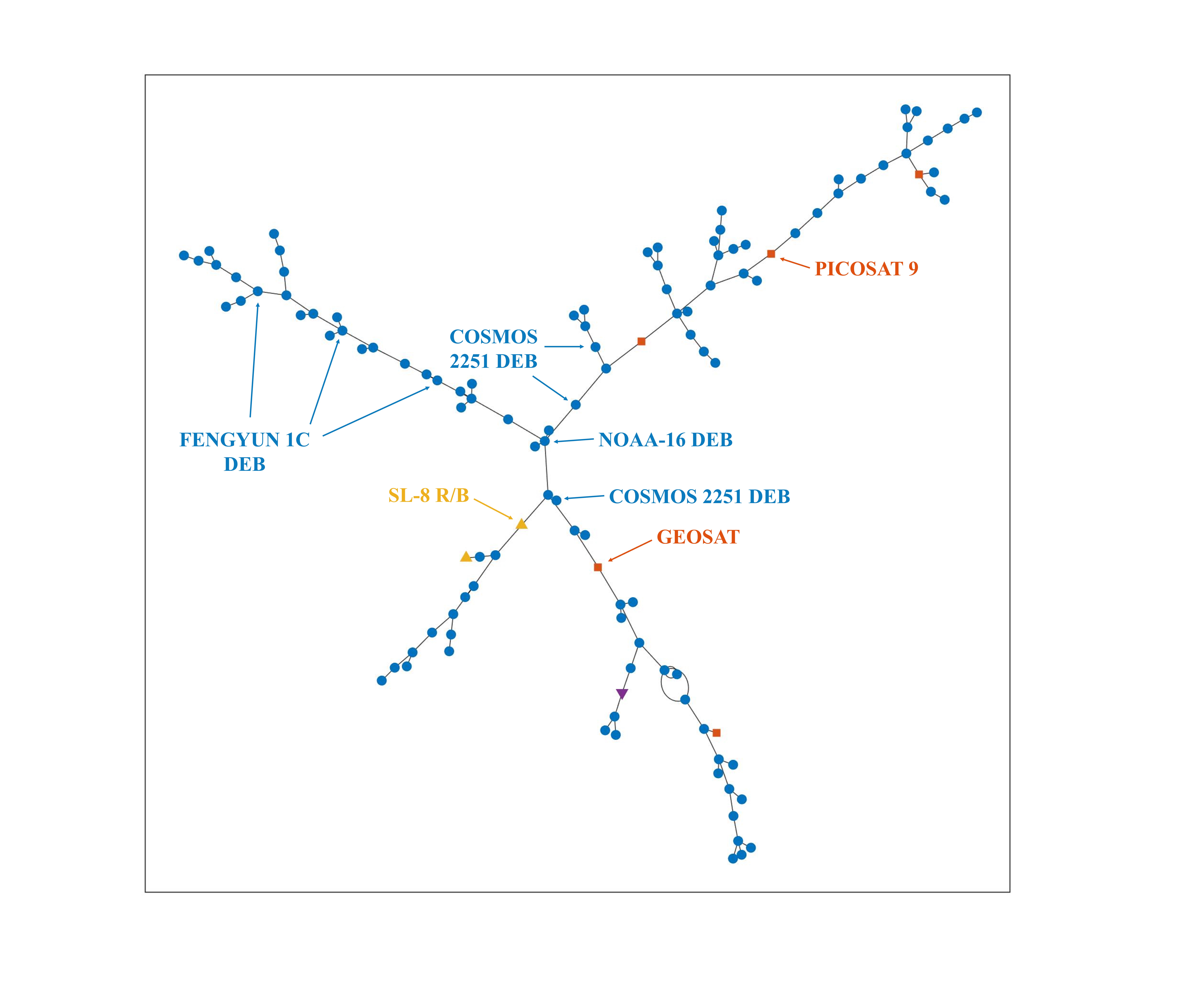}} \\
b) & {\includegraphics[width=0.7\linewidth,trim={4.25cm 3cm 3cm 2cm},clip]{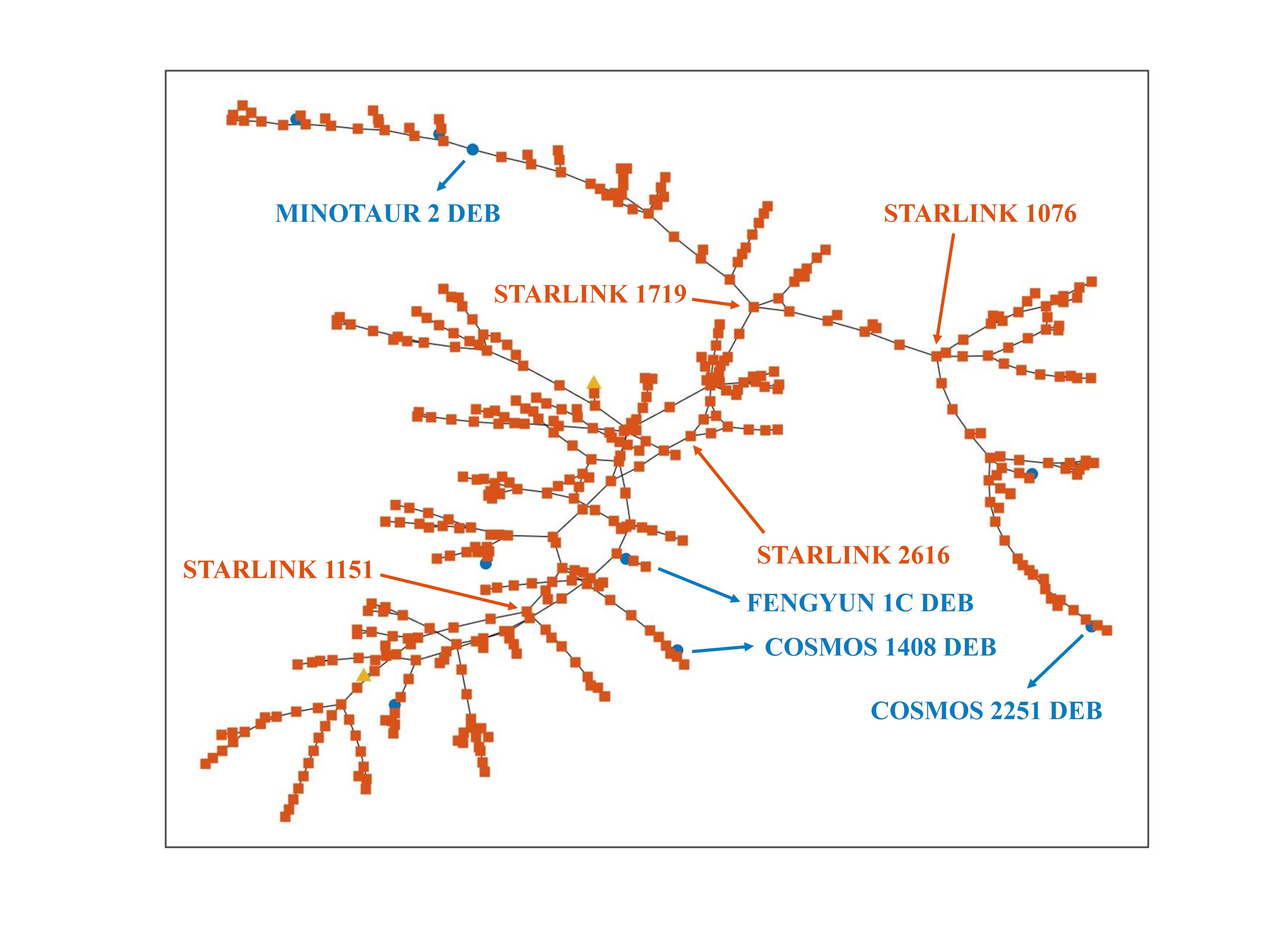}}\\
\end{tabular}
\caption{Largest connected components of the RSONet from CDMs (a) and TLEs (b) over the course of 30 days, using distance thresholds of 1 km and 3 km, respectively.}
\label{fig:TLEvsCMD_largest}
\end{figure}

\begin{figure}[!htp]
\centering
\begin{tabular}{rcrc}
a) & {\includegraphics[width=0.4\textwidth,trim={1.5cm 2cm 3.5cm 3cm},clip]{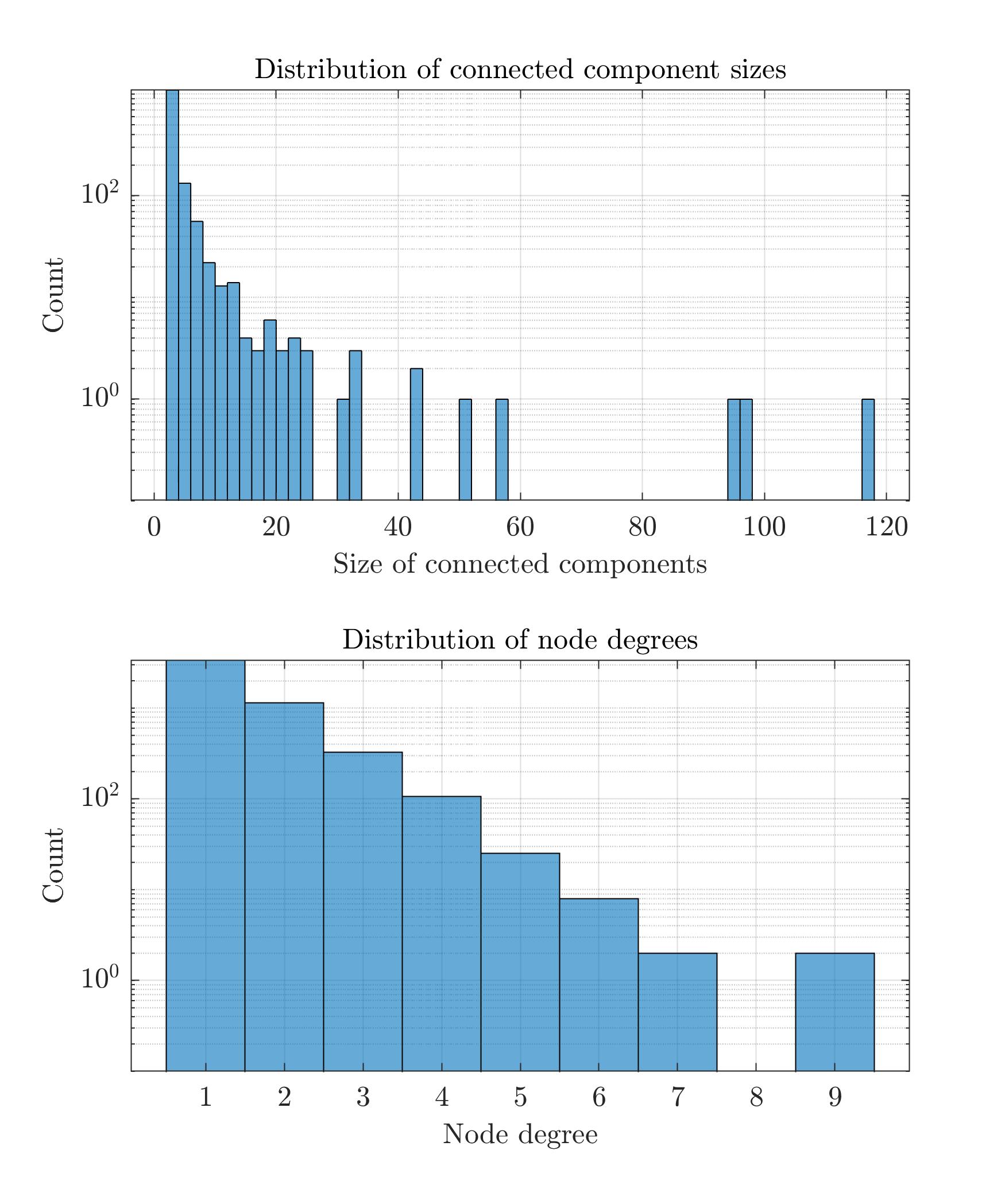}} &
b) & {\includegraphics[width=0.4\textwidth,trim={1.5cm 2cm 3.5cm 3cm},clip]{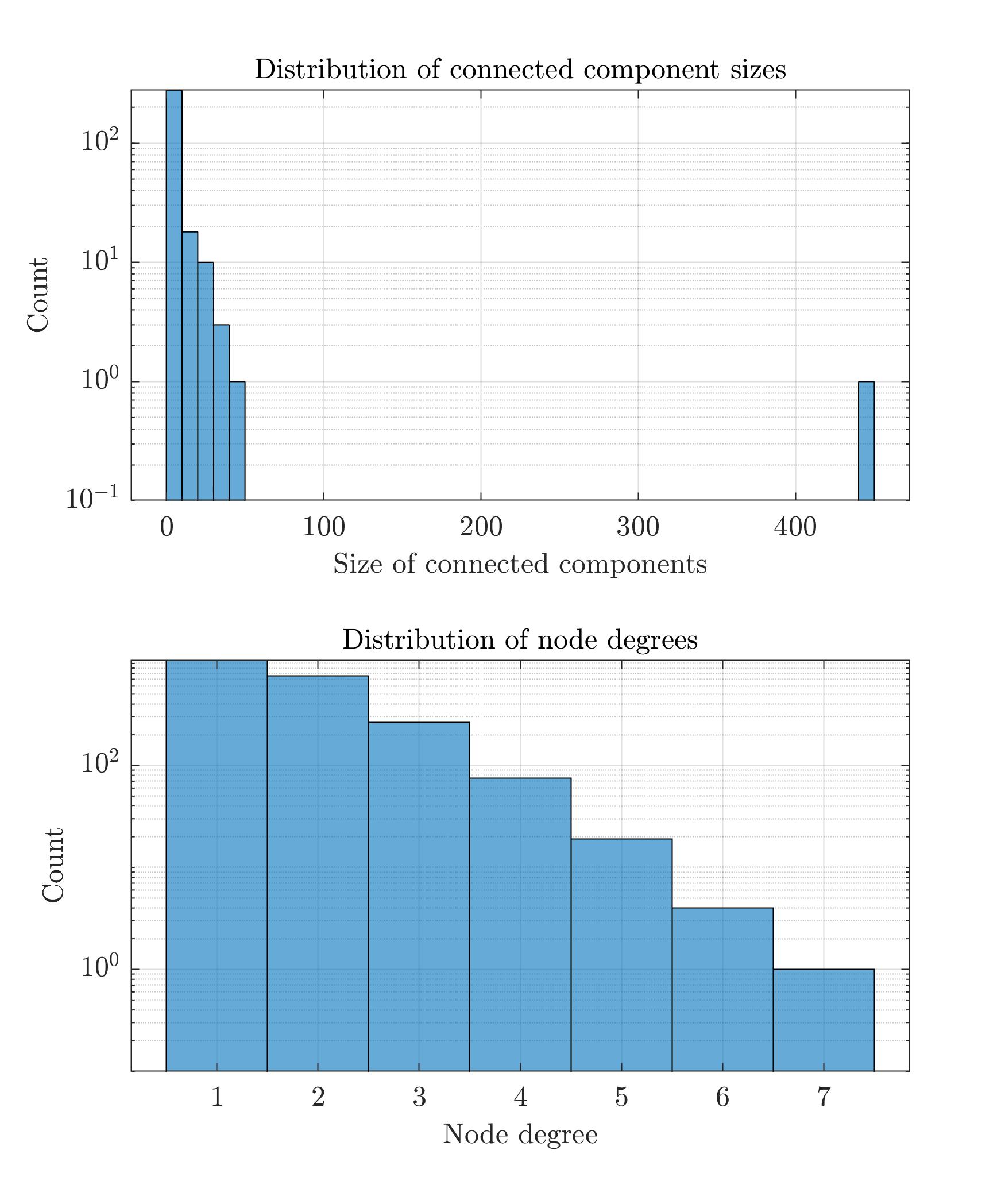}}
\end{tabular}
\caption{Distribution of the size of the connected components and of the node degree in the two cases in the CDM (a) and TLE (b) cases.}
\label{fig:TLEvsCDM_histo}
\end{figure}

Figures \ref{fig:CMD_whole} and \ref{fig:TLE_whole} shows the two network embeddings obtained by combining the conjunction data obtained from CDMs and TLEs over 30 days using $1 \,\mathrm{km}$ and $3 \,\mathrm{km}$ as distance thresholds, respectively. The larger threshold used for the TLE-based embedding was chosen to compensate for the lower precision of the propagation, which reduces the ability to accurately detect conjunctions between RSOs.
As one can observe, the network is sparse and splits into several disjoint connected components (i.e., cluster of interconnected nodes that are disjoint from any larger connected subgraphs), a zoomed-in view of the largest connected component from each network is shown in Fig.\,\ref{fig:TLEvsCMD_largest}. Table\,\ref{table:CDMvsTLE_stats} shows some relevant figures of the networks, such as the number of connected components and their size, the degree, as well as the composition in terms of type of objects and orbital region, while Fig.\,\ref{fig:TLEvsCDM_histo} shows the distribution of the size of the connected components and of the node degree in the two cases.

Let us observe that based on Remark\,\ref{rem:assumptions}, since all interactions involve two objects, most components are composed by only two nodes, representing sporadic single conjunction events, which is reflected by the low values of mean degree, similarly to what had already been observed by Lewis et al. \cite{Lewis2010}. By combining the conjunctions found over longer periods, a more complex structure arises, highlighting new interactions between the members of the population.
In particular, it is visible how components tend to grow by ramification, forming chains which expand as each object encounters another one in rare occasions. This again reflects the underlying foundation of the network relying on pairwise interactions. This aspect will be explored more deeply in Sect.\,\ref{sec:sensitivity}.\\

\begin{table}
\centering
\caption{Comparison of the characteristics of the networks obtained via CDMs and TLEs.}
\label{table:CDMvsTLE_stats} 
\begin{tabular}{lcc} 
\hline
Count 		& CDMs & TLEs \\
\hline\hline
Nodes 		& 4949 & 3562 \\
Edges 		& 3616 & 2590 \\
Connected components	& 1364 & 996 \\
Largest size	& 117 	& 447 \\
Mean size 	& 3.5 	& 3.6 \\
Highest degree 	& 7 		& 9 \\
Mean degree 	& 1.4 	& 1.5 \\
\hline\hline
\# Debris	& 4212 & 880 \\
\# Payload 	& 277 & 2535 \\
\# Rocket stages		& 274 & 76 \\
\hline\hline
Obj. in LEO 	& 4675 & 3546 \\
Obj. in MEO	& 13 & 0 \\
Obj. in GEO 	& 22 & 8 \\
\hline
\end{tabular}
\end{table}

While maintaining several similarities, the source of the initial data has a great impact on the structure and composition of the network.
The TLE-based networks tend to have fewer components with larger size on average with respect to the CDM-based counterpart. This is due both to the precision of the initial data and to the orbital propagation method: indeed, CDMs are obtained from the propagation of high-precision data via a high-fidelity model, used internally by Space-Track. On the contrary, TLEs are inherently less accurate at transmitting orbital data, and SGP4, being an approximate model, also introduces error in the propagation of the orbits.

Thus, different initial data and propagation methods lead to diverging results: different conjunction events involving different objects are detected by the two models, which end up in the representation of different members of the RSO population. This is visible by the contrasting distribution of orbits (several objects outside of LEO appear in the CDM data) and of objects (more debris than payloads appear in CDMs).
The different structure of the two networks is, indeed, also due to the way conjunctions are managed internally, since most encounters between satellites of the same constellations are not displayed in the CDMs from Space-Track. For example, the Starlink constellation accounts for more than $5,000$ satellites as of August 2023; since most of them orbit at close distance, they can trigger algorithms into detecting conjunctions with low impact probability between them, which are then ignored in published CDMs. In the TLE-based network, most connected components are composed entirely of Starlink satellites.\\

Since the networks are made mostly of small connected components, representing all of them would impact the clarity of the images. While the choice to show the entire network in Fig. \ref{fig:CMD_whole} and \ref{fig:TLE_whole} was made to provide context to the reader, from now on the figures will show only the $5$ largest components from each model. Nevertheless, the numerical results will still refer to the entire networks.

\subsection{Relevance score}

This section shows how the score reflects the risk of collision and highlights the relevant nodes even without directly taking into account the actual probability values.

The choice of the value of $p$, while changing the value of the  metric $\mathcal{S}$, does not influence directly the ranking of the various RSOs, since there is proportionality between $\mathcal{S}$ and $p$, thus leaving the ranking mostly unaltered. However, both the values of $\mathcal{S}$ would change if we considered the actual values of the collision probability during each conjunction.

\begin{figure}[!htp]
\centering
\begin{tabular}{ll}
{\includegraphics[width=0.45\textwidth,trim={5cm 3cm 2cm 2cm},clip]{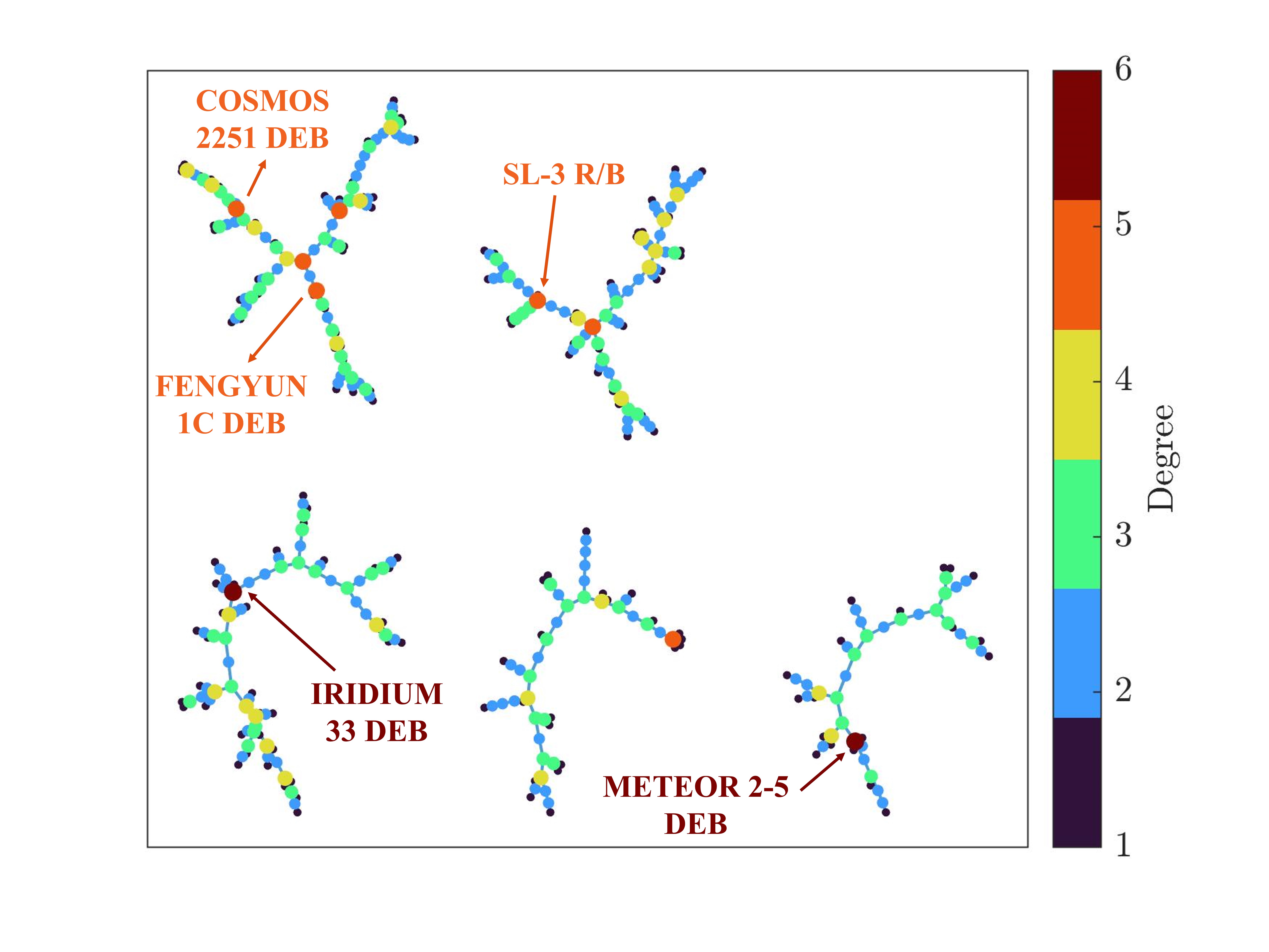}} &
{\includegraphics[width=0.45\textwidth,trim={5cm 3cm 2cm 2cm},clip]{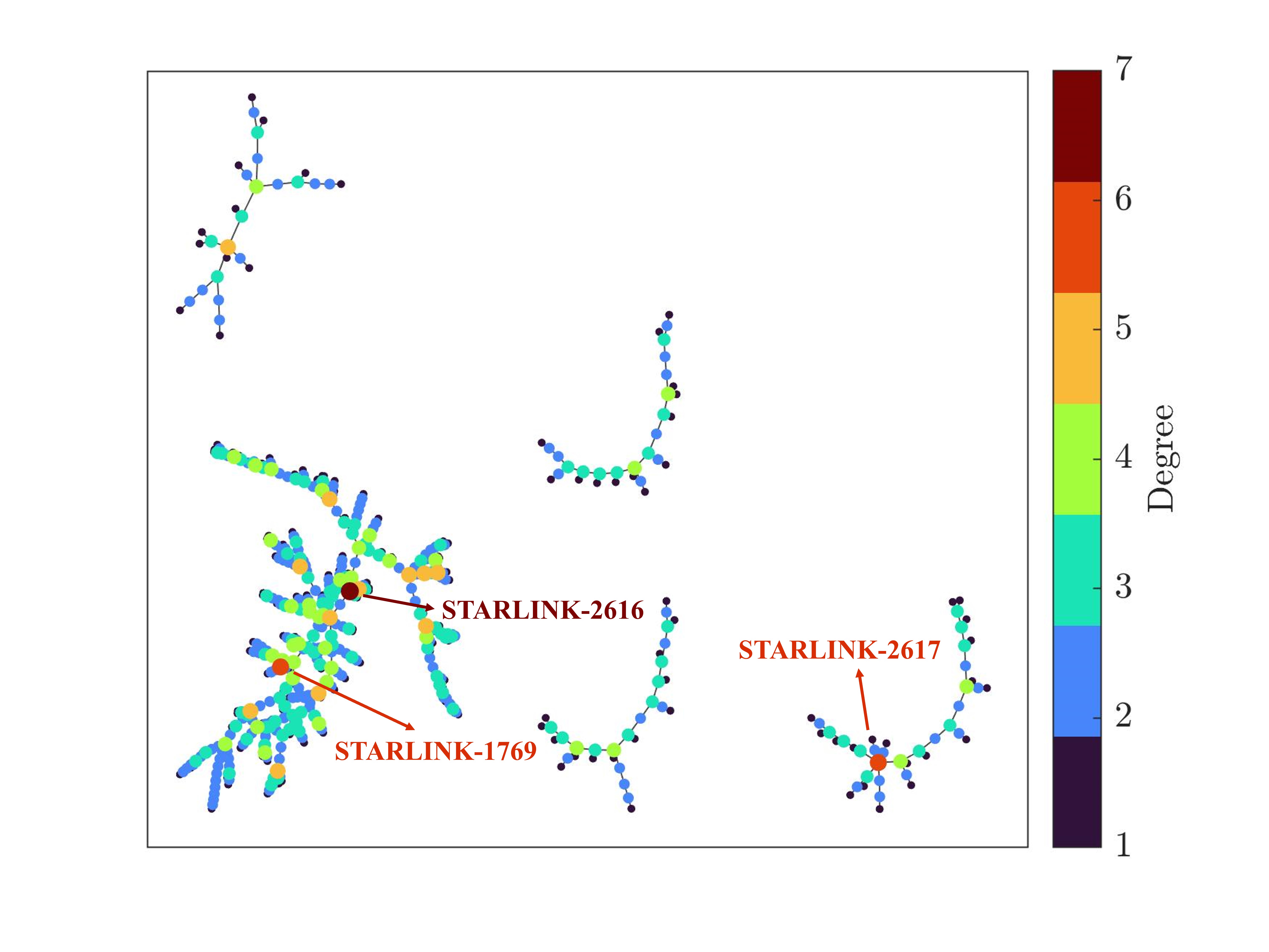}} \\
a) & d)\\
{\includegraphics[width=0.45\textwidth,trim={5cm 3cm 2cm 0.5cm},clip]{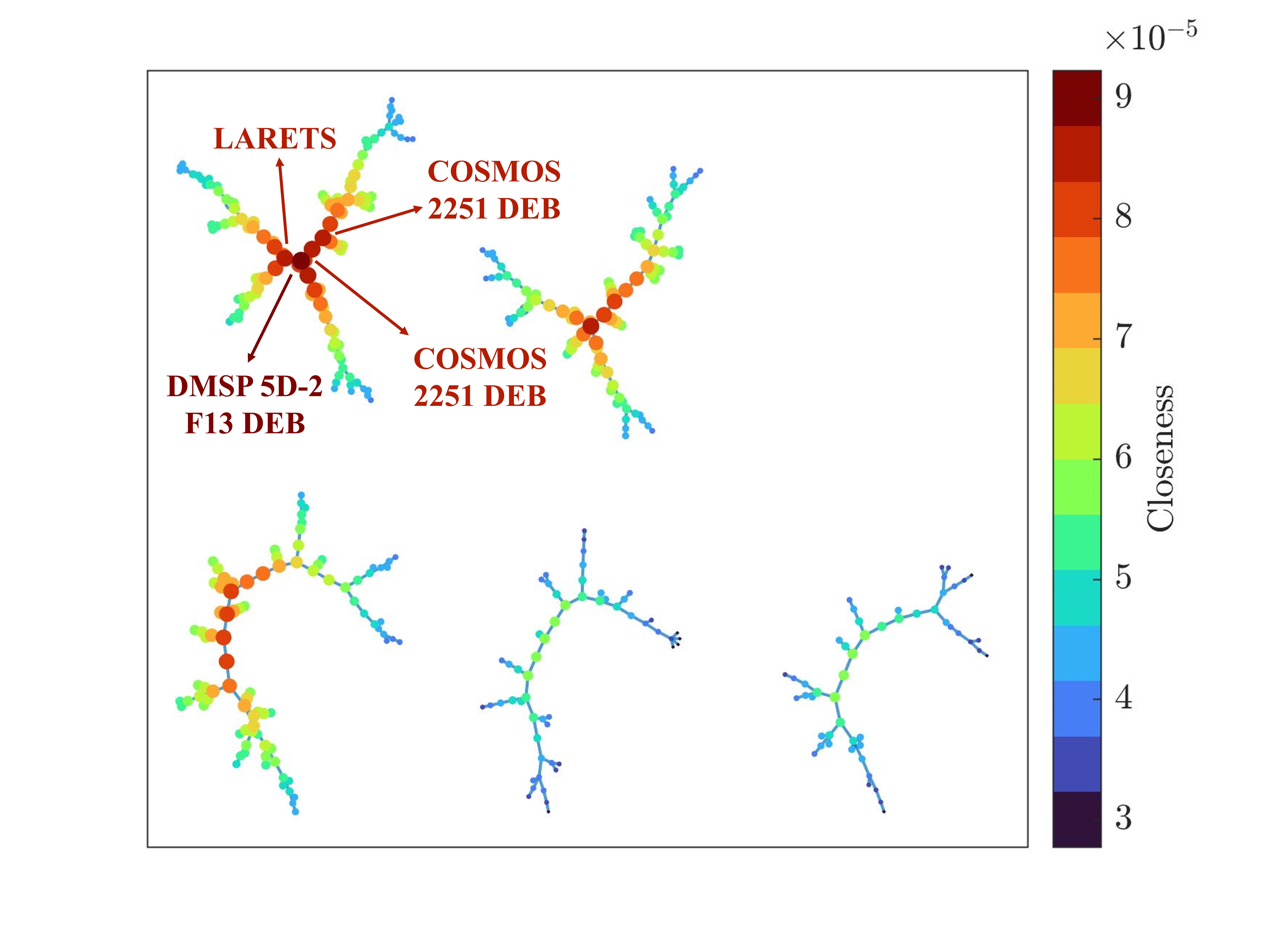}} &
{\includegraphics[width=0.45\textwidth,trim={5cm 3cm 2cm 0.5cm},clip]{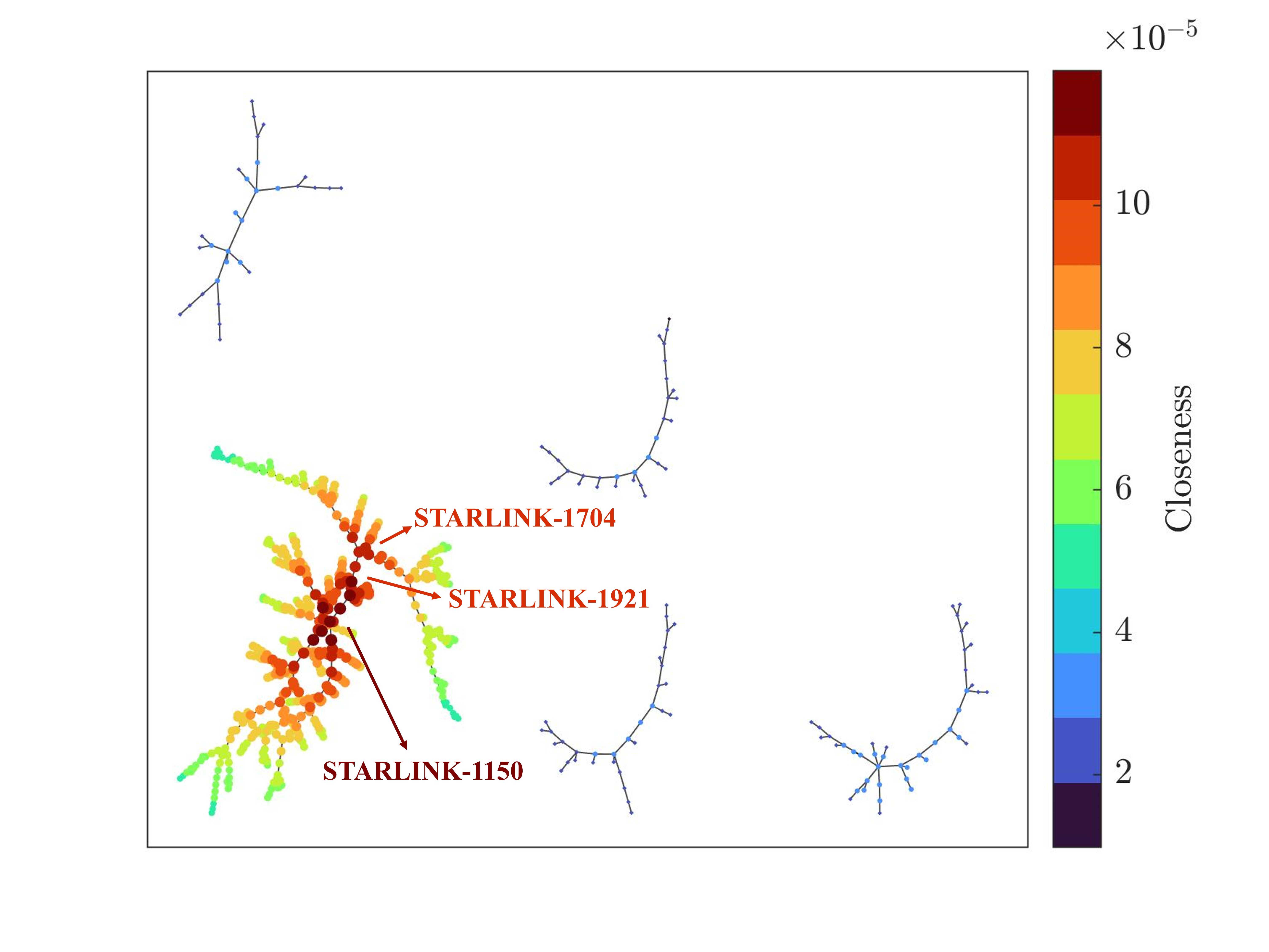}}\\
b) & e)\\
{\includegraphics[width=0.45\textwidth,trim={5cm 3cm 0.5cm 2.5cm},clip]{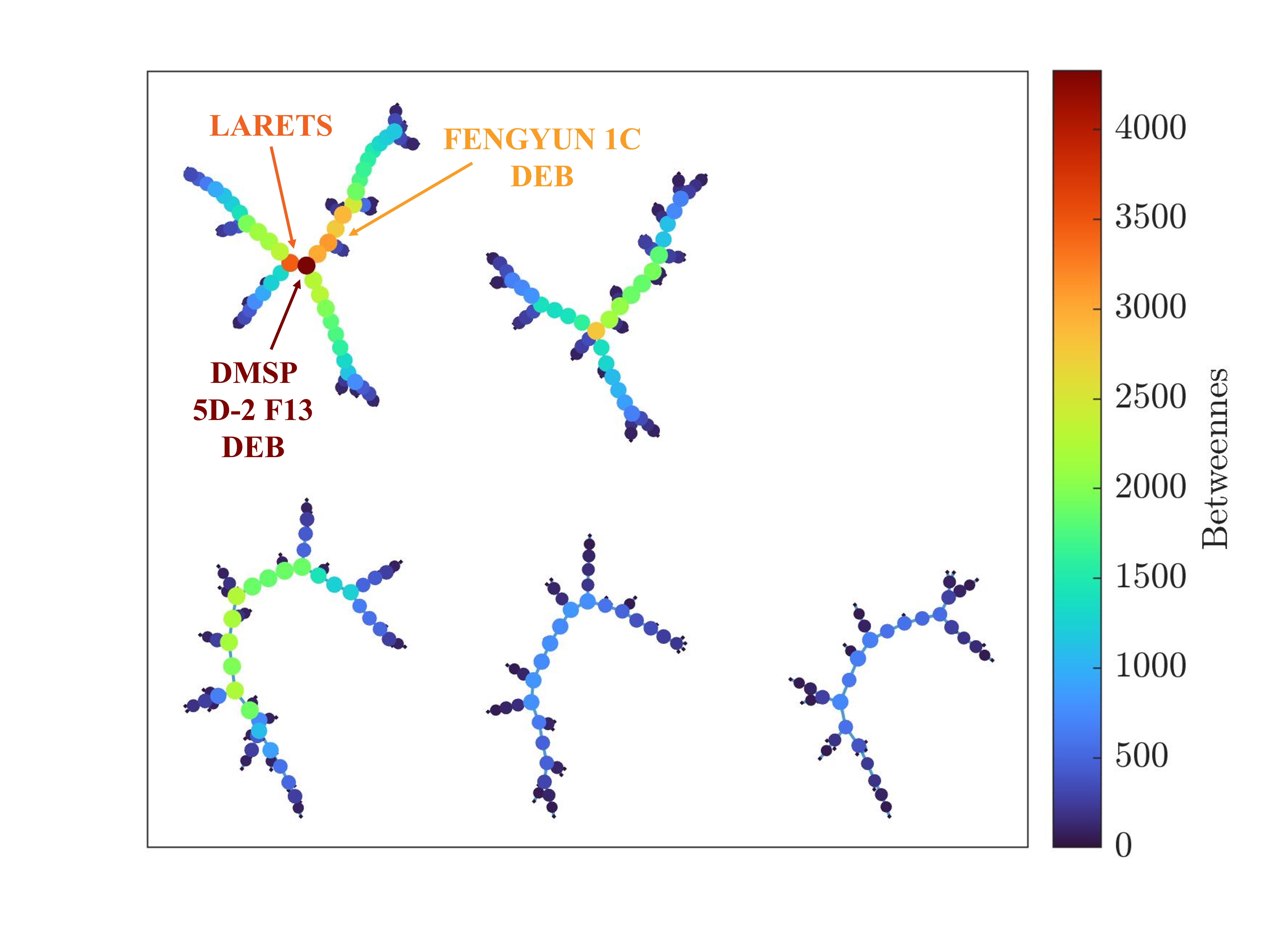}} &
{\includegraphics[width=0.45\textwidth,trim={5cm 3cm 0.5cm 2.5cm},clip]{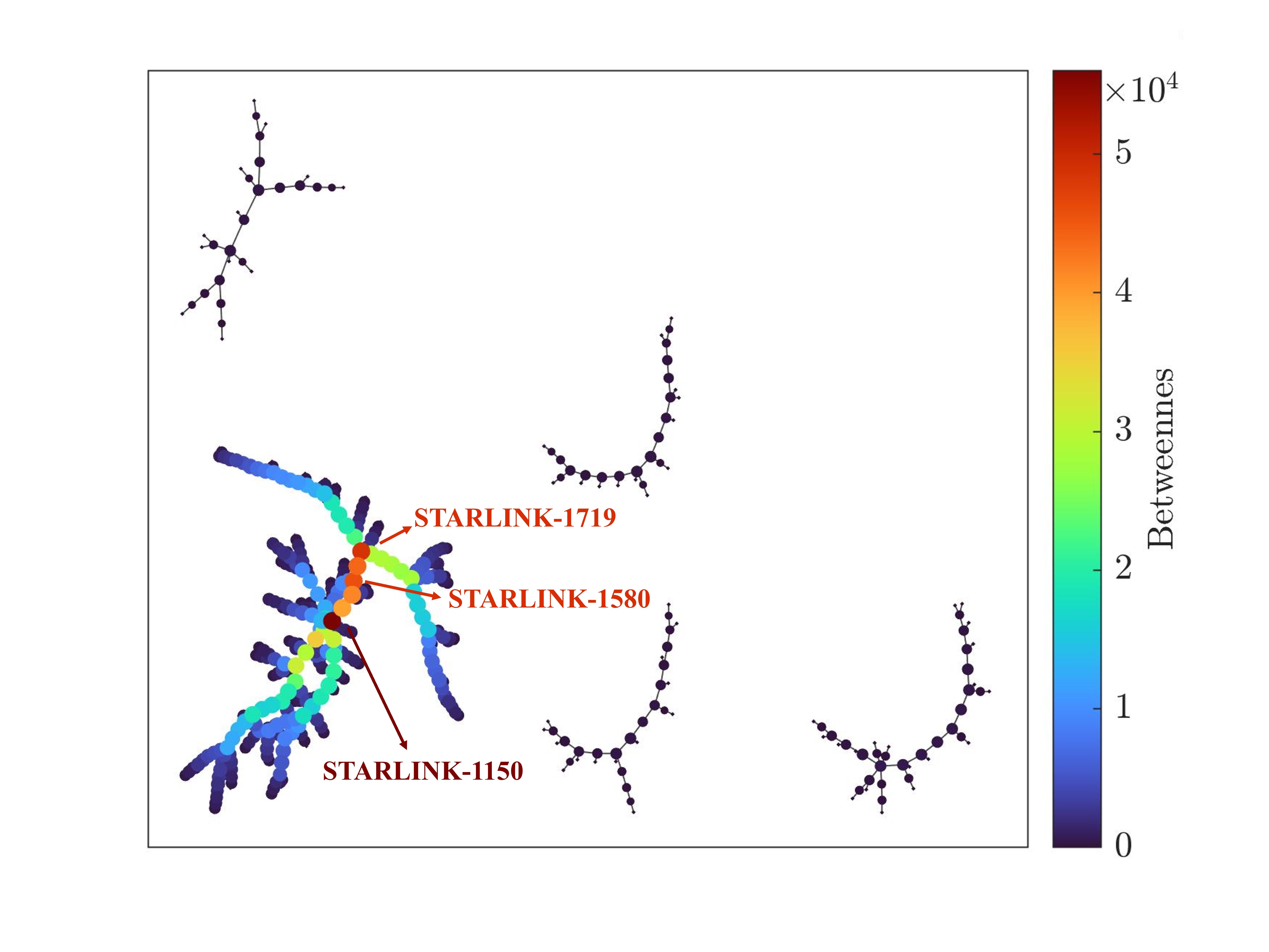}}\\
c) & f)
\end{tabular}
\caption{Network embedding using CDMs ((a) to (c)) and TLEs ((d) to (f)), highlighting various properties: (a,d) node degree, (b,e) closeness, (c,f) betweenness. The values of the properties are mapped using both the size and the colour of the nodes.}
\label{fig:TLEvsCMD_stats}
\end{figure}

\begin{figure}[!htp]
\centering
\begin{tabular}{ll}
{\includegraphics[width=0.45\textwidth,trim={2.5cm 1.5cm 1cm 1cm},clip]{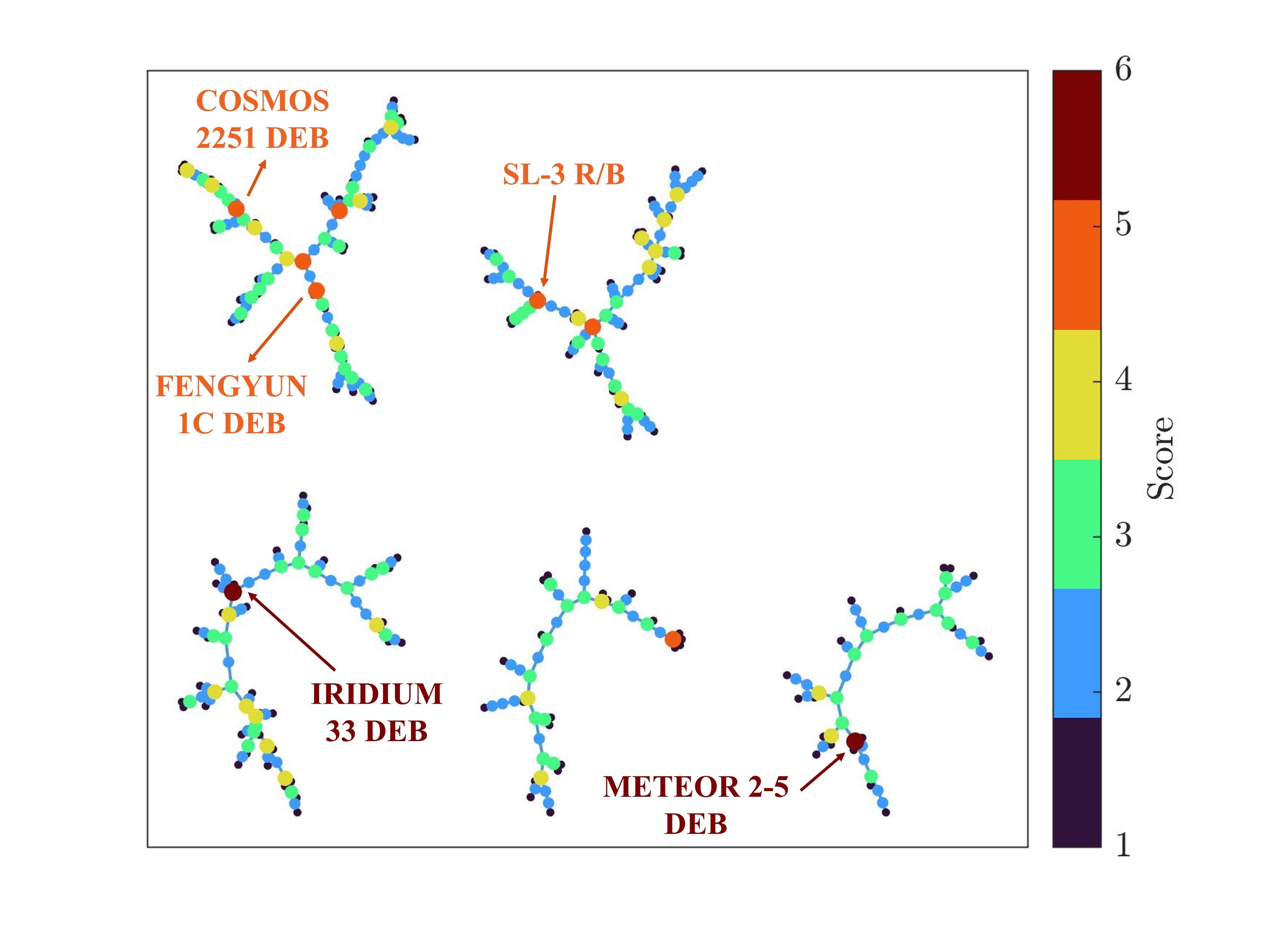}} &
{\includegraphics[width=0.45\textwidth,trim={9cm 6cm 5cm 4cm},clip]{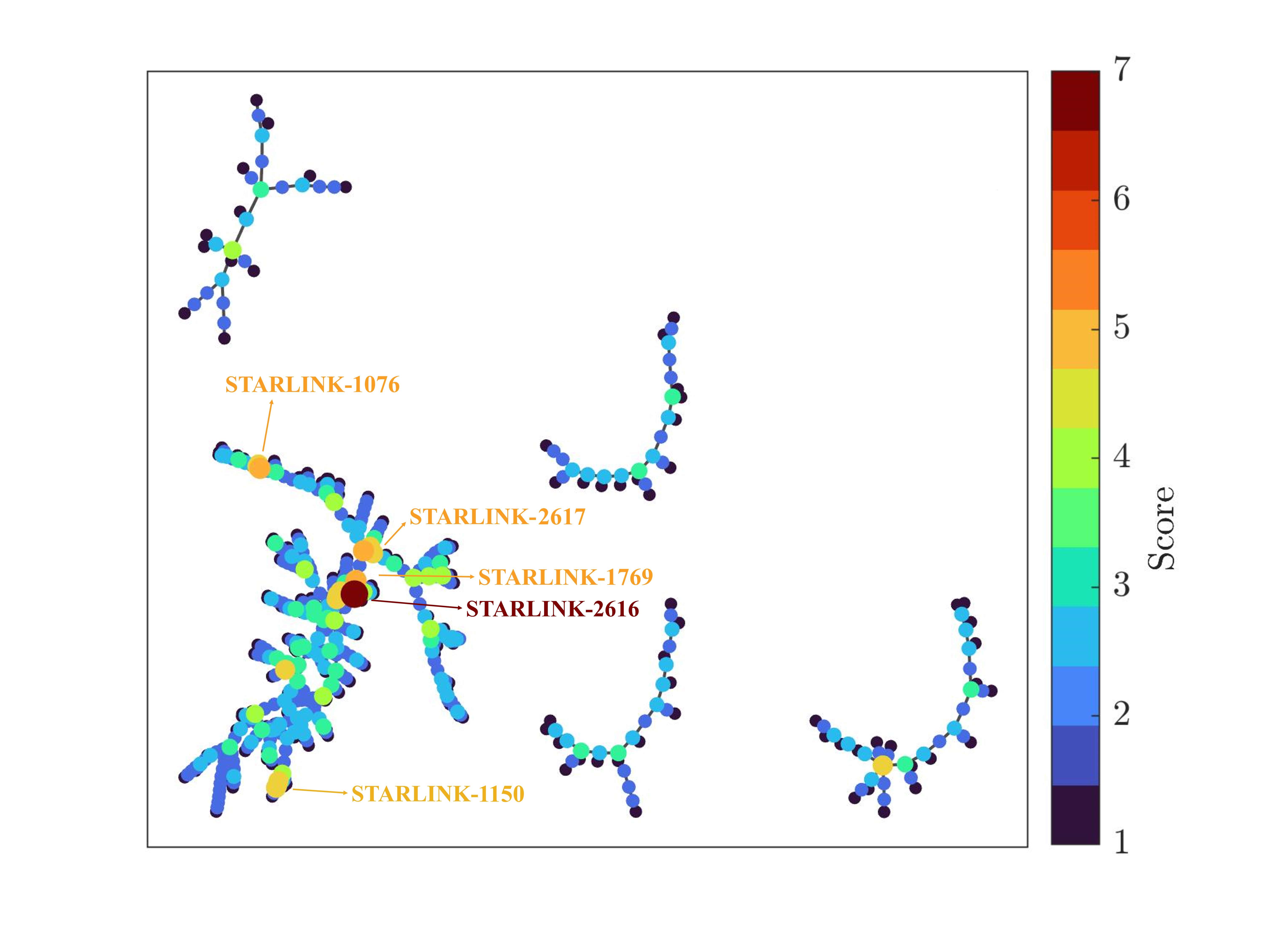}} \\
a) & b)
\end{tabular}
\caption{Network embedding using CDMs (a) and TLEs (b), highlighting relevance score values. The values of the score are mapped using both the size and the colour of the nodes.}
\label{fig:TLEvsCMD_score}
\end{figure}

Figure\,\ref{fig:TLEvsCMD_stats} shows the network embedding obtained from CDMs and TLEs, highlighting the values of the network statistics (degree, closeness, and betweenness, in this order), while Fig.\,\ref{fig:TLEvsCMD_score} highlights the values of the ranking score. Tables\,\ref{table:CDMvsTLE_ranking_degree} to\,\ref{table:CDMvsTLE_ranking_score} report the top 5 objects ranked according to the same network statistics and finally according to the relevance score. \\

\begin{table}
\centering
\caption{Top 5 ranking objects by degree D for the CDM- and TLE-based networks.}
\label{table:CDMvsTLE_ranking_degree} 
\begin{tabular}{ccl|ccl} 
\hline
\multicolumn{3}{c|}{CDMs} 	& \multicolumn{3}{c}{TLEs}\\
\hline
$D$ & NORAD ID & Name	& $D$ & NORAD ID & Name \\
\hline
6 & 38017 & IRIDIUM 33 DEB 	& 7 & 48695 & STARLINK-2616 \\
6 & 39603 & METEOR 2-5 DEB	& 6 & 46382 & STARLINK-1769 \\
5 & 13719 & SL-3 R/B		& 6 & 48689 & STARLINK-2617 \\
5 & 30525 & FENGYUN 1C DEB	& 5 & 44959 & STARLINK-1076 \\
5 & 35705 & COSMOS 2251 DEB	& 5 & 45067 & STARLINK-1150 \\
\hline
\end{tabular}
\end{table}

\begin{table}
\centering
\caption{Top 5 ranking objects by closeness centrality K for the CDM- and TLE-based networks.}
\label{table:CDMvsTLE_ranking_closeness} 
\begin{tabular}{ccl|ccl} 
\hline
\multicolumn{3}{c|}{CDMs} 	& \multicolumn{3}{c}{TLEs}\\
\hline
$K$ & NORAD ID & Name 	& $K$ & NORAD ID & Name \\
\hline
9.2$\cdot 10^{-5}$ & 40681 & DMSP 5D-2 F13 DEB 	& 1.2$\cdot 10^{-4}$ & 45067 & STARLINK-1150 \\
8.9$\cdot 10^{-5}$ & 27944 & LARETS			& 1.1$\cdot 10^{-4}$ & 47625 & STARLINK-1704 \\
8.9$\cdot 10^{-5}$ & 35433 & COSMOS 2251 DEB	& 1.1$\cdot 10^{-4}$ & 46752 & STARLINK-1921 \\
8.6$\cdot 10^{-5}$ & 40291 & CZ-2C DEB		& 1.1$\cdot 10^{-4}$ & 48695 & STARLINK-2616 \\
8.5$\cdot 10^{-5}$ & 31419 & FENGYUN 1C DEB	& 1.0$\cdot 10^{-4}$ & 48148 & STARLINK-2491 \\
\hline
\end{tabular}
\end{table}

\begin{table}
\centering
\caption{Top 5 ranking objects by betweenness centrality B for the CDM- and TLE-based networks.}
\label{table:CDMvsTLE_ranking_betweenness} 
\begin{tabular}{ccl|ccl} 
\hline
\multicolumn{3}{c|}{CDMs} 	& \multicolumn{3}{c}{TLEs}\\
\hline
$B$ & NORAD ID & Name 	& $B$ & NORAD ID & Name \\
\hline
4326 & 40681 & DMSP 5D-2 F13 DEB 	& 55986 & 45067 & STARLINK-1150 \\
3443 & 27944 & LARETS				& 48691 & 46331 & STARLINK-1719 \\
3104 & 31419 & FENGYUN 1C DEB		& 45798 & 46041 & STARLINK-1580 \\
2993 & 35433 & COSMOS 2251 DEB		& 43800 & 48360 & STARLINK-2622 \\
2899 & 43326 & COSMOS 1867 COOLANT	& 41521 & 48695 & STARLINK-2616 \\
\hline
\end{tabular}
\end{table}

\begin{table}
\centering
\caption{Top 5 ranking objects by score $\mathcal{S}$ for the CDM- and TLE-based networks.}
\label{table:CDMvsTLE_ranking_score} 
\begin{tabular}{ccl|ccl} 
\hline
\multicolumn{3}{c|}{CDMs} 	& \multicolumn{3}{c}{TLEs}\\
\hline
$\mathcal{S}$ & NORAD ID & Name 	& $\mathcal{S}$ & NORAD ID & Name \\
\hline
6.0 & 38017 & IRIDIUM 33 DEB 	& 7.0 & 48695 & STARLINK-2616 \\
6.0 & 39603 & METEOR 2-5 DEB	& 6.3 & 46382 & STARLINK-1769 \\
5.0 & 13719 & SL-3 R/B			& 6.3 & 48689 & STARLINK-2617 \\
5.0 & 30525 & FENGYUN 1C DEB	& 6.3 & 44959 & STARLINK-1076 \\
5.0 & 35705 & COSMOS 2251 DEB	& 6.0 & 45067 & STARLINK-1150 \\
\hline
\end{tabular}
\end{table}

By observing the various rankings reported in the Tables, it becomes clear what {was} previously remarked regarding the composition of the networks. Since the initial conditions and propagation models used to generate the two data sets are fundamentally different, the composition of the networks are also different, resulting in distinct rankings across the two cases.

In particular, as already pointed out, the CDM-based rankings tend to emphasize mostly debris, while payloads tend to appear in the TLE-based ones. The prominent presence of the Starlink satellites in the top positions is another reminder that the precision of the TLE-based model (both the initial conditions and the propagator) is an important actor shaping the network. \\

More considerations can be made by observing the values of the various ranking parameters presented in the tables, instead. 

In both cases, closeness and betweenness values tend to favour, as expected, objects in the densest regions (see Fig.\,\ref{fig:TLEvsCMD_stats} (b), (c), (e), and (f)), but their values contribute to the total score less than the degree does, while the degree appears as the main contributor to the relevance score (see Tables\,\ref{table:CDMvsTLE_ranking_degree} and\,\ref{table:CDMvsTLE_ranking_score}).

In fact, the centrality measures appear only in the third contribution of the relevance score, that is, the contribution considering the effects of cascading collisions. This effect is, by definition, smaller, since it represents events which are conditional on others to occur and, thus, is expressed as a power law, which decreases very quickly. On the contrary, the degree contribution represents the effects of direct collisions. 

Moreover, the structure of the networks in the cases under study, composed mostly by branching chains, tends to penalise the two centrality measures and the clustering coefficient, which are more fit to describe dense interconnected networks. A prominent difference among the Betweenness values of the CDM-based and TLE-based networks can be seen in Table\,\ref{table:CDMvsTLE_ranking_betweenness}, which is also an effect of the network structure: since the TLE-based networks has larger components, a higher number of paths pass through each node, increasing the betweenness values.

Another difference visible from Figures\,\ref{fig:TLEvsCMD_stats} and\,\ref{fig:TLEvsCMD_score} is the one between nodes lying near the densest regions of the connected components and the nodes lying on their boundaries. The latter are clearly characterised by low values of relevance score, which is expected since they have low degree (equal to $1$ or $2$ in most cases) and, by definition, low centrality, while the opposite is true for the nodes lying in the densest regions. Once again, these considerations help identify the ``core'' nodes of the network, which contribute the most in terms of risk of collisions. \\

Clustering values are not shown since in most cases they are zero, due to the the branched structure of most connected components which leads to a lack of connected pairs between nodes. \\

Overall, some objects appear in the top rankings regardless of the choice of the resolving parameter (such as FENGYUN 1C DEB, LARETS, COSMOS 2251 DEB, and IRIDIUM 33 DEB), meaning that this score definition is able to capture in some measure how these objects contribute to the overall risk of collisions within the population. Following the example set by McKnight et al. \cite{mcknight2021}, App.\,\ref{app:C} shows the top $50$ objects ranked by relevance score for both test cases. While the prevalence of Starlink satellites in the TLE-based network has already been addressed, the CDM-based one mainly shows mostly debris and a few of objects appearing in this ranking.

\subsection{Sensitivity analysis}
\label{sec:sensitivity}

This section studies the dependence of the network's structure and properties on the main parameters of the analysis, nominally the propagation time $T$ and the conjunction distance threshold $\epsilon$, and finally how they affect the characteristics of the network such as the number of connected components, their size, and the values of some of the key statistics and the relevance score. 

In both CDM and TLE cases, the analysis considers different values of the propagation time and of the conjunction threshold. However, due to the way CDMs are generated, using a fixed distance of $1\, \mathrm{km}$, only the TLE-based network will be analysed with varying distances.

\begin{figure}[!htp]
\centering
\begin{tabular}{rlrl}
a) & {\includegraphics[width=0.25\textwidth,trim={6.5cm 4cm 4.5cm 2.5cm},clip]{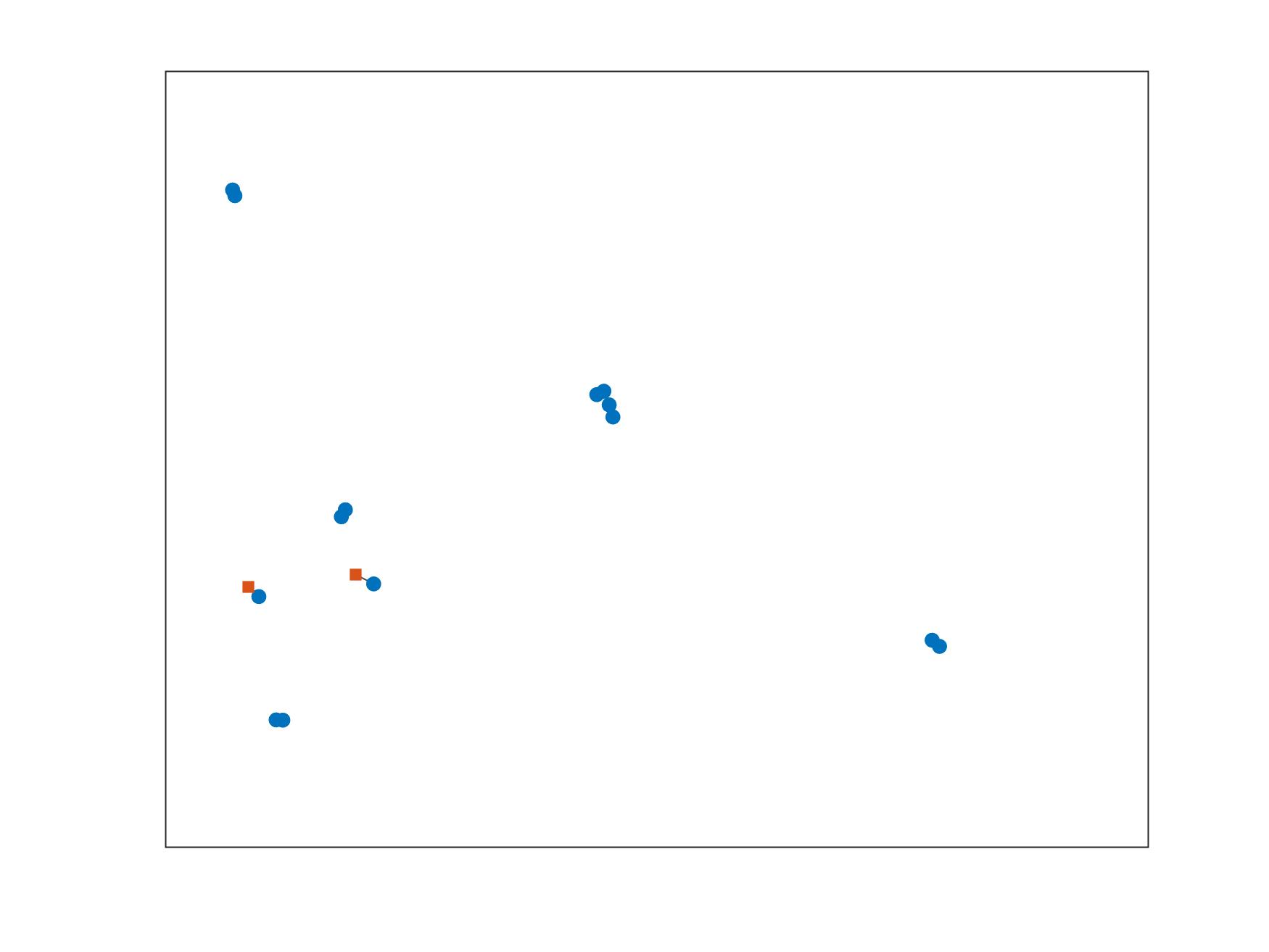}} & e) & {\includegraphics[width=0.25\textwidth,trim={6.5cm 4cm 4.5cm 2.5cm},clip]{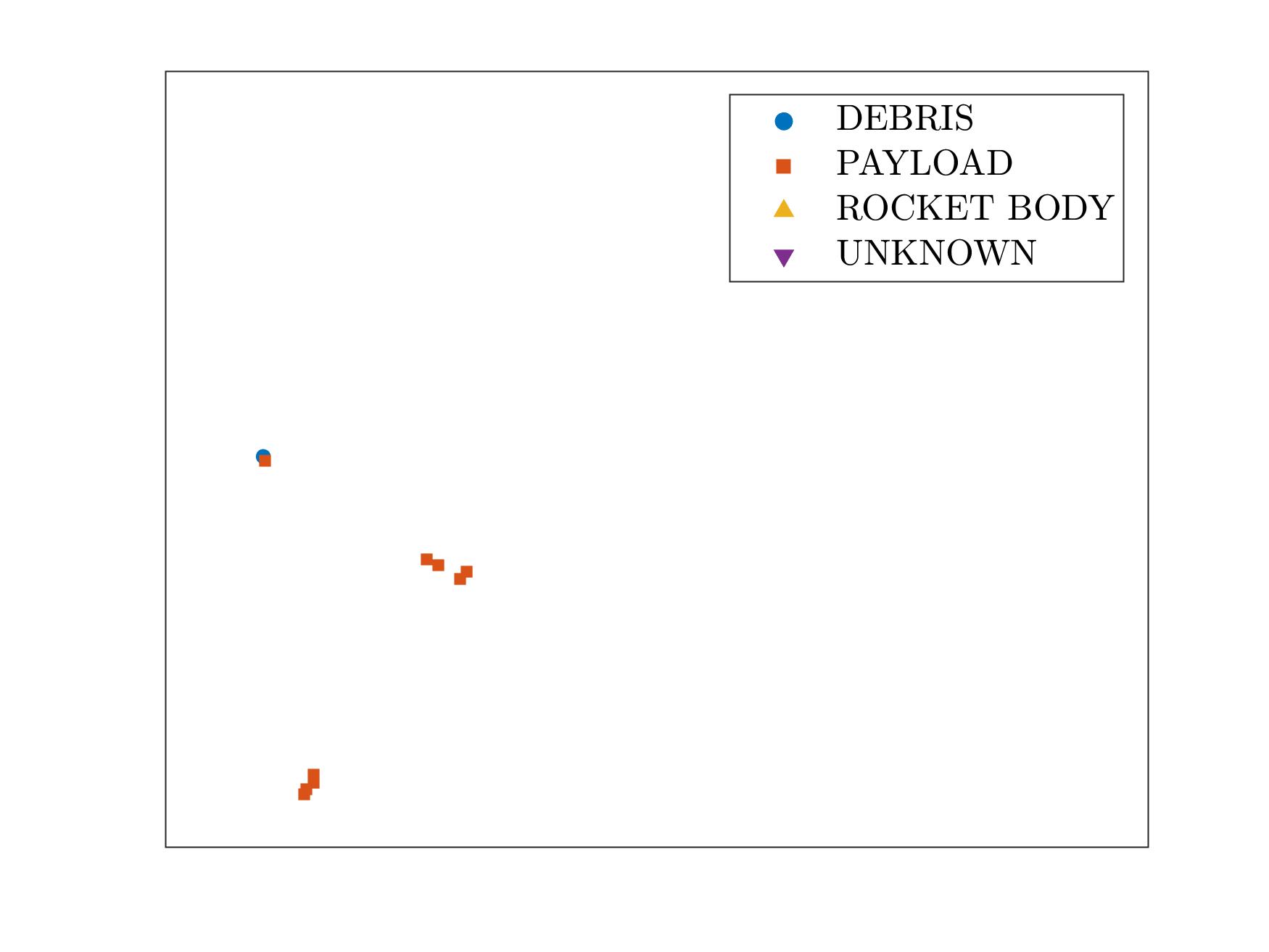}}\\
b) & {\includegraphics[width=0.25\textwidth,trim={6.5cm 4cm 4.5cm 2.5cm},clip]{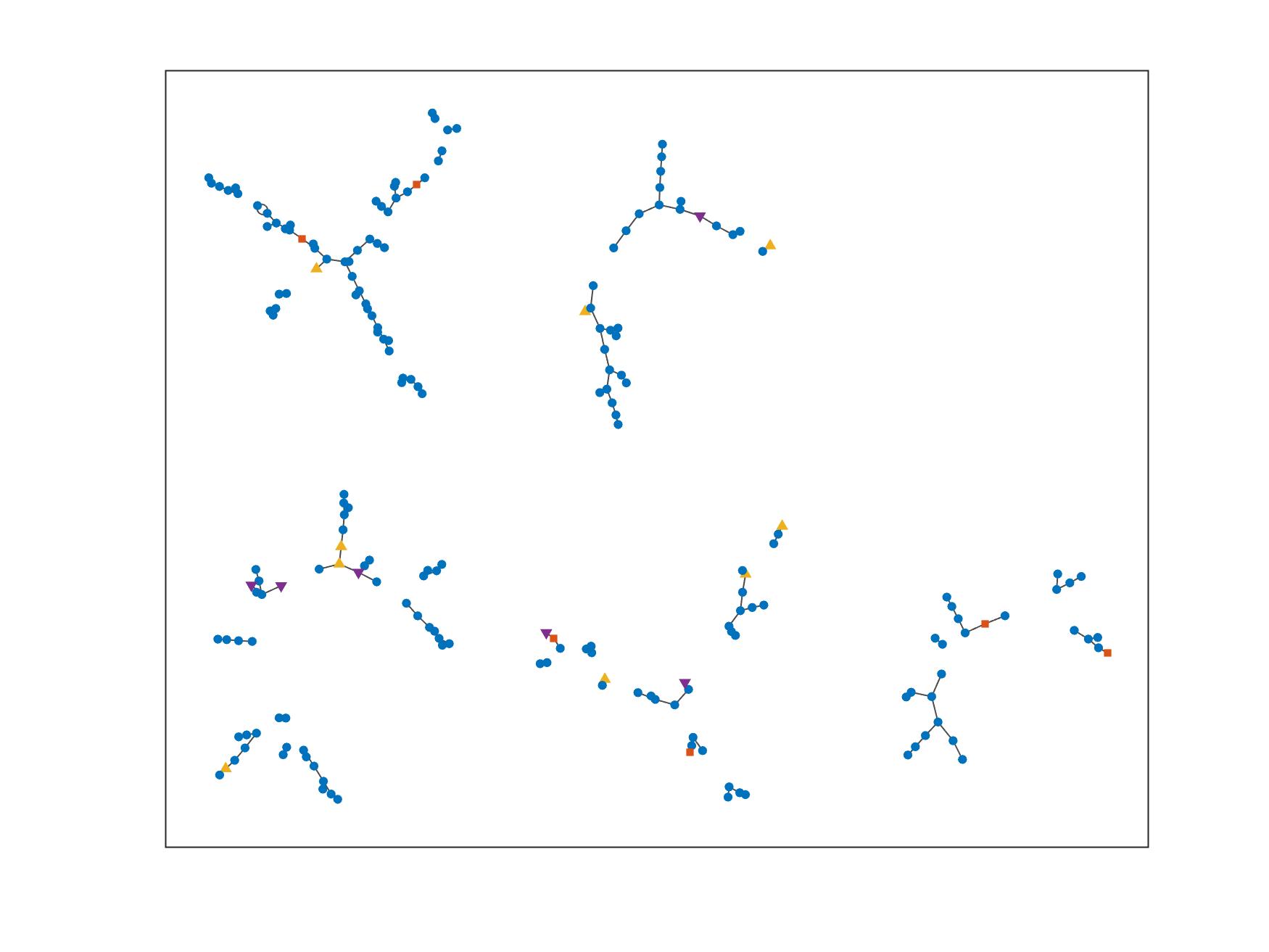}} & f) & {\includegraphics[width=0.25\textwidth,trim={6.5cm 4cm 4.5cm 2.5cm},clip]{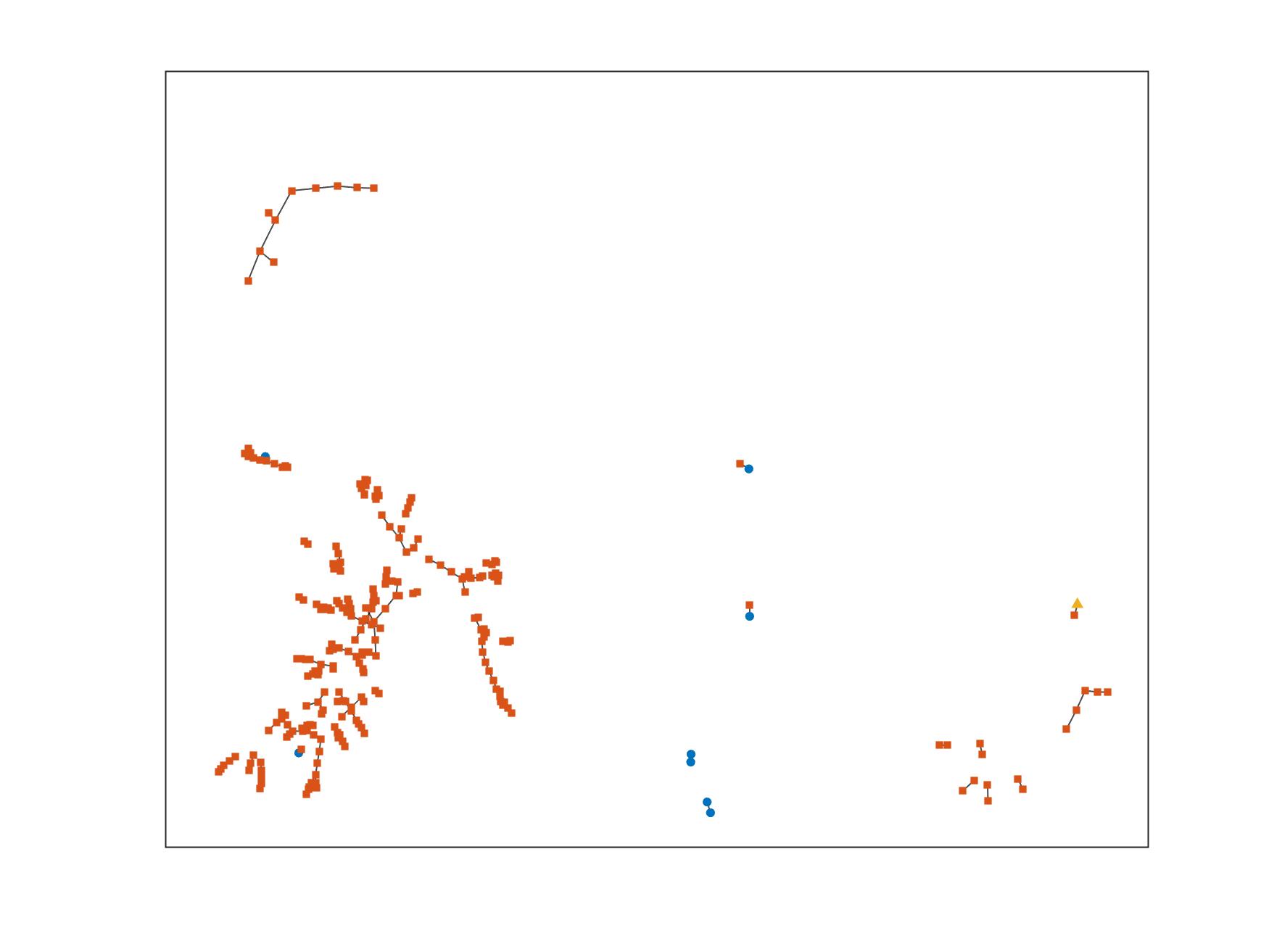}}\\
c) & {\includegraphics[width=0.25\textwidth,trim={6.5cm 4cm 4.5cm 2.5cm},clip]{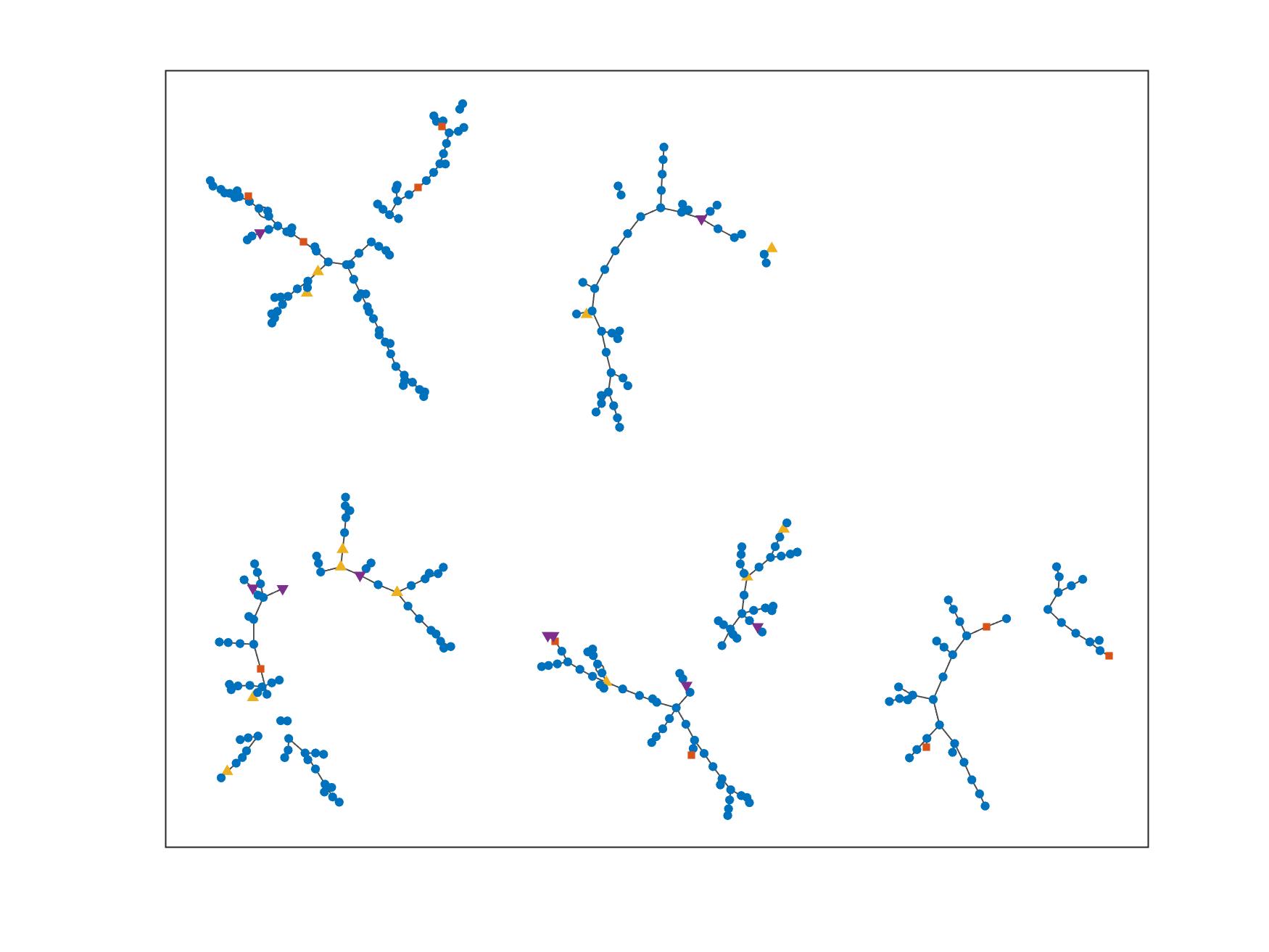}} & g) & {\includegraphics[width=0.25\textwidth,trim={6.5cm 4cm 4.5cm 2.5cm},clip]{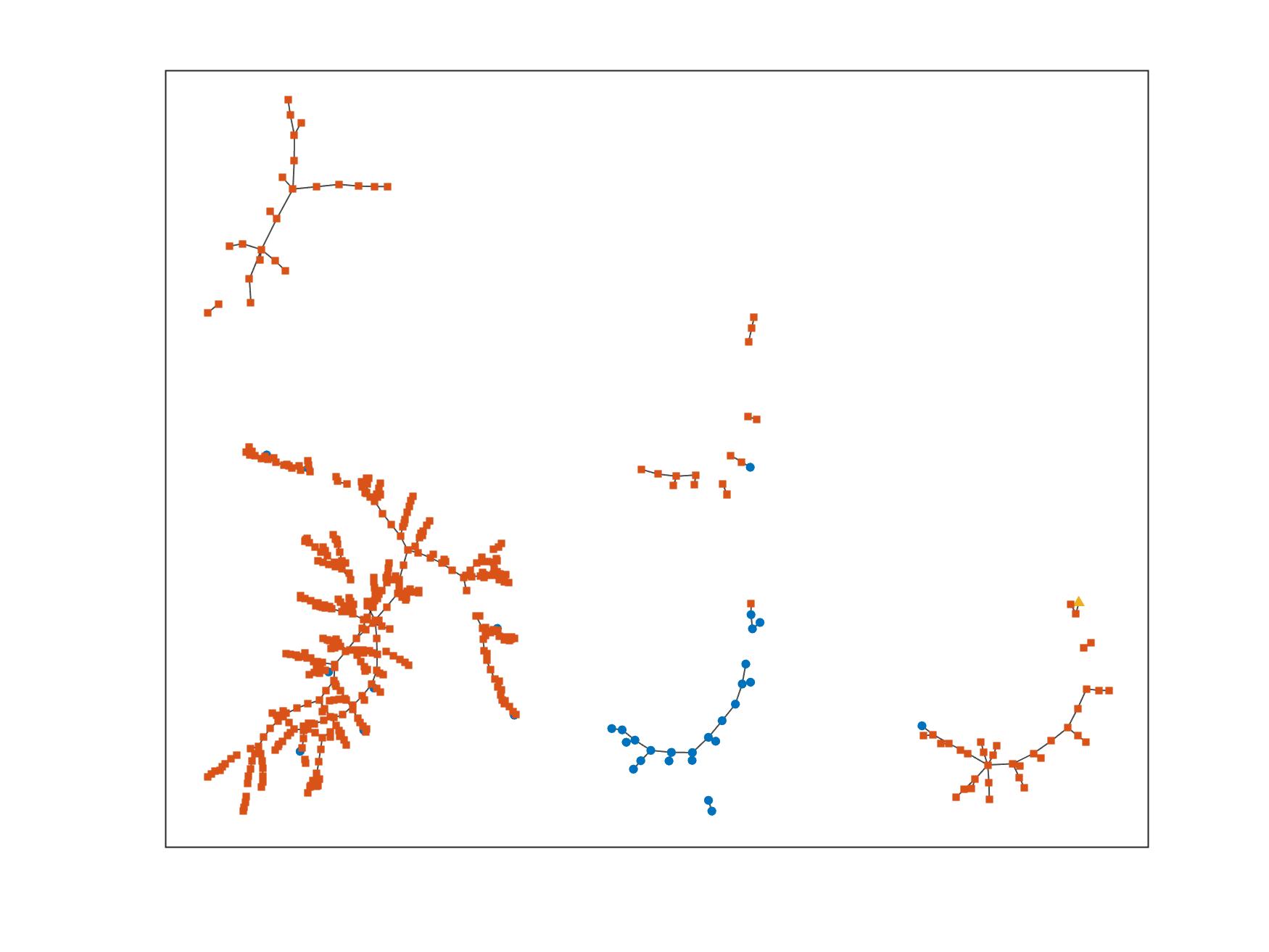}}\\
d) & {\includegraphics[width=0.25\textwidth,trim={6.5cm 4cm 4.5cm 2.5cm},clip]{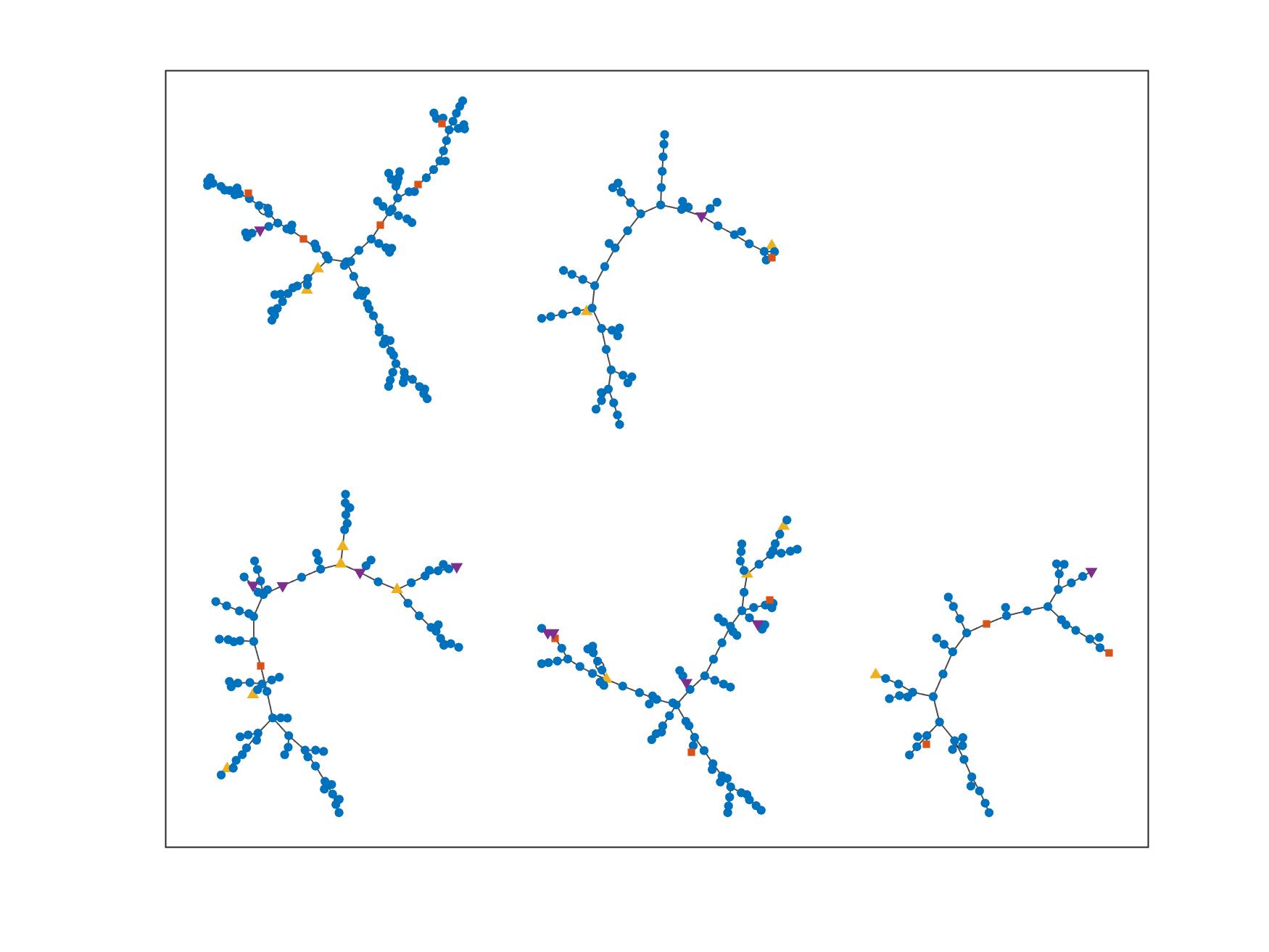}} & h) & {\includegraphics[width=0.25\textwidth,trim={6.5cm 4cm 4.5cm 2.5cm},clip]{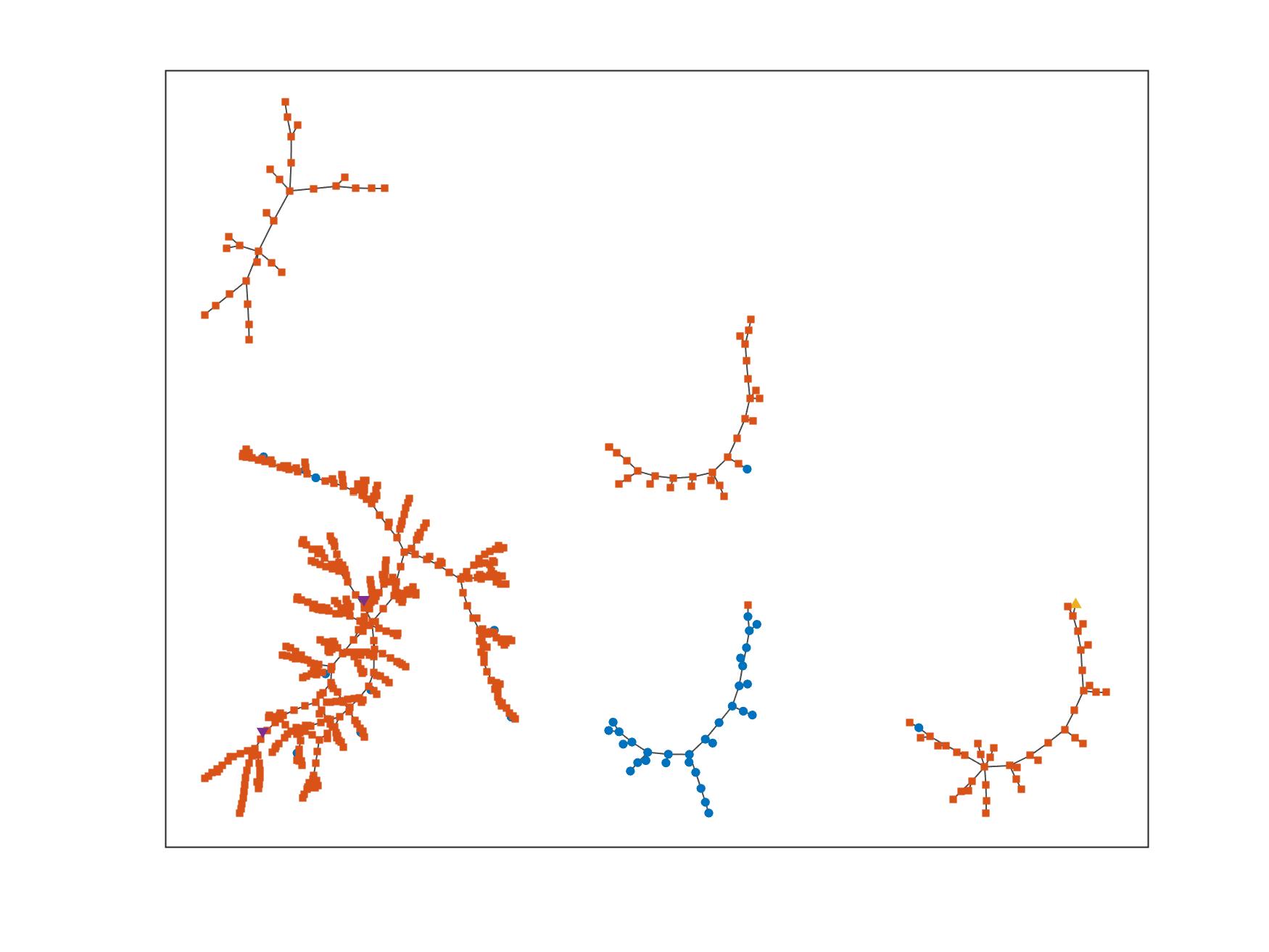}}\\
\end{tabular}
\caption{Variation of the network embeddings from the CDM (a-d) and TLE (e-g) datasets according to the propagation time, respectively set to 1 (a, e), 10 (b, f), 20 (c, h), and 30 (d, g) days, using a fixed threshold distance of 3 km for the TLE-based one.}
\label{fig:TLEvsCMD_sensitivity1}
\end{figure}

\begin{figure}[!htp]
\centering
\begin{tabular}{ll}
a) & {\includegraphics[width=0.45\textwidth,trim={6.5cm 4cm 4.5cm 2.5cm},clip]{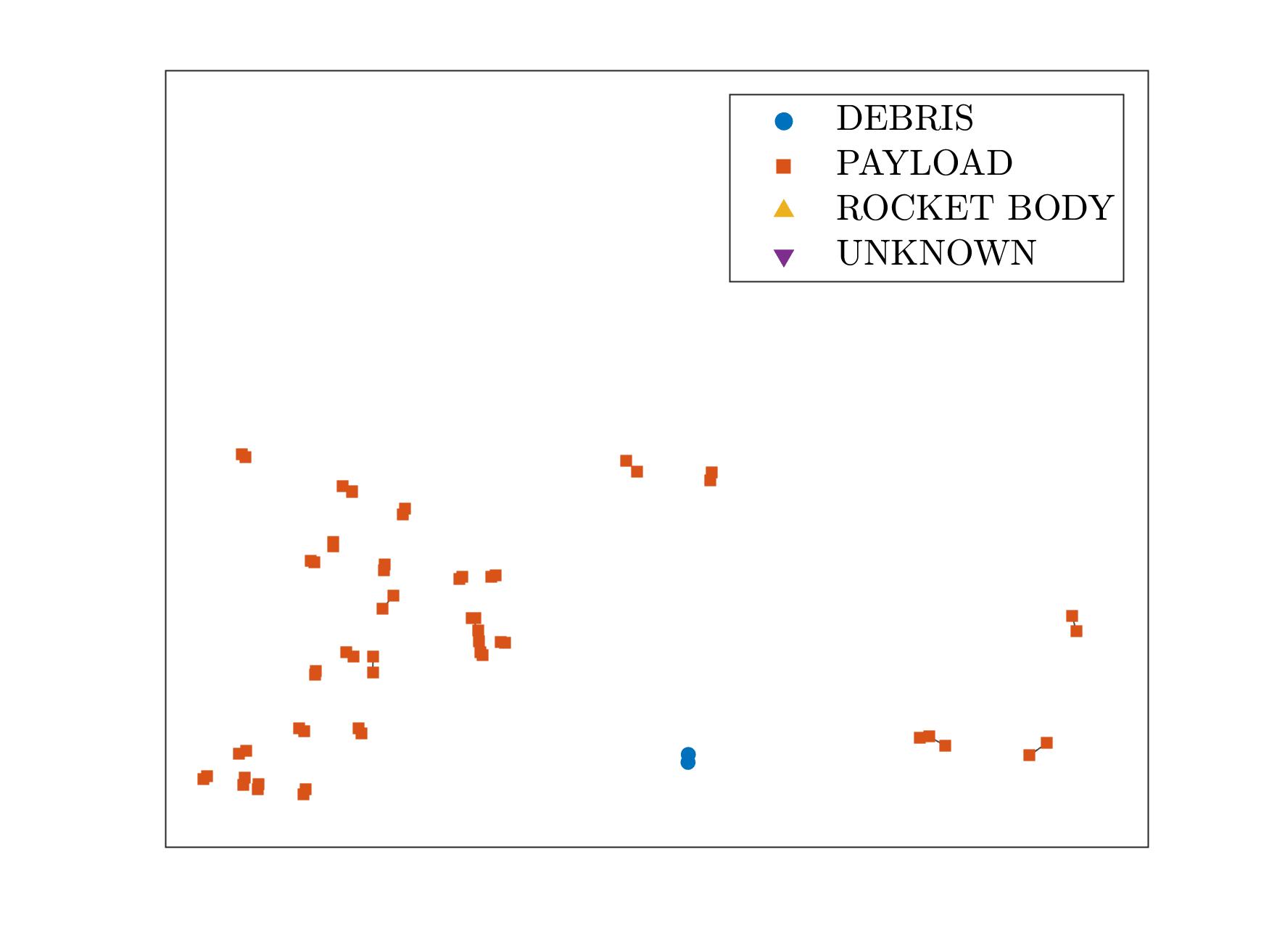}}\\
b) & {\includegraphics[width=0.45\textwidth,trim={6.5cm 4cm 4.5cm 2.5cm},clip]{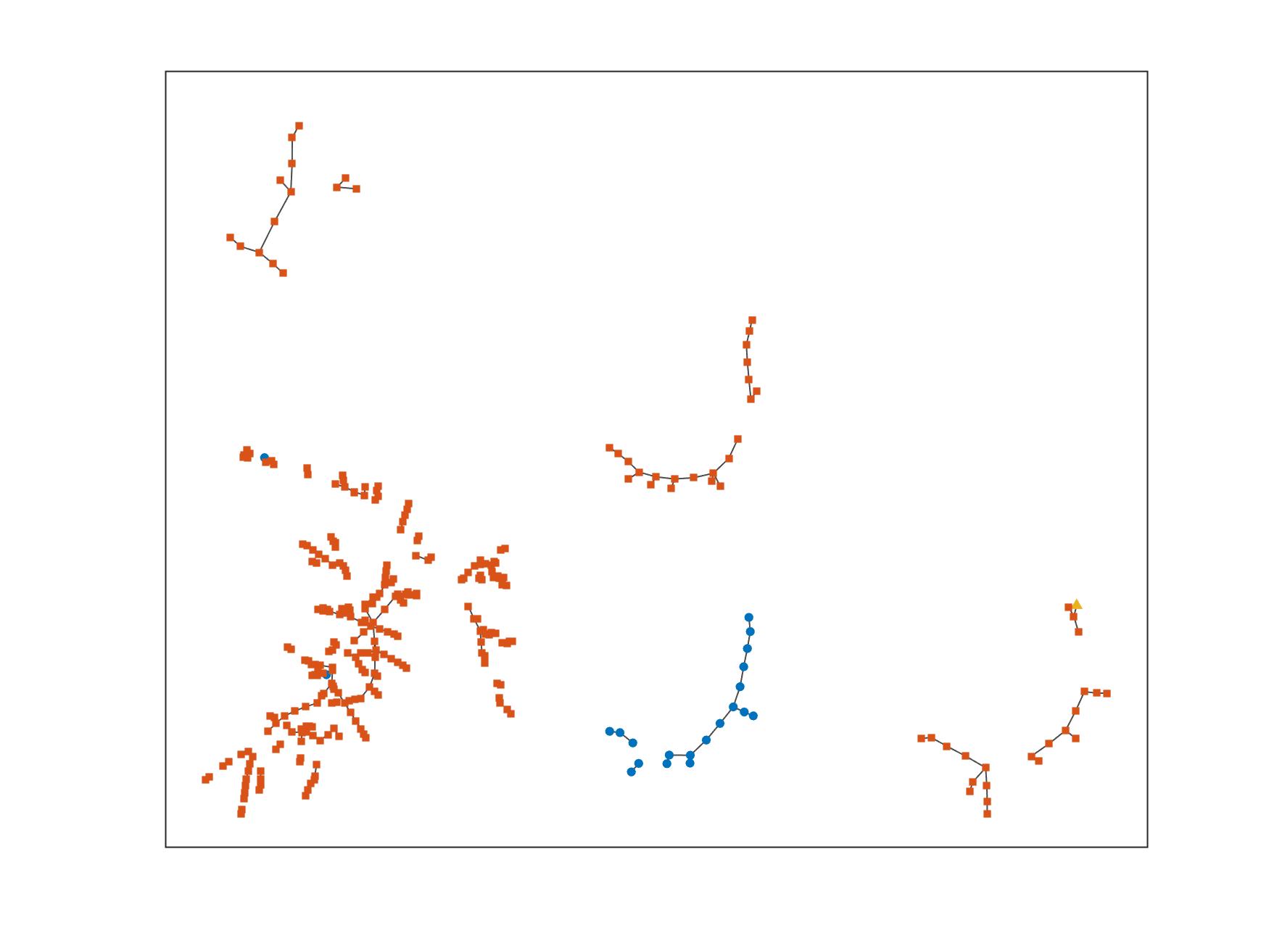}}\\
c) & {\includegraphics[width=0.45\textwidth,trim={6.5cm 4cm 4.5cm 2.5cm},clip]{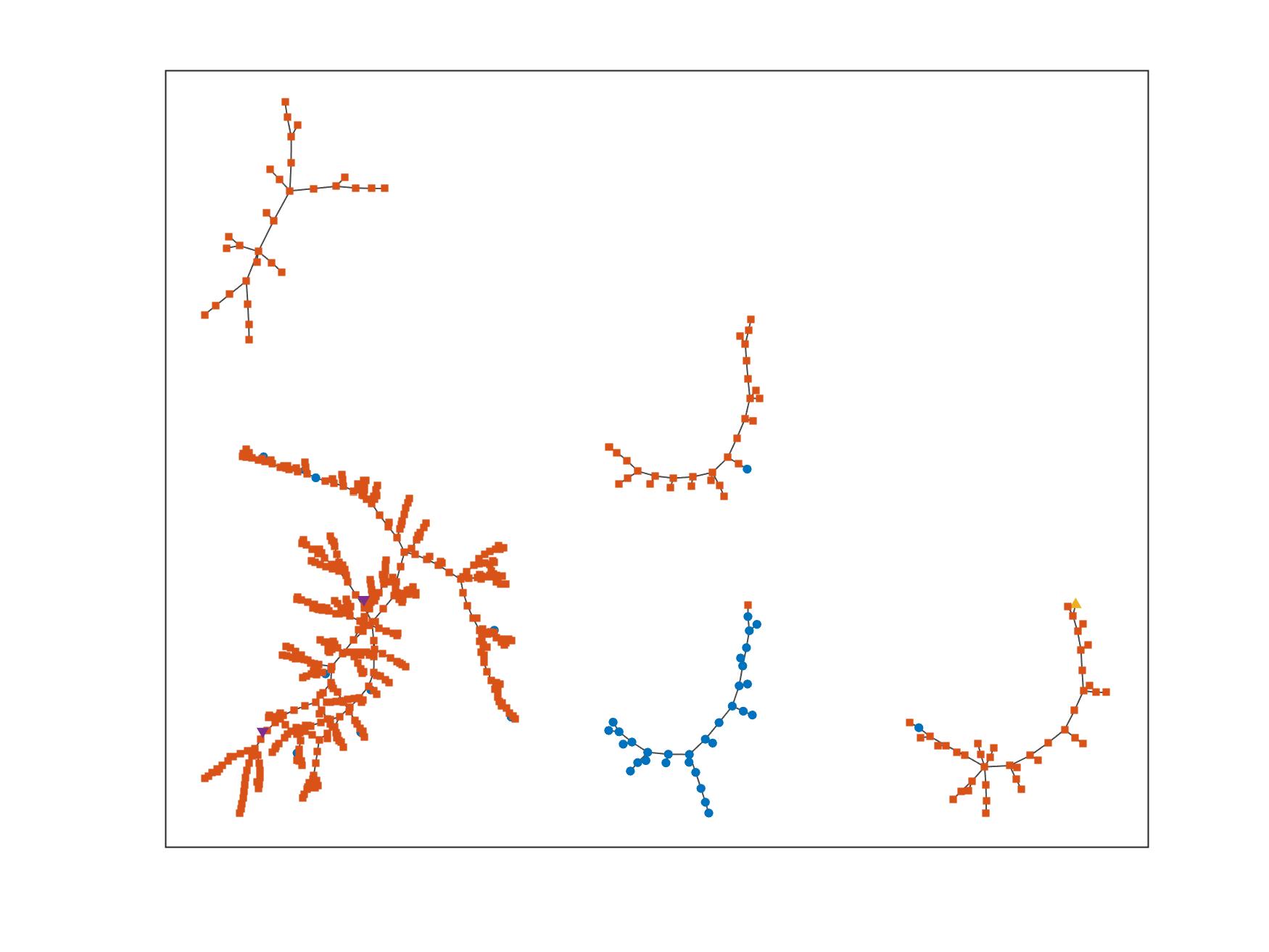}}\\
\end{tabular}
\caption{Variation of the network embeddings from the TLE dataset according to the threshold distance, respectively set to 1, 2, and 3 km from (a) to (c), evaluated after the 30 days propagation.}
\label{fig:TLEvsCMD_sensitivity2}
\end{figure}

Figures\,\ref{fig:TLEvsCMD_sensitivity1} and\,\ref{fig:TLEvsCMD_sensitivity2} show the growth of the networks for increasing values of the propagation time and of the conjunction distance threshold. As the time and distance threshold increase, the network transits from a sparse collection of small separate components to few highly connected clusters. Increasing the propagation time makes detecting a conjunction eventually easier, while increasing the distance threshold makes conjunctions more frequent to detect. Following this change to the extreme, a large enough distance would result in a network made of one fully connected component, since all objects would eventually fall within the threshold between each other. 

\begin{figure}[!htp]
\centering
{\includegraphics[width=0.6\textwidth,trim={3.5cm 2cm 3.5cm 3cm},clip]{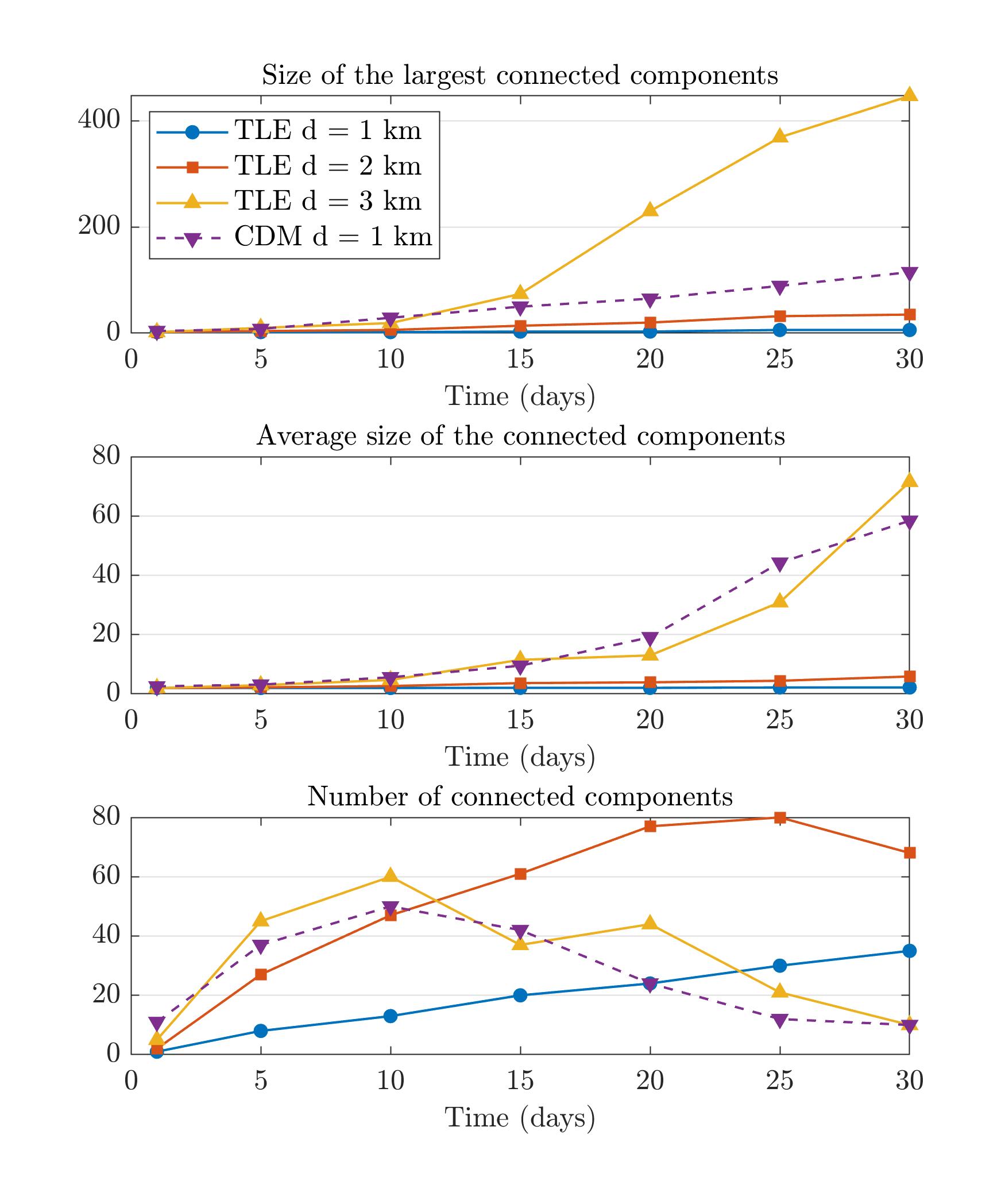}}
\caption{Variation of the number and size of the connected components with respect to the propagation time and conjunction distance.}
\label{fig:may_tles_30day_sens-count}
\end{figure}

\begin{figure}[!htp]
\centering
\begin{tabular}{cc}
{\includegraphics[width=0.45\textwidth,trim={4cm 2.5cm 3.5cm 3cm},clip]{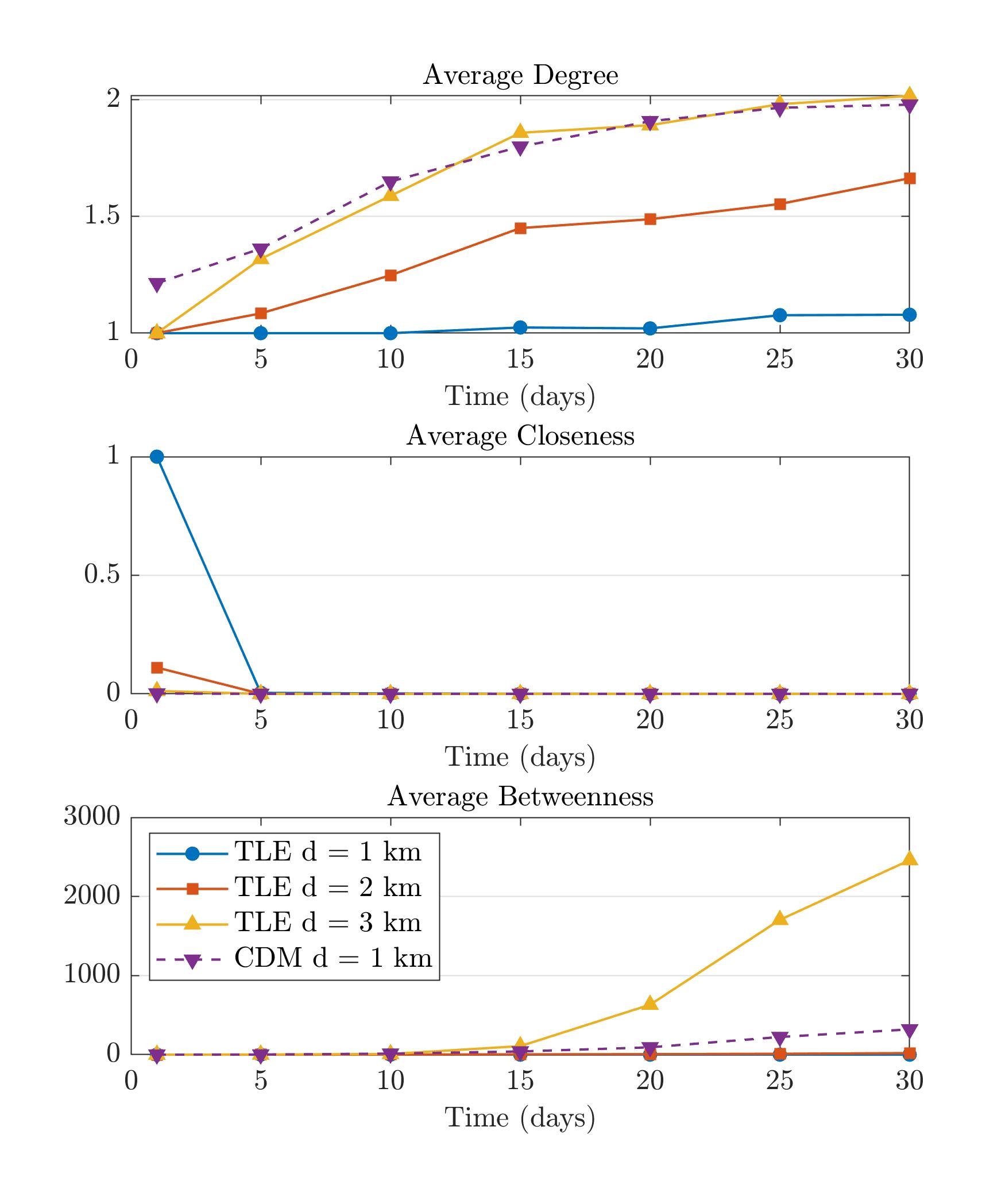}} & {\includegraphics[width=0.45\textwidth,trim={4cm 2.5cm 3.5cm 3cm},clip]{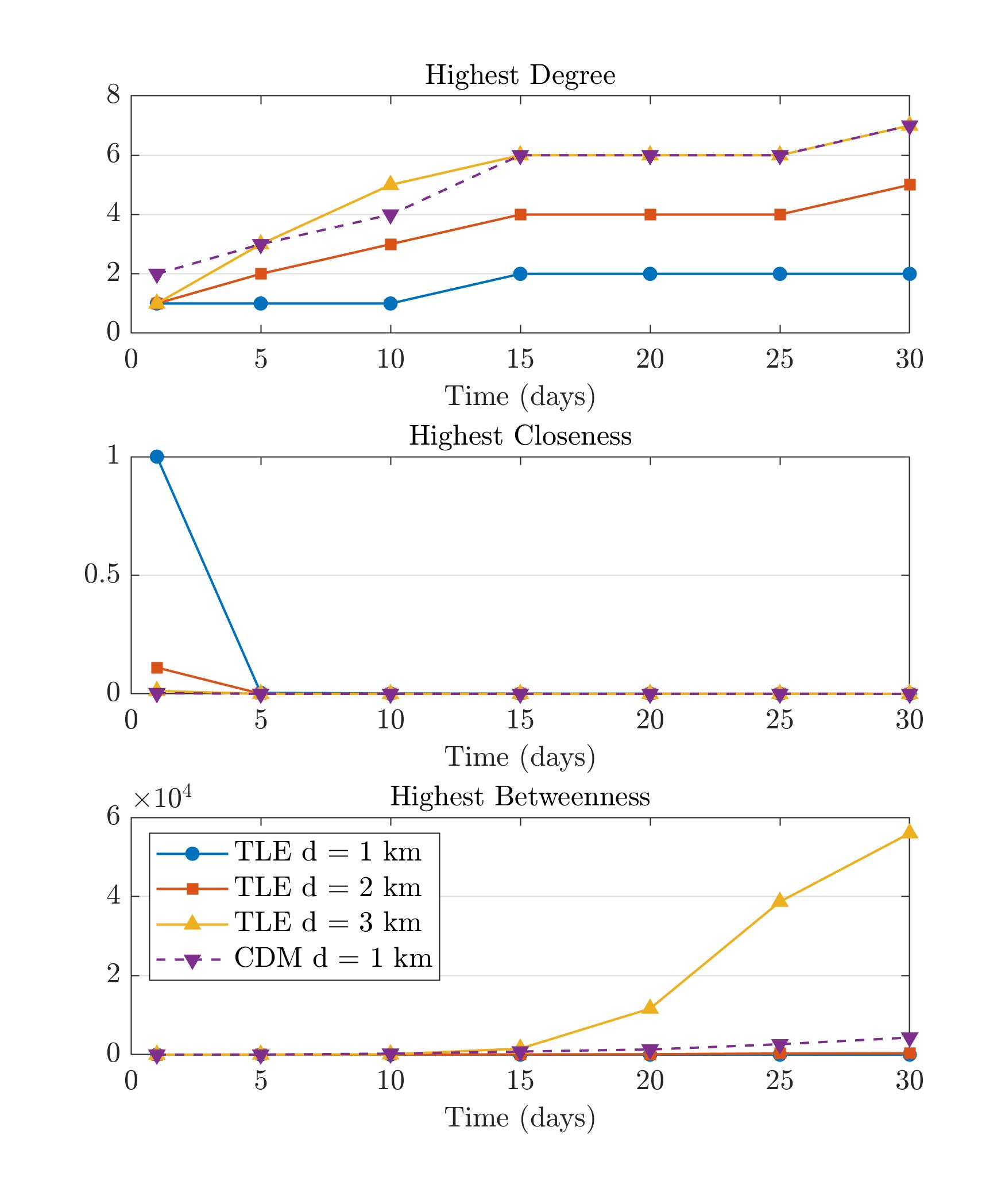}}
\end{tabular}
\caption{Variation of the average (left) and highest (right) statistics of the network (degree, closeness, and betweenness) with respect to the propagation time and conjunction distance.}
\label{fig:may_tles_30day_sens-stats}
\end{figure}

This is visible from Fig.\,\ref{fig:may_tles_30day_sens-count}, which shows that the number of connected components of the network increases with the propagation time for low values of the threshold distance, but starts decreasing again for larger distance values as the components starts connecting to each other, forming few larger clusters. This affects the relevant statistics of the network as well, shown in Fig.\,\ref{fig:may_tles_30day_sens-stats}: as the number of connections between nodes increases, components grow in size and the degree grows as well. Since non-adjacent nodes end up being more easily connected, the closeness coefficient (related to the average distance between nodes) drops, while betweenness (the number of paths passing through a node) show the opposite trend.

\begin{figure}[!htp]
\centering
\begin{tabular}{cc}
{\includegraphics[width=0.45\textwidth,trim={4.5cm 0cm 3.5cm 0cm},clip]{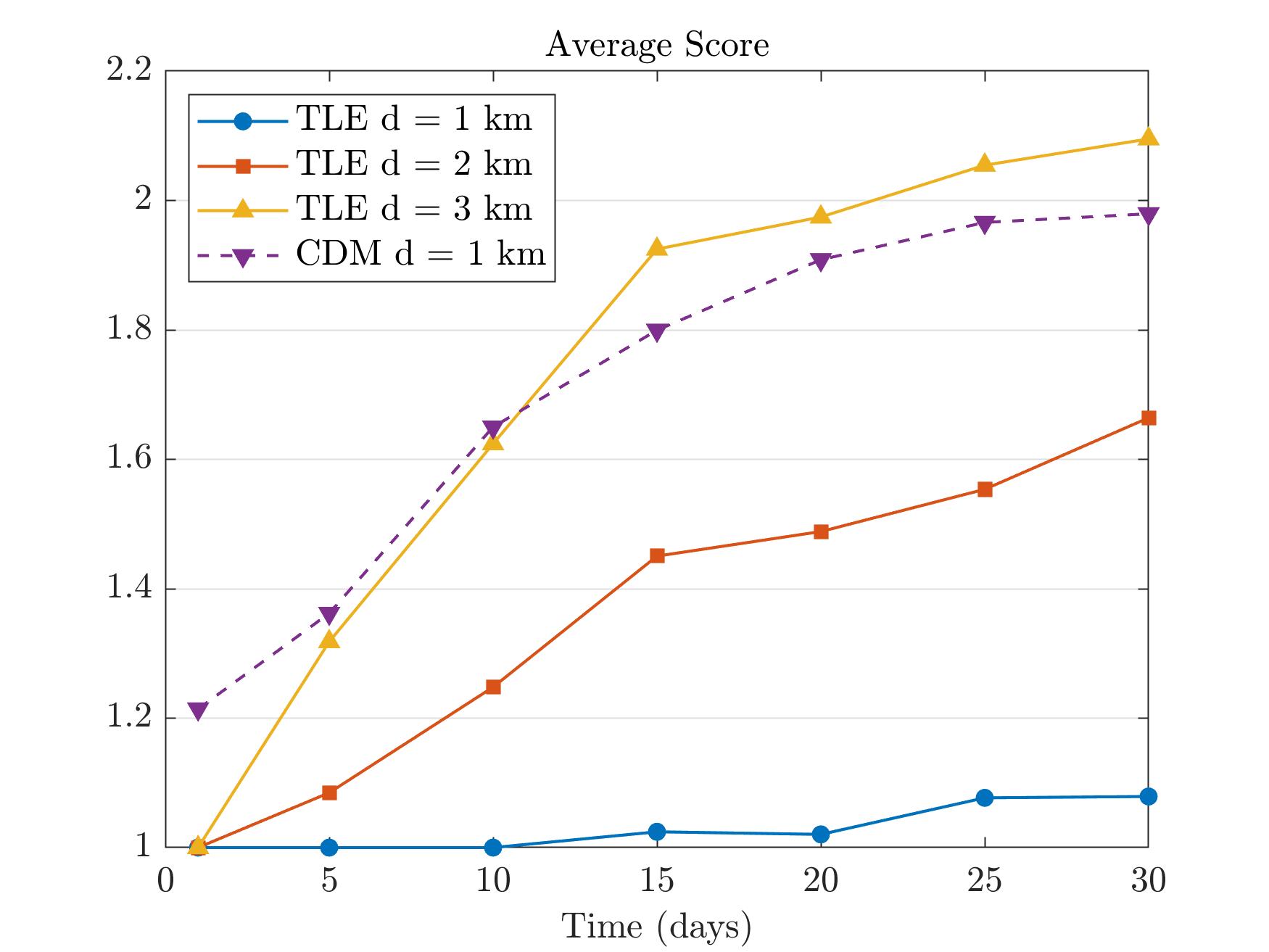}} & {\includegraphics[width=0.45\textwidth,trim={4.5cm 0cm 3.5cm 0cm},clip]{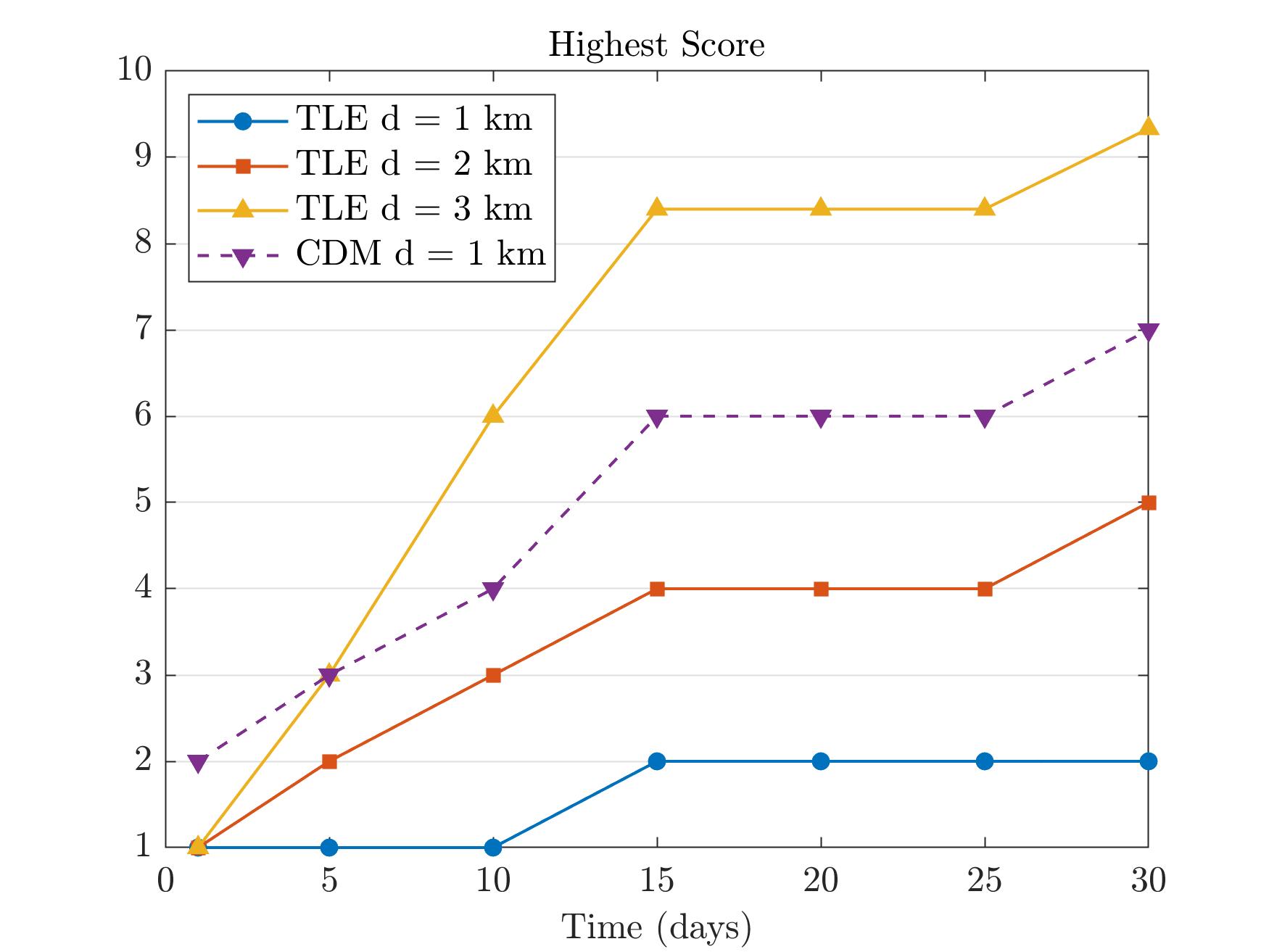}}
\end{tabular}
\caption{Variation of the average (left) and highest (right) relevance score with respect to the propagation time and conjunction distance.}
\label{fig:may_tles_30day_sens-score}
\end{figure}

The relevance score reflects all these changes, as seen in Fig.\,\ref{fig:may_tles_30day_sens-score}. The effects of the dependency of the network from $T$ and $\epsilon$ alter the way the risk of collisions propagates through the population, as even distant nodes (conjunctions that are separated by multiple events) become able to affect each other. Overall, the risk of collisions, direct or indirect, grows.

\section{Conclusions and future work}
\label{sec:conclusions}

This manuscript presented a novel method to address the complexity of the orbital environment around Earth with the use of network theory. The population of RSOs is embedded into a network, the RSO network (RSONet), which represents the interactions between these objects in the form of close conjunctions, allowing to gain insight on and quantify how the risk of collisions is generated and evolves within the population.

These networks are obtained either by propagating {Two-Line Element} data (coupled with a geometrical filtering to detect conjunctions), or by using directly the information obtained from CDMs.
The properties of RSONet are analysed to extract information about its topology and, then, combined to create a metric of comparison among the RSOs. This relevance score incorporates various key parameters of the network, such as the degree, clustering, and centrality measures, to quantify how the RSOs contribute to the risk of collisions within the population and identify the most important ones.

Some test cases have been presented to apply this approach according to different initial data and simulation parameters. The relevance score is able to measure how some objects contribute to the risk of in-orbit collisions, even without directly {taking} into account the actual probability of such events.
The source of the initial data was found to affect the structure and the composition of the RSONet. The difference between CDMs, which are generated starting from high-precision initial condition and orbital propagators, and TLEs, which are inherently less accurate, is reflected in the resulting conjunctions detected by the algorithms introduced here. In particular, TLE-based network appear to include a large number of satellites belonging to the same constellation: this aspect should, thus, be taken into account when evaluating the risk of collisions.
The sensitivity of the embedding and of the score against the time and distance parameters used in the analysis was studied. The choice of these values has a profound effect on the topology of the network, and thus on the values of the score.\\

While the key concepts have been presented in their general formulation, this framework has been built and presented here under some simplifying assumptions that will be further relaxed in forthcoming works focusing on different aspects of this novel approach, {which will include a more realistic model with a proper comparison with existing metrics as well as more precise data}. In particular:
\begin{enumerate}
\item Uncertainties and probability estimations will be taken into account in building the network, in an effort to better capture the reality of the RSO population and its inner interactions.
\item The choice of the orbital propagation will be relaxed to include high-fidelity and semi-analytical models, to maintain accuracy on longer time scales and propagate uncertainties.
\item The definition of the relevance score will be refined to include more accurate values of collision probability and the physical differences between the various RSOs. The new definition will aim to quantify directly the risk of collisions, introducing risk categories to classify the contribution of each object to the overall risk and its level in relation to the other RSOs.
\item Additional numerical campaigns will be performed aiming to comparing the proposed approach against other risk measures {to better explicate the novelty of the proposed approach and its differences with respect to the existing models.}
\item {More precise orbital data than TLEs will be employed to provide more realism to the risk analsysis.}
\end{enumerate}

\backmatter

%
%
%

\section*{Acknowledgements}

This work was funded via the BEWARE programme by the government of the Wallonia region and by the European Commission (Marie Sklodowska Curie Actions grant agreement 847587) as part of the Horizon 2020 research and innovation programme.

\section*{Declarations}

\paragraph{Conflict of interest} The authors declare no conflicts of interest.

%
%
%
%
%
%
%
%

\begin{appendices}

\section{The three-filter algorithm}
\label{app:A}

\begin{figure}[!htp]
\centering
\includegraphics[width=0.8\textwidth]{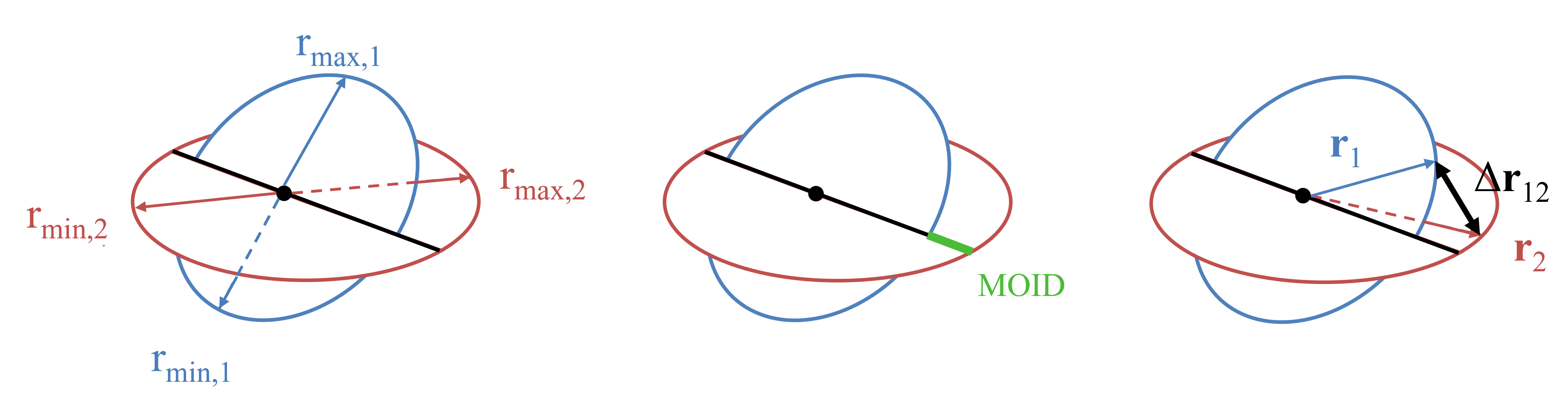}
\caption{{Schematic} geometrical representation of the triple filter approach.}
\label{fig:triple-filter}
\end{figure}

This Appendix presents the three-filter algorithm used in Sect.\,\ref{sec:theory_RSOnetwork}, breaking down each step as shown in Fig.\,\ref{fig:triple-filter}.

\paragraph*{First filter}
The first filter compares the minimum and the maximum geocentric distance reached by the two objects to determine if a crossing between their orbits is possible within the selected time interval. For each orbit with index $1 \leq s \leq N$, the Cartesian position vector along the orbit $\mathbf{r}_s(t) = \lbrace x(t),y(t),z(t) \rbrace$, with $0 \leq t \leq T$, is given and the geocentric distance is defined as 
\begin{equation*}
r_s(t) = \sqrt{x(t)^2+y(t)^2+z(t)^2} .
\end{equation*}
Thus, the absolute minimum and maximum of such function can be set as
\begin{align*}
\text{min}(r_s) &:= \text{min} \lbrace r_s(t), t_1 \leq t \leq t_m \rbrace , \\
\text{max}(r_s) &:= \text{max} \lbrace r_s(t), t_1 \leq t \leq t_m \rbrace .
\end{align*}
Given a pair of objects $(a,b)$ with $1 \leq a < b \leq N$, the following terms are defined:
\begin{align*}
q &:= \text{max} \lbrace \text{min}(r_a), \text{min}(r_b) \rbrace , \\
Q &:= \text{min} \lbrace \text{max}(r_a), \text{max}(r_b) \rbrace .
\end{align*}

\noindent Thus, there is no orbit crossing if the condition $\vert q - Q \vert > \epsilon$ is satisfied, where $\epsilon$ is a fixed threshold distance, and consequently the pair must no longer be considered.

Although the complexity of this operation is of order $N^2$, it is very fast compared to the subsequent filters, thus adding little computational complexity while pruning the data pool of a large number of objects \citep{Casanova2014}.

\paragraph*{Second filter}
The second filter is based on the definition of the local minimal distances introduced by \cite{Gronchi2005}. For the two orbits under consideration, each described by a set of 6 orbital parameters $\mathcal{E}_s, s=1,2$, the Keplerian distance function $d$ is defined as the map 
\begin{equation*}
d(\mathcal{E}_1,\mathcal{E}_2) = \Vert \mathbf{X}_1 - \mathbf{X}_2 \Vert ,
\end{equation*}
where $\mathbf{X}_s = \mathbf{X}(\mathcal{E}_j(t)), s=1,2$ is the Cartesian position of each object along its orbit and $\Vert\cdot\Vert$ is the Euclidean norm in $\mathbb{R}^3$. The critical points of the function $d$ are computed through those of $d^2$ to avoid problems of differentiability when it vanishes using the method developed by \cite{Gronchi2005}, which computes them as the real roots of a 16$^\text{th}$ order complex polynomial. These roots are computed using the algorithm proposed by \cite{bini1996} based on Aberth's iterative method.

The minimal points correspond to the possible close approaches of the two objects. At these points, the sign of the distance map can be changed to obtain a more regular map called distance with sign. For each pair of objects, it is possible to compute their signed distances at any time in the interpolation process. If there is a change of sign in one of the local minimal distances, an orbit crossing occurs, thus the pair passes the second filter; otherwise, the pair is excluded for a possible collision.

\paragraph*{Third filter}
The third filter directly considers the distance between the two objects if they passed the first two filters, meaning that an orbit crossing between the two occurs. Simply, if their relative distance $d(t) = \Vert \mathbf{X}_1 - \mathbf{X}_2 \Vert$ at time $t$ is below a given threshold $\epsilon$, the crossing is counted as a conjunction and they are at risk of a collision. The pair can be excluded if there is never a time when the distance between the objects is less than $\epsilon$.\\

\section{Properties of the network}
\label{app:B}

Considering an undirected network with $N$ nodes or vertices labelled $1,...,n$, a link or an edge between nodes $i$ and $j$ can be denoted by $(i,j)$. The adjacency matrix $\mathbf{A}$ of the network is defined as a $N \times N$ matrix whose elements $a_{ij}$ are equal to 1 if there is a link between nodes $i$ and $j$ and 0 otherwise.

It can be noticed that for a network with no self-edges (that is, no link from a node to itself), the diagonal matrix elements are all zero. Also, the matrix is symmetric, since the link between $i$ and $j$ has no direction.

If the elements $a_{ij}$ assume values different from 1, then the network is defined as \textit{weighted}, with the values of $a_{ij}$ acting as weights to the links.

Different properties can be derived from the conformation of the network and the adjacency matrix. Their definitions are given in the following sections, which can be related to various types of interactions between space objects.

\paragraph*{Node degree}{The degree, or degree centrality, of a node in an unweighted network is defined as the number of links connected to it, representing here the number of conjunctions an object experiences during the selected time frame. In terms of elements of the adjacency matrix, the degree $D_i$ of node $i$ can be defined as \citep{Lewis2010}:
\begin{equation}
D_i = \sum_{j=1}^{n}{a_{ij}} .
\end{equation}
In case of a weighted network, the degree, which in this case takes the name of \textit{strength}, will be equal to the average of the weights on the links connected to $i$.}

\paragraph*{Clustering coefficient}{The clustering coefficient provides a measure of the density of connections between neighbouring vertices. In this work, a high clustering coefficient identifies those objects which are more susceptible to encountering a fragment of a neighbouring object which has undergone a fragmentation. It is related to the number of triangles connected to node $a$ \citep{newman2003} or, alternatively, to the ratio between the number of pairs of neighbours of $a$ that are connected between them and the number of pairs of neighbours of $a$ \cite[pp.199-204]{newman2010}:
\begin{equation}
C_i = \frac{\text{(number of pairs of neighbours of i that are connected)}}{\text{(number of pairs of neighbours of i)}} .
\end{equation}

The local clustering coefficient $C_i$ can be defined by means of a measure called ``redundancy". The redundancy $R_i$ of a node $i$ is the mean number of connections from a neighbour of $i$ to other neighbours of $i$. This implies that the total number of connections between the neighbours of $i$ is thus $\frac{1}{2}D_i R_i$, and since the total number of pairs of neighbours of $i$ is $\frac{1}{2}D_i (D_i-1)$, the local clustering coefficient is the ratio between these two quantities \citep{newman2010}:
\begin{equation}
\label{eq:clusteringR}
C_i = \frac{\frac{1}{2}D_i R_i}{\frac{1}{2}D_i (D_i-1)} .
\end{equation}

The expression of the clustering coefficient can also be derived directly from the elements of the adjacency matrix as \citep{Lewis2010,wang2017}:
\begin{equation}
C_i = \frac{2}{D_i (D_i-1)} \sum_{j \neq k}^{n}{\sum_{k}^{n}{a_{ij} a_{ik} a_{jk}}} .
\end{equation}
{where $\sum_{j=1}^{n}{\sum_{k=1}^{n}{a_{ij} a_{ik} a_{jk}}}$ corresponds to the number of triangles connected to node $i$. It can be demonstrated that this number is equal to half the $i$-th diagonal term of the cube of the adjacency matrix $\mathbf{A}^3$. }

\noindent {Generalising, $ij$-th element of any power $r$ of the adjacency matrix $\mathbf{A}^r$ is equal to the number of paths of length $r$ connecting nodes $i$ and $j$ \cite[pp.136-137]{newman2010}. In case $i=j$ (that is, a diagonal term of $\mathbf{A}^r$), the paths start and end at the same vertex, becoming ``cycles". In case $r=3$, these cycles are triangles of which $i$ is a vertex.}

\noindent {Thus, the local clustering coefficient for node $i$ can be written as}
\begin{equation}
C_i = \frac{\frac{1}{2}[\mathbf{A}^3]_{ii}}{D_i (D_i-1)}
\end{equation}}

\paragraph*{Closeness centrality}{The closeness centrality measures the mean distance from a vertex to other vertices. The mean shortest-path distance from $i$ to $j$, averaged over all $j$ vertices, is
\begin{equation}
\label{eq:meandistance}
\ell_i = \frac{1}{n-1} \sum_{\substack{j=1 \\ j \neq i}}^{n}{d_{ij}} ,
\end{equation}
where $d_{ij}$ is the length of the shortest path connecting nodes $i$ and $j$ in an undirected network, for which $d_{ij}=d_{ji}$. The shortest-path length can be determined numerically.

The closeness $K_i$ for a vertex $i$ is defined as the inverse of $\ell_i$ \cite[pp.181-184]{newman2010}:
\begin{equation}
\label{eq:closeness}
K_i = \frac{1}{\ell_i} = \frac{n-1}{\sum_{\substack{j=1 \\ j \neq i}}^{n}{d_{ij}}} .
\end{equation}}

\paragraph*{Betweenness centrality}{The betweenness centrality, or shortest-path betweenness, quantifies the extent to which a vertex lies on the paths between others: if a vertex is situated on many paths between other vertices then it is said to have high betweenness, and has a role in connecting different parts of the network with the potential of transmitting fragmentation chain reactions. The shortest-path betweenness $B_i$ of a vertex $i$ is defined to be \citep{newman2010}: 
\begin{equation}
B_i = \sum_{s=1}^{n}{\sum_{t=1}^{n}{\sigma_{st}(i)}} .
\end{equation}
where $\sigma_{st}(i)$ represents the total number of shortest paths between nodes $s$ and $t$ passing through node $i$.}

\begin{remark}
Both centrality measures presented here can be related to the ability of an object to propagate a chain of fragmentations, where an object produces debris (whether from a collision or a spontaneous breakup) which hits a neighbouring object which in turn produces new debris which goes on hitting another object and so on. While the scenario of a fragmentation cascade is unlikely, it is still interesting to consider it in the risk evaluation.
With this in mind, Centrality measures the average length of these chains, thus the potential debris production of a fragmentation cascade, while Betweenness indicates those objects which play the most important role as ``bridges" in propagating multiple chains.
\end{remark}

\section{Ranking of top 50 objects}
\label{app:C}
The ranking of the $50$ RSOs with the highest relevance  score (either based on TLEs or CDMs) appears in Tab.\,\ref{tab:Top50}.

\begin{table*}[!hp]
\begin{center}
\caption{Top 50 ranking objects by score $\mathcal{S}$ for the CDM- and TLE-based networks.}
\label{tab:Top50}
\begin{tabular}{ccl|ccl} 
\hline
\multicolumn{3}{c|}{CDMs} 	& \multicolumn{3}{c}{TLEs}\\
\hline
$\mathcal{S}$ & NORAD ID & Name 	& $\mathcal{S}$ & NORAD ID & Name \\
\hline
6.0 & 38017 & IRIDIUM 33 DEB 	& 7.0 & 48695 & STARLINK-2616 \\
6.0 & 39603 & METEOR 2-5 DEB	& 6.3 & 46382 & STARLINK-1769 \\
5.0 & 13719 & SL-3 R/B			& 6.3 & 48689 & STARLINK-2617 \\
5.0 & 30525 & FENGYUN 1C DEB	& 6.3 & 44959 & STARLINK-1076 \\
5.0 & 35705 & COSMOS 2251 DEB	& 6.0 & 45067 & STARLINK-1150 \\
5.0 & 36274 & FENGYUN 1C DEB		& 6.0 & 45706 & STARLINK-1411  \\
5.0 & 38069 & COSMOS 2251 DEB		& 6.0 & 45713 & STARLINK-1436  \\
5.0 & 40681 & DMSP 5D-2 F13 DEB	& 6.0 & 46156 & STARLINK-1545  \\
5.0 & 43326 & COSMOS 1867 COOLANT	& 6.0 & 46173 & STARLINK-1640  \\
4.0 & 8924 & SL-8 R/B		& 5.0 & 46746 & STARLINK-1905  \\
4.0 & 17621 & COSMOS 1275 DEB		& 5.0 & 47602 & STARLINK-2007  \\
4.0 & 22455 & SL-16 DEB		& 5.0 & 47770 & STARLINK-2193  \\
4.0 & 27944 & LARETS		& 5.0 & 48285 & STARLINK-2548  \\
4.0 & 29525 & DMSP 5D-3 F17 DEB	& 5.0 & 48566 & STARLINK-2214  \\
4.0 & 30938 & FENGYUN 1C DEB		& 5.0 & 48598 & STARLINK-2256  \\
4.0 & 30960 & FENGYUN 1C DEB		& 5.0 & 46041 & STARLINK-1580  \\
4.0 & 31231 & FENGYUN 1C DEB		& 5.0 & 46331 & STARLINK-1719  \\
4.0 & 31568 & FENGYUN 1C DEB		& 4.0 & 47164 & STARLINK-1879  \\
4.0 & 33716 & FENGYUN 1C DEB		& 4.0 & 45058 & STARLINK-1162  \\
4.0 & 34477 & COSMOS 2251 DEB		& 4.0 & 45060 & STARLINK-1166  \\
4.0 & 34979 & COSMOS 2251 DEB		& 4.0 & 45073 & STARLINK-1170  \\
4.0 & 35229 & FENGYUN 1C DEB		& 4.0 & 45082 & STARLINK-1160  \\
4.0 & 35391 & ERS 2 DEB		& 4.0 & 45368 & STARLINK-1276  \\
4.0 & 36259 & FENGYUN 1C DEB		& 4.0 & 45380 & STARLINK-1207  \\
4.0 & 36271 & FENGYUN 1C DEB		& 4.0 & 45419 & STARLINK-1308  \\
4.0 & 37053 & FENGYUN 1C DEB		& 4.0 & 45551 & STARLINK-1294  \\
4.0 & 39302 & SL-16 DEB		& 4.0 & 45583 & STARLINK-1340  \\
4.0 & 39554 & COSMOS 2251 DEB		& 4.0 & 45669 & STARLINK-1452  \\
4.0 & 39985 & SL-14 DEB		& 4.0 & 46034 & STARLINK-1557  \\
4.0 & 40672 & DMSP 5D-2 F13 DEB	& 4.0 & 46363 & STARLINK-1739  \\
4.0 & 41153 & NOAA 16 DEB		& 4.0 & 46572 & STARLINK-1531  \\
4.0 & 41247 & NOAA 16 DEB		& 4.0 & 46578 & STARLINK-1683  \\
4.0 & 41657 & NOAA 16 DEB		& 4.0 & 46785 & STARLINK-1883  \\
4.0 & 46997 & FENGYUN 1C DEB *		& 4.0 & 46796 & STARLINK-1944  \\
4.0 & 54600 & CZ-6A DEB		& 4.0 & 47552 & STARLINK-1940  \\
3.0 & 4718 & THORAD AGENA D DEB	& 4.0 & 47622 & STARLINK-1645  \\
3.0 & 5024 & THORAD AGENA D DEB	& 4.0 & 47640 & STARLINK-2018  \\
3.0 & 5063 & THORAD AGENA D DEB	& 4.0 & 47728 & STARLINK-2131  \\
3.0 & 7209 & METEOR 1-16		& 4.0 & 48027 & STARLINK-2300  \\
3.0 & 12169 & DELTA 1 DEB		& 4.0 & 48122 & STARLINK-2463  \\
3.0 & 12285 & DELTA 1 DEB		& 4.0 & 48123 & STARLINK-2464  \\
3.0 & 12294 & DELTA 1 DEB		& 4.0 & 48305 & STARLINK-2519  \\
3.0 & 12693 & COSMOS 1275 DEB		& 4.0 & 48481 & STARLINK-2706  \\
3.0 & 13464 & COSMOS 1275 DEB		& 4.0 & 48584 & STARLINK-2238  \\
3.0 & 15950 & SCOUT G-1 DEB		& 4.0 & 48587 & STARLINK-2242  \\
3.0 & 17719 & THORAD AGENA D DEB	& 4.0 & 48644 & STARLINK-2695   \\
3.0 & 17768 & DELTA 1 DEB		& 4.0 & 48660 & STARLINK-2688  \\
3.0 & 18095 & COSMOS 1850		& 4.0 & 48666 & STARLINK-2666  \\
3.0 & 18552 & SL-8 DEB		& 4.0 & 53611 & STARLINK-4650  \\
3.0 & 20433 & SL-8 R/B		& 4.0 & 54280 & CZ-6A DEB \\
\hline
\end{tabular}
\end{center}
\end{table*}

\section{The CDM format}
\label{app:D}
{This Appendix provides the reader with a breakdown of CDM format. CDMs usually change depending on the provider. However, the public format used by Space-Track follows the CCSDS Recommended Standard 508.0-B-1 \cite{ccsds2013}.}

\begin{figure}[!h]
\centering
\includegraphics[width=0.8\textwidth]{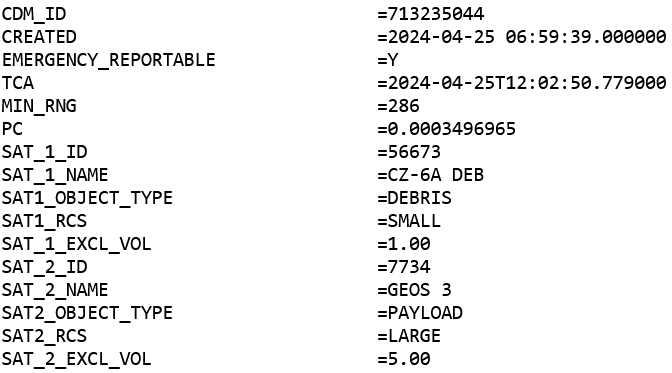}
\caption{Representation of the CDM format.}
\label{fig:cdm}
\end{figure}

{Figure\,\ref{fig:cdm} shows an example of CDM, while Table\,\ref{table:cdm} explains how data is stored.}

\begin{table*}[!h]
\begin{center}
\caption{Explanation of the TLE format.}
\label{table:cdm} 
\begin{tabular}{ccl} 
\hline
Field & Example & Description \\
\hline
CDM\_ID & 713235044 & Identification number of the CDM\\
CREATED & \begin{minipage}[c]{0.25\textwidth}2024-04-25 06:59:39.000000\end{minipage} & Creation date and time of the CDM \\
EMERGENCY\_REPORTABLE & Y & \begin{minipage}[l]{0.3\textwidth}Whether (Y/N) the conjunction constitutes an emergency (that is, the collision probability is above the $10^{-4}$ threshold) or not\end{minipage}\\
TCA & 2024-04-25T12:02:50.779000 & Epoch of closest approach\\
MIN\_RNG & 286 & Distance at closest approach (in metres)\\
PC & 0.0003496965 & Probability of collision \\
SAT\_1\_ID & 56673 & NORAD ID of first object\\
SAT\_1\_NAME & CZ-6A DEB & Catalogue name of first object\\
SAT1\_OBJECT\_TYPE & DEBRIS & Classification of first object\\
SAT1\_RCS & SMALL & \begin{minipage}[l]{0.3\textwidth}Size of the RCS volume of the first object\end{minipage}\\
SAT\_1\_EXCL\_VOL & 1.00 & \begin{minipage}[l]{0.3\textwidth}Exclusion volume of the first object (in metres)\end{minipage}\\
SAT\_2\_ID & 7734 & NORAD ID of second object\\
SAT\_2\_NAME & GEOS 3 & Catalogue name of second object\\
SAT2\_OBJECT\_TYPE & PAYLOAD & Classification of second object\\
SAT2\_RCS & LARGE & \begin{minipage}[l]{0.3\textwidth}Size of the RCS volume of the second object\end{minipage}\\
SAT\_2\_EXCL\_VOL & 5.00 & \begin{minipage}[l]{0.3\textwidth}Exclusion volume of the second object (in metres)\end{minipage}\\
\hline
\end{tabular}
\end{center}
\end{table*}

\end{appendices}


\clearpage

\bibliography{references.bib}


\begin{thebibliography}{67}
\ifx \bisbn   \undefined \def \bisbn  #1{ISBN #1}\fi
\ifx \binits  \undefined \def \binits#1{#1}\fi
\ifx \bauthor  \undefined \def \bauthor#1{#1}\fi
\ifx \batitle  \undefined \def \batitle#1{#1}\fi
\ifx \bjtitle  \undefined \def \bjtitle#1{#1}\fi
\ifx \bvolume  \undefined \def \bvolume#1{\textbf{#1}}\fi
\ifx \byear  \undefined \def \byear#1{#1}\fi
\ifx \bissue  \undefined \def \bissue#1{#1}\fi
\ifx \bfpage  \undefined \def \bfpage#1{#1}\fi
\ifx \blpage  \undefined \def \blpage #1{#1}\fi
\ifx \burl  \undefined \def \burl#1{\textsf{#1}}\fi
\ifx \doiurl  \undefined \def \doiurl#1{\url{https://doi.org/#1}}\fi
\ifx \betal  \undefined \def \betal{\textit{et al.}}\fi
\ifx \binstitute  \undefined \def \binstitute#1{#1}\fi
\ifx \binstitutionaled  \undefined \def \binstitutionaled#1{#1}\fi
\ifx \bctitle  \undefined \def \bctitle#1{#1}\fi
\ifx \beditor  \undefined \def \beditor#1{#1}\fi
\ifx \bpublisher  \undefined \def \bpublisher#1{#1}\fi
\ifx \bbtitle  \undefined \def \bbtitle#1{#1}\fi
\ifx \bedition  \undefined \def \bedition#1{#1}\fi
\ifx \bseriesno  \undefined \def \bseriesno#1{#1}\fi
\ifx \blocation  \undefined \def \blocation#1{#1}\fi
\ifx \bsertitle  \undefined \def \bsertitle#1{#1}\fi
\ifx \bsnm \undefined \def \bsnm#1{#1}\fi
\ifx \bsuffix \undefined \def \bsuffix#1{#1}\fi
\ifx \bparticle \undefined \def \bparticle#1{#1}\fi
\ifx \barticle \undefined \def \barticle#1{#1}\fi
\bibcommenthead
\ifx \bconfdate \undefined \def \bconfdate #1{#1}\fi
\ifx \botherref \undefined \def \botherref #1{#1}\fi
\ifx \url \undefined \def \url#1{\textsf{#1}}\fi
\ifx \bchapter \undefined \def \bchapter#1{#1}\fi
\ifx \bbook \undefined \def \bbook#1{#1}\fi
\ifx \bcomment \undefined \def \bcomment#1{#1}\fi
\ifx \oauthor \undefined \def \oauthor#1{#1}\fi
\ifx \citeauthoryear \undefined \def \citeauthoryear#1{#1}\fi
\ifx \endbibitem  \undefined \def \endbibitem {}\fi
\ifx \bconflocation  \undefined \def \bconflocation#1{#1}\fi
\ifx \arxivurl  \undefined \def \arxivurl#1{\textsf{#1}}\fi
\csname PreBibitemsHook\endcsname

\bibitem[\protect\citeauthoryear{{ESA}}{2023}]{ESA2023a}
\begin{botherref}
\oauthor{\bsnm{{ESA}}}:
{ESA}'s annual space environment report.
{GEN-DB-LOG-00288-OPS-SD}. Available online at
  \url{https://www.iadc-home.org/documents_public/file_down/id/5432} (accessed
  01/10/2023)
(2023)
\end{botherref}
\endbibitem

\bibitem[\protect\citeauthoryear{Kelso}{2007}]{kelso2007}
\begin{bchapter}
\bauthor{\bsnm{Kelso}, \binits{T.}}:
\bctitle{Analysis of the 2007 chinese asat test and the impact of its debris on
  the space environment}.
In: \bbtitle{8th Advanced Maui Optical and Space Surveillance Technologies
  Conference, Maui, HI},
vol. \bseriesno{7}
(\byear{2007})
\end{bchapter}
\endbibitem

\bibitem[\protect\citeauthoryear{Pardini and Anselmo}{2007}]{pardini2007}
\begin{bchapter}
\bauthor{\bsnm{Pardini}, \binits{C.}},
\bauthor{\bsnm{Anselmo}, \binits{L.}}:
\bctitle{Evolution of the debris cloud generated by the fengyun-1c
  fragmentation event}.
In: \bbtitle{Proceedings of the 20th International Symposium on Space Flight
  Dynamics}
(\byear{2007})
\end{bchapter}
\endbibitem

\bibitem[\protect\citeauthoryear{Lambert}{2018}]{lambert2018}
\begin{bchapter}
\bauthor{\bsnm{Lambert}, \binits{J.}}:
\bctitle{Fengyun-1c debris cloud evolution over one decade}.
In: \bbtitle{Proceedings of the Advanced Maui Optical and Space Surveillance
  (AMOS) Technology Conference}
(\byear{2018})
\end{bchapter}
\endbibitem

\bibitem[\protect\citeauthoryear{Kelso et~al.}{2009}]{kelso2009}
\begin{barticle}
\bauthor{\bsnm{Kelso}, \binits{T.}}, \betal:
\batitle{Analysis of the iridium 33-cosmos 2251 collision}.
\bjtitle{Advances in the Astronautical Sciences}
\bvolume{135}(\bissue{2}),
\bfpage{1099}--\blpage{1112}
(\byear{2009})
\end{barticle}
\endbibitem

\bibitem[\protect\citeauthoryear{Anselmo and Pardini}{2009}]{anselmo2009}
\begin{bchapter}
\bauthor{\bsnm{Anselmo}, \binits{L.}},
\bauthor{\bsnm{Pardini}, \binits{C.}}:
\bctitle{Analysis of the consequences in low earth orbit of the collision
  between cosmos 2251 and iridium 33}.
In: \bbtitle{Proceedings of the 21st International Symposium on Space Flight
  Dynamics},
pp. \bfpage{2009}--\blpage{294}
(\byear{2009}).
\bcomment{Centre nationale d'etudes spatiales Paris, France}
\end{bchapter}
\endbibitem

\bibitem[\protect\citeauthoryear{Pardini and Anselmo}{2017}]{pardini2017}
\begin{barticle}
\bauthor{\bsnm{Pardini}, \binits{C.}},
\bauthor{\bsnm{Anselmo}, \binits{L.}}:
\batitle{Revisiting the collision risk with cataloged objects for the iridium
  and cosmo-skymed satellite constellations}.
\bjtitle{Acta Astronautica}
\bvolume{134},
\bfpage{23}--\blpage{32}
(\byear{2017})
\end{barticle}
\endbibitem

\bibitem[\protect\citeauthoryear{Weber et~al.}{2023}]{weber2023}
\begin{barticle}
\bauthor{\bsnm{Weber}, \binits{D.}},
\bauthor{\bsnm{Letizia}, \binits{F.}},
\bauthor{\bsnm{{Bastida Virgili}}, \binits{B.}},
\bauthor{\bsnm{Lemmens}, \binits{S.}}:
\batitle{Statistical analysis of conjunctions in low earth orbit}.
\bjtitle{Advances in Space Research}
\bvolume{72}(\bissue{7}),
\bfpage{2578}--\blpage{2584}
(\byear{2023})
\doiurl{10.1016/j.asr.2022.06.062} .
\bcomment{Space Environment Management and Space Sustainability}
\end{barticle}
\endbibitem

\bibitem[\protect\citeauthoryear{McDowell}{2020}]{mcdowell2020}
\begin{barticle}
\bauthor{\bsnm{McDowell}, \binits{J.}}:
\batitle{The low earth orbit satellite population and impacts of the spacex
  starlink constellation}.
\bjtitle{The Astrophysical Journal Letters}
\bvolume{892}(\bissue{2}),
\bfpage{36}
(\byear{2020})
\doiurl{10.3847/2041-8213/ab8016}
\end{barticle}
\endbibitem

\bibitem[\protect\citeauthoryear{Massey et~al.}{2020}]{massey2020}
\begin{barticle}
\bauthor{\bsnm{Massey}, \binits{R.}},
\bauthor{\bsnm{Lucatello}, \binits{S.}},
\bauthor{\bsnm{Benvenuti}, \binits{P.}}:
\batitle{The challenge of satellite megaconstellations}.
\bjtitle{Nature Astronomy}
\bvolume{4},
\bfpage{1022}--\blpage{1023}
(\byear{2020})
\doiurl{10.1038/s41550-020-01224-9}
\end{barticle}
\endbibitem

\bibitem[\protect\citeauthoryear{Acciarini et~al.}{2023}]{acciarini2023}
\begin{botherref}
\oauthor{\bsnm{Acciarini}, \binits{G.}},
\oauthor{\bsnm{Baresi}, \binits{N.}},
\oauthor{\bsnm{Bridges}, \binits{C.}},
\oauthor{\bsnm{Felicetti}, \binits{L.}},
\oauthor{\bsnm{Hobbs}, \binits{S.}},
\oauthor{\bsnm{Günes~Baydin}, \binits{A.}}:
Observation strategies and megaconstellations impact on current {LEO}
  population.
Proceedings of the 2nd NEO and Debris Detection Conference (NEOSST2)
(2023)
\end{botherref}
\endbibitem

\bibitem[\protect\citeauthoryear{Martinez}{2021}]{martinez2021copuos}
\begin{barticle}
\bauthor{\bsnm{Martinez}, \binits{P.}}:
\batitle{The un copuos guidelines for the long-term sustainability of outer
  space activities}.
\bjtitle{Journal of Space Safety Engineering}
\bvolume{8}(\bissue{1}),
\bfpage{98}--\blpage{107}
(\byear{2021})
\end{barticle}
\endbibitem

\bibitem[\protect\citeauthoryear{Wormnes et~al.}{2013}]{wormnes2013}
\begin{bchapter}
\bauthor{\bsnm{Wormnes}, \binits{K.}},
\bauthor{\bsnm{Le~Letty}, \binits{R.}},
\bauthor{\bsnm{Summerer}, \binits{L.}},
\bauthor{\bsnm{Schonenborg}, \binits{R.}},
\bauthor{\bsnm{Dubois-Matra}, \binits{O.}},
\bauthor{\bsnm{Luraschi}, \binits{E.}},
\bauthor{\bsnm{Cropp}, \binits{A.}},
\bauthor{\bsnm{Krag}, \binits{H.}},
\bauthor{\bsnm{Delaval}, \binits{J.}}:
\bctitle{{ESA} technologies for space debris remediation}.
In: \bbtitle{6th European Conference on Space Debris},
vol. \bseriesno{1},
pp. \bfpage{1}--\blpage{8}
(\byear{2013}).
\bcomment{{ESA} Communications ESTEC Noordwijk, The Netherlands}
\end{bchapter}
\endbibitem

\bibitem[\protect\citeauthoryear{Cattani et~al.}{2021}]{cattani2021}
\begin{bchapter}
\bauthor{\bsnm{Cattani}, \binits{B.}},
\bauthor{\bsnm{Soares}, \binits{T.}},
\bauthor{\bsnm{Serrano}, \binits{S.M.}},
\bauthor{\bsnm{Briot}, \binits{D.}},
\bauthor{\bsnm{Serra}, \binits{S.V.}},
\bauthor{\bsnm{Thiry}, \binits{N.}}:
\bctitle{The impact of space debris mitigation requirements on mission design
  choices: an overview from {ESA} clean space}.
In: \bbtitle{Proceedings of 8th European Conference on Space Debris (virtual
  Edition, {SDC8})}
(\byear{2021})
\end{bchapter}
\endbibitem

\bibitem[\protect\citeauthoryear{Letizia et~al.}{2023a}]{letizia2023a}
\begin{barticle}
\bauthor{\bsnm{Letizia}, \binits{F.}},
\bauthor{\bsnm{Virgili}, \binits{B.B.}},
\bauthor{\bsnm{Lemmens}, \binits{S.}}:
\batitle{Assessment of orbital capacity thresholds through long-term
  simulations of the debris environment}.
\bjtitle{Advances in Space Research}
\bvolume{72}(\bissue{7}),
\bfpage{2552}--\blpage{2569}
(\byear{2023})
\end{barticle}
\endbibitem

\bibitem[\protect\citeauthoryear{Letizia et~al.}{2023b}]{letizia2023b}
\begin{barticle}
\bauthor{\bsnm{Letizia}, \binits{F.}},
\bauthor{\bsnm{Sanvido}, \binits{S.}},
\bauthor{\bsnm{Lemmens}, \binits{S.}},
\bauthor{\bsnm{Merz}, \binits{K.}},
\bauthor{\bsnm{Southworth}, \binits{R.}},
\bauthor{\bsnm{Sousa}, \binits{B.}}:
\batitle{{ESA}'s current approaches to end-of-life strategies for {HEO}
  missions}.
\bjtitle{Journal of Space Safety Engineering}
\bvolume{10},
\bfpage{407}--\blpage{413}
(\byear{2023})
\end{barticle}
\endbibitem

\bibitem[\protect\citeauthoryear{Ely and Howell}{1996}]{ely1996}
\begin{barticle}
\bauthor{\bsnm{Ely}, \binits{T.}},
\bauthor{\bsnm{Howell}, \binits{K.}}:
\batitle{Long-term evolution of artificial satellite orbits due to resonant
  tesseral harmonics}.
\bjtitle{Journal of the Astronautical Sciences}
\bvolume{44},
\bfpage{167}--\blpage{190}
(\byear{1996})
\end{barticle}
\endbibitem

\bibitem[\protect\citeauthoryear{Breiter}{1999}]{breiter1999}
\begin{barticle}
\bauthor{\bsnm{Breiter}, \binits{S.}}:
\batitle{Lunisolar apsidal resonances at low satellite orbits}.
\bjtitle{Celestial Mechanics and Dynamical Astronomy}
\bvolume{74},
\bfpage{253}--\blpage{274}
(\byear{1999})
\end{barticle}
\endbibitem

\bibitem[\protect\citeauthoryear{Gkolias et~al.}{2018}]{gkolias2018}
\begin{barticle}
\bauthor{\bsnm{Gkolias}, \binits{I.}},
\bauthor{\bsnm{Colombo}, \binits{C.}}, \betal:
\batitle{Disposal design for geosynchronous satellites revisited}.
\bjtitle{Advances in the Astronautical Sciences}
\bvolume{167},
\bfpage{1843}--\blpage{1857}
(\byear{2018})
\end{barticle}
\endbibitem

\bibitem[\protect\citeauthoryear{Skoulidou et~al.}{2018}]{skoulidou2018}
\begin{barticle}
\bauthor{\bsnm{Skoulidou}, \binits{D.K.}},
\bauthor{\bsnm{Rosengren}, \binits{A.J.}},
\bauthor{\bsnm{Tsiganis}, \binits{K.}},
\bauthor{\bsnm{Voyatzis}, \binits{G.}}:
\batitle{Dynamical lifetime survey of geostationary transfer orbits}.
\bjtitle{Celestial Mechanics and Dynamical Astronomy}
\bvolume{130},
\bfpage{1}--\blpage{18}
(\byear{2018})
\end{barticle}
\endbibitem

\bibitem[\protect\citeauthoryear{Rossi et~al.}{2018}]{rossi2018}
\begin{barticle}
\bauthor{\bsnm{Rossi}, \binits{A.}},
\bauthor{\bsnm{Colombo}, \binits{C.}},
\bauthor{\bsnm{Tsiganis}, \binits{K.}},
\bauthor{\bsnm{Beck}, \binits{J.}},
\bauthor{\bsnm{Rodriguez}, \binits{J.B.}},
\bauthor{\bsnm{Walker}, \binits{S.}},
\bauthor{\bsnm{Letterio}, \binits{F.}},
\bauthor{\bsnm{Dalla~Vedova}, \binits{F.}},
\bauthor{\bsnm{Schaus}, \binits{V.}},
\bauthor{\bsnm{Popova}, \binits{R.}}, \betal:
\batitle{Redshift: A global approach to space debris mitigation}.
\bjtitle{Aerospace}
\bvolume{5}(\bissue{2}),
\bfpage{64}
(\byear{2018})
\end{barticle}
\endbibitem

\bibitem[\protect\citeauthoryear{Rossi et~al.}{2019}]{rossi2019}
\begin{bchapter}
\bauthor{\bsnm{Rossi}, \binits{A.}},
\bauthor{\bsnm{Alessi}, \binits{E.M.}},
\bauthor{\bsnm{Schaus}, \binits{V.}}:
\bctitle{Assessing the effectiveness of resonant corridors in passive debris
  disposal}.
In: \bbtitle{First International Orbital Debris Conference},
vol. \bseriesno{2109},
p. \bfpage{6022}
(\byear{2019})
\end{bchapter}
\endbibitem

\bibitem[\protect\citeauthoryear{Gondelach et~al.}{2019}]{gondelach2019}
\begin{barticle}
\bauthor{\bsnm{Gondelach}, \binits{D.J.}},
\bauthor{\bsnm{Armellin}, \binits{R.}},
\bauthor{\bsnm{Wittig}, \binits{A.}}:
\batitle{On the predictability and robustness of galileo disposal orbits}.
\bjtitle{Celestial Mechanics and Dynamical Astronomy}
\bvolume{131}(\bissue{12}),
\bfpage{60}
(\byear{2019})
\end{barticle}
\endbibitem

\bibitem[\protect\citeauthoryear{Maclay and McKnight}{2021}]{maclay2021}
\begin{barticle}
\bauthor{\bsnm{Maclay}, \binits{T.}},
\bauthor{\bsnm{McKnight}, \binits{D.}}:
\batitle{Space environment management: Framing the objective and setting
  priorities for controlling orbital debris risk}.
\bjtitle{Journal of Space Safety Engineering}
\bvolume{8}(\bissue{1}),
\bfpage{93}--\blpage{97}
(\byear{2021})
\doiurl{10.1016/j.jsse.2020.11.002}
\end{barticle}
\endbibitem

\bibitem[\protect\citeauthoryear{Liou}{2020}]{liou2020}
\begin{botherref}
\oauthor{\bsnm{Liou}, \binits{J.-C.}}:
The 2019 {US} government orbital debris mitigation standard practices.
Technical report
(2020).
\url{https://orbitaldebris.jsc.nasa.gov/library/usg_orbital_debris_mitigation_standard_practices_november_2019.pdf}
\end{botherref}
\endbibitem

\bibitem[\protect\citeauthoryear{{ESA}}{2023}]{esa2023b}
\begin{botherref}
\oauthor{\bsnm{{ESA}}}:
{ESA} space debris mitigation compliance verification guidelines.
{ESSB-HB-U-002}. Available online at
  \url{https://esamultimedia.esa.int/docs/spacesafety/ESSB-HB-U-002-Issue2(14February2023).pdf}
  (accessed 30/04/2024)
(2023)
\end{botherref}
\endbibitem

\bibitem[\protect\citeauthoryear{Valk et~al.}{2009}]{valk2009}
\begin{barticle}
\bauthor{\bsnm{Valk}, \binits{S.}},
\bauthor{\bsnm{Delsate}, \binits{N.}},
\bauthor{\bsnm{Lema{\^\i}tre}, \binits{A.}},
\bauthor{\bsnm{Carletti}, \binits{T.}}:
\batitle{Global dynamics of high area-to-mass ratios geo space debris by means
  of the megno indicator}.
\bjtitle{Advances in Space Research}
\bvolume{43}(\bissue{10}),
\bfpage{1509}--\blpage{1526}
(\byear{2009})
\end{barticle}
\endbibitem

\bibitem[\protect\citeauthoryear{Lara et~al.}{2011}]{mLa11}
\begin{barticle}
\bauthor{\bsnm{Lara}, \binits{M.}},
\bauthor{\bsnm{San-Juan}, \binits{J.F.}},
\bauthor{\bsnm{Folcik}, \binits{Z.J.}},
\bauthor{\bsnm{Cefola}, \binits{P.}}:
\batitle{Deep resonant {GPS}-dynamics due to the geopotential}.
\bjtitle{The Journal of the Astronautical Sciences}
\bvolume{58}(\bissue{4}),
\bfpage{661}--\blpage{676}
(\byear{2011})
\end{barticle}
\endbibitem

\bibitem[\protect\citeauthoryear{Lara et~al.}{2014}]{mLa14}
\begin{barticle}
\bauthor{\bsnm{Lara}, \binits{M.}},
\bauthor{\bsnm{San-Juan}, \binits{J.F.}},
\bauthor{\bsnm{L{\'o}pez-Ochoa}, \binits{L.M.}},
\bauthor{\bsnm{Cefola}, \binits{P.}}:
\batitle{Long-term evolution of {G}alileo operational orbits by canonical
  perturbation theory}.
\bjtitle{Acta Astronautica}
\bvolume{94}(\bissue{2}),
\bfpage{646}--\blpage{655}
(\byear{2014})
\end{barticle}
\endbibitem

\bibitem[\protect\citeauthoryear{Celletti and Gales}{2018}]{aCe18}
\begin{barticle}
\bauthor{\bsnm{Celletti}, \binits{A.}},
\bauthor{\bsnm{Gales}, \binits{C.}}:
\batitle{Dynamics of resonances and equilibria of {L}ow {E}arth {O}bjects}.
\bjtitle{SIAM Journal on Applied Dynamical Systems}
\bvolume{17}(\bissue{1}),
\bfpage{203}--\blpage{235}
(\byear{2018})
\end{barticle}
\endbibitem

\bibitem[\protect\citeauthoryear{Alessi et~al.}{2018}]{emAl18}
\begin{barticle}
\bauthor{\bsnm{Alessi}, \binits{E.M.}},
\bauthor{\bsnm{Schettino}, \binits{G.}},
\bauthor{\bsnm{Rossi}, \binits{A.}},
\bauthor{\bsnm{Valsecchi}, \binits{G.B.}}:
\batitle{Natural highways for end-of-life solutions in the {LEO} region}.
\bjtitle{Celestial Mechanics and Dynamical Astronomy}
\bvolume{130}(\bissue{5}),
\bfpage{34}
(\byear{2018})
\end{barticle}
\endbibitem

\bibitem[\protect\citeauthoryear{Daquin et~al.}{2022}]{jDa22}
\begin{barticle}
\bauthor{\bsnm{Daquin}, \binits{J.}},
\bauthor{\bsnm{Legnaro}, \binits{E.}},
\bauthor{\bsnm{Gkolias}, \binits{I.}},
\bauthor{\bsnm{Efthymiopoulos}, \binits{C.}}:
\batitle{A deep dive into the 2g+h resonance: separatrices, manifolds and phase
  space structure of navigation satellites}.
\bjtitle{Celestial Mechanics and Dynamical Astronomy}
\bvolume{134}(\bissue{1}),
\bfpage{6}
(\byear{2022})
\end{barticle}
\endbibitem

\bibitem[\protect\citeauthoryear{Skinner et~al.}{2022}]{skinner2022}
\begin{barticle}
\bauthor{\bsnm{Skinner}, \binits{M.A.}},
\bauthor{\bsnm{Oltrogge}, \binits{D.}},
\bauthor{\bsnm{Strah}, \binits{M.}},
\bauthor{\bsnm{Rovetto}, \binits{R.J.}},
\bauthor{\bsnm{Lacroix}, \binits{A.}},
\bauthor{\bsnm{Kumar}, \binits{A.K.A.}},
\bauthor{\bsnm{Grattan}, \binits{K.}},
\bauthor{\bsnm{Francillout}, \binits{L.}},
\bauthor{\bsnm{Alonso}, \binits{I.}}:
\batitle{Space traffic management terminology}.
\bjtitle{Journal of Space Safety Engineering}
\bvolume{9}(\bissue{4}),
\bfpage{644}--\blpage{648}
(\byear{2022})
\doiurl{10.1016/j.jsse.2022.09.001}
\end{barticle}
\endbibitem

\bibitem[\protect\citeauthoryear{IADC}{2023}]{IADC2023}
\begin{botherref}
\oauthor{\bsnm{IADC}}:
Iadc report on the status of the space debris environment.
Available online at \url{https://www.iadc-home.org/documents_public} (accessed
  01/10/2023)
(2023)
\end{botherref}
\endbibitem

\bibitem[\protect\citeauthoryear{Landi et~al.}{2018}]{landi2018}
\begin{barticle}
\bauthor{\bsnm{Landi}, \binits{P.}},
\bauthor{\bsnm{Minoarivelo}, \binits{H.O.}},
\bauthor{\bsnm{Br{\"a}nnstr{\"o}m}, \binits{{\AA}.}},
\bauthor{\bsnm{Hui}, \binits{C.}},
\bauthor{\bsnm{Dieckmann}, \binits{U.}}:
\batitle{Complexity and stability of ecological networks: a review of the
  theory}.
\bjtitle{Population Ecology}
\bvolume{60},
\bfpage{319}--\blpage{345}
(\byear{2018})
\end{barticle}
\endbibitem

\bibitem[\protect\citeauthoryear{de~Kemmeter et~al.}{2023}]{dekemmeter2023}
\begin{botherref}
\oauthor{\bsnm{Kemmeter}, \binits{J.-F.}},
\oauthor{\bsnm{Gallo}, \binits{L.}},
\oauthor{\bsnm{Boncoraglio}, \binits{F.}},
\oauthor{\bsnm{Latora}, \binits{V.}},
\oauthor{\bsnm{Carletti}, \binits{T.}}:
Complex contagion in social systems with distrust.
arXiv preprint arXiv:2305.03879
(2023)
\end{botherref}
\endbibitem

\bibitem[\protect\citeauthoryear{Muolo et~al.}{2019}]{muolo2019}
\begin{barticle}
\bauthor{\bsnm{Muolo}, \binits{R.}},
\bauthor{\bsnm{Asllani}, \binits{M.}},
\bauthor{\bsnm{Fanelli}, \binits{D.}},
\bauthor{\bsnm{Maini}, \binits{P.K.}},
\bauthor{\bsnm{Carletti}, \binits{T.}}:
\batitle{Patterns of non-normality in networked systems}.
\bjtitle{Journal of theoretical biology}
\bvolume{480},
\bfpage{81}--\blpage{91}
(\byear{2019})
\end{barticle}
\endbibitem

\bibitem[\protect\citeauthoryear{Newman}{2003}]{newman2003}
\begin{barticle}
\bauthor{\bsnm{Newman}, \binits{M.E.J.}}:
\batitle{The structure and function of complex networks}.
\bjtitle{SIAM Review}
\bvolume{45}(\bissue{2}),
\bfpage{167}--\blpage{256}
(\byear{2003})
\doiurl{10.1137/S003614450342480}
\end{barticle}
\endbibitem

\bibitem[\protect\citeauthoryear{Lewis et~al.}{2010}]{Lewis2010}
\begin{barticle}
\bauthor{\bsnm{Lewis}, \binits{H.}},
\bauthor{\bsnm{Newland}, \binits{R.}},
\bauthor{\bsnm{Swinerd}, \binits{G.}},
\bauthor{\bsnm{Saunders}, \binits{A.}}:
\batitle{A new analysis of debris mitigation and removal using networks}.
\bjtitle{Acta Astronautica - ACTA ASTRONAUT}
\bvolume{66},
\bfpage{257}--\blpage{268}
(\byear{2010})
\doiurl{10.1016/j.actaastro.2009.05.010}
\end{barticle}
\endbibitem

\bibitem[\protect\citeauthoryear{Newland}{2012}]{newland2012}
\begin{botherref}
\oauthor{\bsnm{Newland}, \binits{R.J.}}:
Assessing the use of network theory as a method for developing a targeted
  approach to active debris removal.
PhD thesis,
University of Southampton
(2012)
\end{botherref}
\endbibitem

\bibitem[\protect\citeauthoryear{Acciarini and Vasile}{2020}]{Acciarini2020}
\begin{bchapter}
\bauthor{\bsnm{Acciarini}, \binits{G.}},
\bauthor{\bsnm{Vasile}, \binits{M.}}:
\bctitle{A multi-layer temporal network model of the space environment}.
In: \bbtitle{71st International Astronautical Congress}
(\byear{2020}).
\burl{https://strathprints.strath.ac.uk/74276/}
\end{bchapter}
\endbibitem

\bibitem[\protect\citeauthoryear{Acciarini and Vasile}{2021}]{Acciarini2021}
\begin{bchapter}
\bauthor{\bsnm{Acciarini}, \binits{G.}},
\bauthor{\bsnm{Vasile}, \binits{M.}}:
\bctitle{A network-based evolutionary model of the space environment}.
In: \bbtitle{8th European Conference on Space Debris},
\bconflocation{DEU},
pp. \bfpage{1}--\blpage{9}
(\byear{2021}).
\burl{https://strathprints.strath.ac.uk/76218/}
\end{bchapter}
\endbibitem

\bibitem[\protect\citeauthoryear{Wang et~al.}{2023}]{wang2023}
\begin{bchapter}
\bauthor{\bsnm{Wang}, \binits{Y.}},
\bauthor{\bsnm{Wilson}, \binits{C.}},
\bauthor{\bsnm{Vasile}, \binits{M.}}:
\bctitle{Multi-layer temporal network model of the space environment}.
In: \bbtitle{2023 AAS/AIAA Astrodynamics Specialist Conference, Big Sky,
  Montana, USA}
(\byear{2023})
\end{bchapter}
\endbibitem

\bibitem[\protect\citeauthoryear{Stevenson et~al.}{2022}]{stevenson2022}
\begin{bchapter}
\bauthor{\bsnm{Stevenson}, \binits{E.}},
\bauthor{\bsnm{Rodriguez-Fernandez}, \binits{V.}},
\bauthor{\bsnm{Urrutxua}, \binits{H.}}:
\bctitle{Towards graph-based machine learning for conjunction assessment}.
In: \bbtitle{Proceedings of the Advanced Maui Optical and Space Surveillance
  {(AMOS)} Technologies Conference 2022}
(\byear{2022})
\end{bchapter}
\endbibitem

\bibitem[\protect\citeauthoryear{Casanova et~al.}{2014}]{Casanova2014}
\begin{barticle}
\bauthor{\bsnm{Casanova}, \binits{D.}},
\bauthor{\bsnm{Tardioli}, \binits{C.}},
\bauthor{\bsnm{Lemaitre}, \binits{A.}}:
\batitle{Space debris collision avoidance using a three-filter sequence}.
\bjtitle{Monthly Notices of the Royal Astronomical Society}
\bvolume{442},
\bfpage{3235}--\blpage{3242}
(\byear{2014})
\doiurl{10.1093/mnras/stu1065}
\end{barticle}
\endbibitem

\bibitem[\protect\citeauthoryear{McKnight et~al.}{2021}]{mcknight2021}
\begin{barticle}
\bauthor{\bsnm{McKnight}, \binits{D.}},
\bauthor{\bsnm{Witner}, \binits{R.}},
\bauthor{\bsnm{Letizia}, \binits{F.}},
\bauthor{\bsnm{Lemmens}, \binits{S.}},
\bauthor{\bsnm{Anselmo}, \binits{L.}},
\bauthor{\bsnm{Pardini}, \binits{C.}},
\bauthor{\bsnm{Rossi}, \binits{A.}},
\bauthor{\bsnm{Kunstadter}, \binits{C.}},
\bauthor{\bsnm{Kawamoto}, \binits{S.}},
\bauthor{\bsnm{Aslanov}, \binits{V.}},
\bauthor{\bsnm{{Dolado Perez}}, \binits{J.-C.}},
\bauthor{\bsnm{Ruch}, \binits{V.}},
\bauthor{\bsnm{Lewis}, \binits{H.}},
\bauthor{\bsnm{Nicolls}, \binits{M.}},
\bauthor{\bsnm{Jing}, \binits{L.}},
\bauthor{\bsnm{Dan}, \binits{S.}},
\bauthor{\bsnm{Dongfang}, \binits{W.}},
\bauthor{\bsnm{Baranov}, \binits{A.}},
\bauthor{\bsnm{Grishko}, \binits{D.}}:
\batitle{Identifying the 50 statistically-most-concerning derelict objects in
  leo}.
\bjtitle{Acta Astronautica}
\bvolume{181},
\bfpage{282}--\blpage{291}
(\byear{2021})
\doiurl{10.1016/j.actaastro.2021.01.021}
\end{barticle}
\endbibitem

\bibitem[\protect\citeauthoryear{McKnight et~al.}{2017}]{mcknight2017}
\begin{bchapter}
\bauthor{\bsnm{McKnight}, \binits{D.}},
\bauthor{\bsnm{Matney}, \binits{M.}},
\bauthor{\bsnm{Walbert}, \binits{K.}},
\bauthor{\bsnm{Behrend}, \binits{S.}},
\bauthor{\bsnm{Casey}, \binits{P.}},
\bauthor{\bsnm{Speaks}, \binits{S.}}:
\bctitle{Preliminary analysis of two years of the massive collision monitoring
  activity}.
In: \bbtitle{International Astronautical Congress 2017}
(\byear{2017})
\end{bchapter}
\endbibitem

\bibitem[\protect\citeauthoryear{McKnight et~al.}{2018}]{mcknight2018}
\begin{bchapter}
\bauthor{\bsnm{McKnight}, \binits{D.}},
\bauthor{\bsnm{Speaks}, \binits{S.}},
\bauthor{\bsnm{Macdonald}, \binits{J.}},
\bauthor{\bsnm{Ebright}, \binits{K.}}:
\bctitle{Assessing potential for cross-contaminating breakup events from leo to
  geo}.
In: \bbtitle{69th International Astronautical Congress. Presented Paper.
  Bremen, Germany}
(\byear{2018})
\end{bchapter}
\endbibitem

\bibitem[\protect\citeauthoryear{Anselmo and Pardini}{2017}]{anselmo2017}
\begin{bchapter}
\bauthor{\bsnm{Anselmo}, \binits{L.}},
\bauthor{\bsnm{Pardini}, \binits{C.}}:
\bctitle{An index for ranking active debris removal targets in leo}.
In: \bbtitle{7th European Conference on Space Debris. ESA Space Debris Office,
  Darmstadt-ESOC},
pp. \bfpage{18}--\blpage{21}
(\byear{2017})
\end{bchapter}
\endbibitem

\bibitem[\protect\citeauthoryear{Letizia et~al.}{2017}]{letizia2017}
\begin{bchapter}
\bauthor{\bsnm{Letizia}, \binits{F.}},
\bauthor{\bsnm{Colombo}, \binits{C.}},
\bauthor{\bsnm{Lewis}, \binits{H.}},
\bauthor{\bsnm{Krag}, \binits{H.}}:
\bctitle{Extending the {ECOB} space debris index with fragmentation risk
  estimation}.
In: \bbtitle{7th European Conference on Space Debris 2017, {ESA} Space Debris
  Office, Darmstadt-ESOC}
(\byear{2017})
\end{bchapter}
\endbibitem

\bibitem[\protect\citeauthoryear{Rossi et~al.}{2009}]{rossi2009}
\begin{bchapter}
\bauthor{\bsnm{Rossi}, \binits{A.}},
\bauthor{\bsnm{Anselmo}, \binits{L.}},
\bauthor{\bsnm{Pardini}, \binits{C.}},
\bauthor{\bsnm{Jehn}, \binits{R.}},
\bauthor{\bsnm{Valsecchi}, \binits{G.}}:
\bctitle{The new space debris mitigation (sdm 4.0) long term evolution code}.
In: \bbtitle{Proceedings of the Fifth European Conference on Space Debris, ESA
  SP-672, CD-ROM, ESA Communication Production Office, Noordwijk, The
  Netherlands}
(\byear{2009})
\end{bchapter}
\endbibitem

\bibitem[\protect\citeauthoryear{Lewis}{2020}]{lewis2020}
\begin{barticle}
\bauthor{\bsnm{Lewis}, \binits{H.G.}}:
\batitle{Understanding long-term orbital debris population dynamics}.
\bjtitle{Journal of Space Safety Engineering}
\bvolume{7}(\bissue{3}),
\bfpage{164}--\blpage{170}
(\byear{2020})
\end{barticle}
\endbibitem

\bibitem[\protect\citeauthoryear{Ruch and Revelin}{2020}]{ruch2020}
\begin{bchapter}
\bauthor{\bsnm{Ruch}, \binits{V.}},
\bauthor{\bsnm{Revelin}, \binits{B.}}:
\bctitle{Space environment index at cnes}.
In: \bbtitle{8th Satellites End of Life and Sustainable Technologies Workshop,
  CNES},
pp. \bfpage{01}--\blpage{03}
(\byear{2020})
\end{bchapter}
\endbibitem

\bibitem[\protect\citeauthoryear{Muciaccia et~al.}{2023}]{muciaccia2023}
\begin{bchapter}
\bauthor{\bsnm{Muciaccia}, \binits{A.}},
\bauthor{\bsnm{Giudici}, \binits{L.}},
\bauthor{\bsnm{Trisolini}, \binits{M.}},
\bauthor{\bsnm{Colombo}, \binits{C.}},
\bauthor{\bsnm{Campo~Lopez}, \binits{B.}},
\bauthor{\bsnm{Letizia}, \binits{F.}},
\bauthor{\bsnm{Lemmens}, \binits{S.}}, \betal:
\bctitle{Space environment investigation using a space debris index}.
In: \bbtitle{9th Annual Space Traffic Management Conference},
pp. \bfpage{1}--\blpage{8}
(\byear{2023})
\end{bchapter}
\endbibitem

\bibitem[\protect\citeauthoryear{Vallado and Cefola}{2012}]{Vallado2012}
\begin{barticle}
\bauthor{\bsnm{Vallado}, \binits{D.}},
\bauthor{\bsnm{Cefola}, \binits{P.}}:
\batitle{Two-line element sets - practice and use}.
\bjtitle{Proceedings of the International Astronautical Congress, IAC}
\bvolume{7},
\bfpage{5812}--\blpage{5825}
(\byear{2012})
\end{barticle}
\endbibitem

\bibitem[\protect\citeauthoryear{Kelso et~al.}{1988}]{kelso1988}
\begin{botherref}
\oauthor{\bsnm{Kelso}, \binits{T.}},
\oauthor{\bsnm{Hoots}, \binits{F.}},
\oauthor{\bsnm{Roehrich}, \binits{R.}}:
Spacetrack report no. 3 - models for propagation of {NORAD} element sets.
NASA, Tech. Rep
(1988)
\end{botherref}
\endbibitem

\bibitem[\protect\citeauthoryear{Berry}{2014}]{berry2014}
\begin{bchapter}
\bauthor{\bsnm{Berry}, \binits{D.}}:
\bctitle{Using {CCSDS} standards for space situational awareness}.
In: \bbtitle{SpaceOps 2014 Conference},
p. \bfpage{1894}
(\byear{2014})
\end{bchapter}
\endbibitem

\bibitem[\protect\citeauthoryear{Berry and Oltrogge}{2018}]{berry2018}
\begin{bchapter}
\bauthor{\bsnm{Berry}, \binits{D.}},
\bauthor{\bsnm{Oltrogge}, \binits{D.L.}}:
\bctitle{The evolution of the ccsds orbit data messages}.
In: \bbtitle{2018 SpaceOps Conference},
p. \bfpage{2456}
(\byear{2018})
\end{bchapter}
\endbibitem

\bibitem[\protect\citeauthoryear{Hoots et~al.}{1984}]{Hoots1984}
\begin{barticle}
\bauthor{\bsnm{Hoots}, \binits{F.R.}},
\bauthor{\bsnm{Crawford}, \binits{L.L.}},
\bauthor{\bsnm{Roehrich}, \binits{R.L.}}:
\batitle{An analytic method to determine future close approaches between
  satellites}.
\bjtitle{Celestial Mechanics}
\bvolume{33},
\bfpage{143}--\blpage{158}
(\byear{1984})
\doiurl{10.1007/BF01234152}
\end{barticle}
\endbibitem

\bibitem[\protect\citeauthoryear{Vallado et~al.}{2006}]{Vallado2006b}
\begin{bchapter}
\bauthor{\bsnm{Vallado}, \binits{D.}},
\bauthor{\bsnm{Crawford}, \binits{P.}},
\bauthor{\bsnm{Hujsak}, \binits{R.}},
\bauthor{\bsnm{Kelso}, \binits{T.S.}}:
\bctitle{Revisiting spacetrack report no. 3: Rev 2}.
(\byear{2006})
\end{bchapter}
\endbibitem

\bibitem[\protect\citeauthoryear{Vallado and Crawford}{2008}]{vallado2008}
\begin{bchapter}
\bauthor{\bsnm{Vallado}, \binits{D.}},
\bauthor{\bsnm{Crawford}, \binits{P.}}:
\bctitle{{SGP4} orbit determination}.
In: \bbtitle{AIAA/AAS Astrodynamics Specialist Conference and Exhibit},
p. \bfpage{6770}
(\byear{2008})
\end{bchapter}
\endbibitem

\bibitem[\protect\citeauthoryear{Seago and Vallado}{2000}]{seago2000}
\begin{bchapter}
\bauthor{\bsnm{Seago}, \binits{J.}},
\bauthor{\bsnm{Vallado}, \binits{D.}}:
\bctitle{Coordinate frames of the us space object catalogs}.
In: \bbtitle{Astrodynamics Specialist Conference},
p. \bfpage{4025}
(\byear{2000})
\end{bchapter}
\endbibitem

\bibitem[\protect\citeauthoryear{Gronchi}{2005}]{Gronchi2005}
\begin{barticle}
\bauthor{\bsnm{Gronchi}, \binits{G.F.}}:
\batitle{An algebraic method to compute the critical points of the distance
  function between two keplerian orbits}.
\bjtitle{Celestial Mechanics and Dynamical Astronomy}
\bvolume{93}(\bissue{1}),
\bfpage{295}--\blpage{329}
(\byear{2005})
\end{barticle}
\endbibitem

\bibitem[\protect\citeauthoryear{Bini}{1996}]{bini1996}
\begin{barticle}
\bauthor{\bsnm{Bini}, \binits{D.A.}}:
\batitle{Numerical computation of polynomial zeros by means of aberth's
  method}.
\bjtitle{Numerical algorithms}
\bvolume{13}(\bissue{2}),
\bfpage{179}--\blpage{200}
(\byear{1996})
\end{barticle}
\endbibitem

\bibitem[\protect\citeauthoryear{Newman}{2010}]{newman2010}
\begin{bbook}
\bauthor{\bsnm{Newman}, \binits{M.}}:
\bbtitle{{Networks: An Introduction}}.
\bpublisher{Oxford University Press},
\blocation{Oxford}
(\byear{2010}).
\doiurl{10.1093/acprof:oso/9780199206650.001.0001} .
\bcomment{1st edition}.
\burl{https://doi.org/10.1093/acprof:oso/9780199206650.001.0001}
\end{bbook}
\endbibitem

\bibitem[\protect\citeauthoryear{Wang et~al.}{2017}]{wang2017}
\begin{barticle}
\bauthor{\bsnm{Wang}, \binits{Y.}},
\bauthor{\bsnm{Ghumare}, \binits{E.}},
\bauthor{\bsnm{Vandenberghe}, \binits{R.}},
\bauthor{\bsnm{Dupont}, \binits{P.}}:
\batitle{{Comparison of Different Generalizations of Clustering Coefficient and
  Local Efficiency for Weighted Undirected Graphs}}.
\bjtitle{Neural Computation}
\bvolume{29}(\bissue{2}),
\bfpage{313}--\blpage{331}
(\byear{2017})
\end{barticle}
\endbibitem

\bibitem[\protect\citeauthoryear{}{2013}]{ccsds2013}
\begin{botherref}
Conjunction Data Message, CCSDS 508.0-B-1, Blue Book, Issue 1
(2013).
\url{https://public.ccsds.org/Pubs/508x0b1e2c2.pdf}
\end{botherref}
\endbibitem

\end{thebibliography}

\end{document}